\documentclass[journal=jctcce]{achemso}
\setkeys{acs}{maxauthors=5,etalmode=truncate,chaptertitle =true,articletitle=true}
\usepackage[utf8]{inputenc}
\usepackage{amsmath}
\usepackage{natbib}
\usepackage{xcolor}
\usepackage{graphicx}
\usepackage{csvsimple}
\usepackage{siunitx}

\usepackage{subcaption}
\newcommand{\bra}[1]{\left\langle #1 \right|}
\newcommand{\ket}[1]{\left|#1\right\rangle}

\title{Third order M\o ller-Plesset theory made more useful? The role of density functional theory orbitals}

\author{Adam Rettig}
\email{adam_rettig@berkeley.edu}
\altaffiliation{These authors contributed equally to this work.}
\affiliation
{{Kenneth S. Pitzer Center for Theoretical Chemistry, Department of Chemistry, University of California, Berkeley, California 94720, USA}}
\alsoaffiliation{Chemical Sciences Division, Lawrence Berkeley National Laboratory, Berkeley, California 94720, USA}

\author{Diptarka Hait}
\email{diptarka@berkeley.edu}
\altaffiliation{These authors contributed equally to this work.}
\affiliation
{{Kenneth S. Pitzer Center for Theoretical Chemistry, Department of Chemistry, University of California, Berkeley, California 94720, USA}}
\alsoaffiliation{Chemical Sciences Division, Lawrence Berkeley National Laboratory, Berkeley, California 94720, USA}

\author{Luke W. Bertels}
\affiliation
{{Kenneth S. Pitzer Center for Theoretical Chemistry, Department of Chemistry, University of California, Berkeley, California 94720, USA}}
\altaffiliation{Current address: Department of Chemistry, Virginia Tech, Blacksburg, VA 24061, USA}
\author{Martin Head-Gordon}
\email{mhg@cchem.berkeley.edu}
\affiliation
{{Kenneth S. Pitzer Center for Theoretical Chemistry, Department of Chemistry, University of California, Berkeley, California 94720, USA}}
\alsoaffiliation{Chemical Sciences Division, Lawrence Berkeley National Laboratory, Berkeley, California 94720, USA}

\begin{document}

\begin{abstract}
The practical utility of M\o ller-Plesset (MP) perturbation theory is severely constrained by the use of Hartree-Fock (HF) orbitals. It has recently been shown that use of regularized orbital-optimized MP2 orbitals and scaling of MP3 energy could lead to a significant reduction in MP3 error (J. Phys. Chem. Lett. 10, 4170, 2019). In this work we examine whether density functional theory (DFT) optimized orbitals can be similarly employed to improve the performance of MP theory at both the MP2 and MP3 levels. We find that use of DFT orbitals leads to significantly improved performance for prediction of thermochemistry, barrier heights, non-covalent interactions, and dipole moments relative to standard HF based MP theory. Indeed MP3 (with or without scaling) with DFT orbitals is found to surpass the accuracy of coupled cluster singles and doubles (CCSD) for several datasets. We also found that the results are not particularly functional sensitive in most cases, (although range-separated hybrid functionals with low delocalization error perform the best). 
MP3 based on DFT orbitals thus appears to be an efficient, non-iterative $O(N^6)$ scaling wave function approach for single-reference electronic structure computations. Scaled MP2 with DFT orbitals is also found to be quite accurate in many cases, although modern double hybrid functionals are likely to be considerably more accurate. 
\end{abstract}

\maketitle

\section{Introduction}
Perturbative approaches offer a straightforward route for inexpensively improving predictions from existing quantum chemistry approximations\cite{moller1934note,cremer2011moller,Raghavachari1989,andersson1990second,andersson1992second,harrison1991approximating,angeli2001introduction,angeli2001n,gwaltney2001second,guo2018perturbative}.  Perhaps the best known perturbation technique is M\o ller-Plesset (MP) theory\cite{moller1934note,szabo2012modern,cremer2011moller}, which acts upon a single Slater determinant $\ket{\Phi}$. The zeroth-order Hamiltonian in MP theory is the mean-field one-particle Fock operator $\mathbf{F}$ corresponding to $\ket{\Phi}$ while the remaining terms of the true many-body Hamiltonian $\mathbf{H}$ are treated as a perturbative fluctuation potential $\mathbf{U}$. The resulting theory is size-consistent at all orders in the fluctuation\cite{szabo2012modern}. The energies at all orders are also invariant to rotations between degenerate orbitals, which is not typically true for many other perturbation theories (like Epstein-Nesbet theory\cite{epstein1926,nesbet1955}). Such alternative theories can consequently predict dramatically different energies for different orbital representations\cite{malrieu1979possible}, making MP theory the preferred single reference perturbative approach.

Typically, the reference zeroth-order wave function $\ket{\Phi}$ is a solution to the Hartree-Fock (HF) equations, which leads to the sum of the zeroth and first-order (i.e. MP0 and MP1) energies to be equal to the HF energy $E_\textrm{HF}$. Correlation effects thus arise from higher-order terms by definition, with second-order MP2 perhaps being the simplest wave function based dynamical correlation approach. The historical popularity of MP2 owes a great deal to its relatively low computational cost, which is asymptotically dominated by the formally $O(N^5)$ scaling cost\cite{cremer2011moller} of building two electron integrals in the molecular orbital (MO) basis ($N$ being the number of basis functions).  MP2 like expressions also arise in density functional theory (DFT) via G{\"o}rling-Levy perturbation theory\cite{gorling1994exact}. This has led to development of double hybrid density functionals\cite{grimme2006semiempirical,zhang2009doubly,peverati2013orbital,goerigk2014double} (employing MP2-like expressions within the exchange-correlation contribution) that are amongst the most accurate DFT approximations known to date\cite{mardirossian2018survival,santra2019minimally}.

Higher order terms in the MP series are however not as widely used in practice. This is largely a consequence of the slow convergence of the MP series for even apparently single reference problems\cite{cremer2011moller}. Spin-unrestricted MP theory has long been known to converge extremely slowly, on account of spin-contamination in the UHF orbitals\cite{handy1985convergence,nobes1987slow,gill1988does}. However, oscillatory behavior around the exact value and even divergent behavior are known to arise for closed-shell systems without spin-symmetry breaking\cite{olsen1996surprising,leininger2000mo,cremer1996sixth}, even for systems as simple as a Ne atom. In practice therefore, predictions from MP theory are not guaranteed to be systematically improvable beyond MP2. However, addition of only a fraction of the third-order energy to MP2 is empirically often found to be effective, leading to an equal interpolation between MP2 and MP3 to gain some popularity as ``MP2.5" \cite{Pitonak2009}. Nonetheless, MP3 based approaches are less popular than coupled-cluster singles and doubles (CCSD), despite the latter requiring multiple $O(N^6)$ iterations (vs a single $O(N^6)$ scaling step for MP3). This is a consequence of  CCSD being also exact to third-order in $\mathbf{U}$ and much better behaved in practice. MP4 is even less competitive due to $O(N^7)$ scaling, although its form inspired the development of the (T) triples correction\cite{Raghavachari1989} to CCSD, with the resulting CCSD(T) method being widely considered to be the ``gold-standard" of single-reference quantum chemistry. 

In practice, the use of MP2 has also been hindered by the use of reference HF orbitals\cite{cremer2011moller}. HF has a propensity for artificially breaking spin-symmetry even for mostly single reference problems, leading to extremely slow convergence of the spin-unrestricted MP series (and subsequent poor performance of MP2)\cite{handy1985convergence,nobes1987slow,gill1988does}. HF also overly localizes electron density due to lack of correlation\cite{li2017piecewise,hait2018delocalization}, which can adversely affect MP2 performance by providing it with a poor starting point. 
This has led to the development of orbital optimized MP2 (OOMP2) approaches\cite{lochan2007orbital,neese2009assessment,kurlancheek2012exploring,bozkaya2011quadratically,bozkaya2014orbital}, where the orbitals are optimized in the presence of MP2 correlation to ameliorate artificial symmetry breaking or overlocalization errors in the reference Slater determinant. Indeed, OOMP2 has been shown to both significantly reduce spin-contamination and greatly improve upon results relative to standard MP2 for systems where HF solutions are heavily spin-contaminated
\cite{bozkaya2014analytic}. In addition, spin-component scaled MP2 methods\cite{grimme2003improved,jung2004scaled,lochan2005scaled} have also been developed to improve prediction quality, though such approaches cannot be readily extended to spin-general orbitals. Extensions of both orbital optimization and spin-component scaling have also been considered at the MP3 level\cite{grimme2003improvedmp3,bozkaya2011orbital,soydas2013assessment,soydas2015assessment} .

Very recently, some of us have shown that orbitals obtained from a regularized OOMP2 (namely $\kappa-$OOMP2\cite{lee2018regularized}) could be employed to dramatically improve performance of MP3\cite{bertels2019third}. Specifically, use of $\kappa$-OOMP2 orbitals and scaling the third-order energy $E_\textrm{MP3}$ by 0.8 greatly improved  prediction of thermochemistry, barrier heights and non-covalent interactions by over a factor of 3 in most cases (relative to standard MP3/MP2.5), leading to better performance than CCSD over several datasets. 
The success of this MP2.8:$\kappa-$OOMP2 approach raises the question as to whether this improvement is largely the consequence of optimizing orbitals in the presence of MP2 correlation, or if other high quality reference orbitals would yield similar results (with or without scaling). Specifically, it is interesting to consider whether DFT approximations with a low penchant for spin-contamination\cite{sherrill1999performance} and tunable delocalization error\cite{perdew1982density,mori2006many,hait2018accurate} can yield reasonable results. DFT orbitals have been shown to improve performance of CC approaches in many difficult cases\cite{beran2003approaching,fang2016use,fang2017prediction,bertels2020polishing} and it seems plausible that they would have an even larger impact on MP theory due to lack of iterative singles amplitudes (that mimic orbital relaxation in projected CC theories) in the latter. We consequently examine the performance of both scaled and unscaled MP2 and MP3 with DFT orbitals in this work, focusing on the ability to predict chemically relevant energy differences and dipole moments.%

\section{Methods}
We computed MP energies with DFT orbitals (MP:DFT) via the following protocol:
\begin{enumerate}
    \item A DFT calculation was run to obtain converged (spin-unrestricted) orbitals, with a stability analysis to guarantee that a minimum in orbital space is reached. 
    \item $\mathbf{F}$ was built from the converged  DFT density matrix, using the HF functional. The HF energy $E_\textrm{HF}$ is also found similarly. 
    \item $\mathbf{F}$ was semi-canonicalized by diagonalizing the occupied-occupied and virtual-virtual blocks separately. Note that the reference Slater determinant/density matrix is unaffected by this process, and the occupied-virtual block of $\mathbf{F}$ remains non-zero in general. 
    \item MP2 and MP3 energies were computed using the semi-canonicalized orbitals $\ket{\phi_p}$ and the corresponding orbital ``energies" $\epsilon_p=\bra{\phi_p}\mathbf{F}\ket{\phi_p}$. The effects of the occupied-virtual block of $\mathbf{F}$ was computed at the MP2 level (the so-called non-Brillouin singles contribution) but not the MP3 level, similar to Ref \citenum{bertels2019third}. Complete exclusion of singles was considered, but is not presented due to suboptimal behavior, as discussed later (and demonstrated in the Supporting Information). 
\end{enumerate}
This approach is superficially similar to xDH double hybrid functionals\cite{zhang2009doubly,su2016xyg3}, where a lower rung functional is employed to generate reference orbitals and a higher rung functional (here MP2/MP3 with/without scaling) is used to compute the final energy from the previously converged orbitals (and their energies) without further optimization. However, we stress that the lower rung orbital energies were not used to compute the MP energies in our protocol, unlike typical xDH functionals. The entire purpose of the DFT calculation here is to provide a good set of occupied orbitals, which are then used to form a HF derived $\mathbf{F}$ operator from which semi-canonical orbitals (and their energies) can be obtained for MP theory. Our energy functional is thus purely wave function based, albeit acting upon a DFT generated reference Slater determinant. This distinction is important due to the relatively small fundamental gaps predicted by many density functionals (leading to MP overcorrelation in a normal xDH approach\cite{goerigk2011efficient}), whose effect should be reduced by use of $\mathbf{F}$ orbital energies obtained from HF. 

Based on previous success of spin-component scaled MP2s, MP2.5 and the results of Ref \citenum{bertels2019third}, we consider four MP models in the present work:
\begin{enumerate}
    \item MP2: $E=E_\textrm{HF}+E_\textrm{MP2}$.
    \item Scaled MP2 (sMP2): $E=E_\textrm{HF}+c_2E_\textrm{MP2}$ \label{scamp2}
    \item MP3: $E=E_{HF}+E_{MP2}+E_{MP3}$
    \item Scaled MP3 (sMP3): $E=E_\textrm{HF}+E_\textrm{MP2}+c_3E_\textrm{MP3}$ \label{scamp3}
\end{enumerate}
where the scaling parameters $c_{2,3}$ are found from fitting to the training set (here the non multi-reference subset of the W4-11\cite{karton2011w4} thermochemistry dataset).

A fifth model with $E=E_\textrm{HF}+c_2E_\textrm{MP2}+c_3E_\textrm{MP3}$ was also considered, but yielded results very similar to model \ref{scamp3} and was thus not considered further (since it involves two empirical parameters, as opposed to one for model \ref{scamp3}). Further information about this approach is supplied in the supporting information. 

We investigated the performance of MP:DFT in two regimes. The first was an extensive assessment of performance vs CCSD(T) benchmarks for reasonably sized triple zeta basis sets (similar to to Ref \citenum{bertels2019third}). The relative computational inexpensiveness of this approach permitted us to assess orbitals obtained from several density functionals like SPW92\cite{Slater,PW92}, PBE\cite{PBE}, BLYP\cite{b88,lyp}, B97M-V\cite{b97mv}, SCAN\cite{SCAN}, revM06-L\cite{revm06l}, TPSS\cite{tpss}, B3LYP\cite{b3lyp}, PBE0\cite{pbe0}, MN15\cite{MN15}, CAM-B3LYP\cite{camb3lyp}, $\omega$B97X-V\cite{wb97xv} and $\omega$B97M-V\cite{wB97MV}. We also considered the behavior of pure Slater exchange\cite{Slater} and HFLYP (100\% HF exchange + 100\% LYP correlation\cite{lyp}) over the training set, to understand limiting behavior.
The relevant ground state energy calculations were done using the aug-cc-pVTZ basis set \cite{dunning1989, Kendall1992, Woon1993}, while the aug-cc-pCVTZ\cite{dunning1989, Kendall1992,woon1995gaussian} basis set was employed for dipole moments, in order to be consistent with the CCSD(T)/aug-cc-pCVTZ reference numbers reported in Ref \citenum{hait2018accurate}. The dipole moments were computed via a centered two point finite difference formula (the same protocol as Ref \citenum{hait2018accurate}), using a field strength of magnitude $10^{-4}$ a.u.
The frozen core approximation was not employed for any of these calculations.

The second regime entailed assessment against benchmarks at the complete basis set (CBS) limit, for direct comparison to DFT approaches. Only MP:$\omega$B97M-V and MP:HF were examined, with the CBS limit being estimated via the following protocol:
\begin{enumerate}
    \item The aug-cc-pV5Z basis $E_\textrm{HF}$ was treated as the CBS limit. 
    \item CBS limit $E_\textrm{MP2}$ and  $E_\textrm{MP3}$ were found from basis set extrapolation using frozen-core aug-cc-pVTZ and aug-cc-pVQZ results. The $E_{X}=E_{\infty}+AX^{-3}$ extrapolation formula\cite{helgaker1997basis,halkier1998basis} (where $X$ is basis set cardinality) was employed.
    \item The frozen-core correction to $E_\textrm{MP2}$ and  $E_\textrm{MP3}$ were estimated from aug-cc-pCVTZ basis results. Lack of aug-cc-pCVTZ for Br necessitated the use of aug-cc-pwCVQZ basis (obtained from the basis set exchange\cite{pritchard2019new}) for Br (and aug-cc-pCVQZ for other atoms in the relevant systems). The K shell of third period elements was kept frozen throughout (K and L shells for Br).
    \item Orbital stability analysis was initially only carried out at the aug-cc-pVTZ level. Stability analysis in larger basis sets was only repeated for species that were found to initially yield unstable solutions at the aug-cc-pVTZ level. 
\end{enumerate}

All calculations were done with a development version of the Q-Chem 5.2 package \cite{Shao2015}. The RI approximation\cite{feyereisen1993use,bernholdt1996large} was not employed for any MP calculations.
\newpage 
\section{Triple Zeta Basis Results}
\subsection{Training Set}

\begin{table}[htb!]
    \centering
    \begin{tabular}{| c | S[round-precision=4] S[round-precision=4] |}
        \hline
        \textbf{Name} & \textbf{c$_{2}$} (sMP2/Model 2) & \textbf{c$_{3}$} (sMP3/Model 4) \\
        \hline
        \begin{filecontents*}{newTables/W4.csv}
Name,mptwormse,mptwomse,mptwomaxae,scamptwoctwo,scamptwormse,scamptwomse,scamptwomaxae,mpthreermse,mpthreemse,mpthreemaxae,scampthreecthree,scampthreermse,scampthreemse,scampthreemaxae,scamptwothreectwo,scamptwothreecthree,scamptwothreermse,scamptwothreemse,scamptwothreemaxae,nbsmptwormse,nbsmptwomse,nbsmptwomaxae,fitnbsmptwoctwo,fitnbsmptwormse,fitnbsmptwomse,fitnbsmptwomaxae,nbsmpthreermse,nbsmpthreemse,nbsmpthreemaxae,fitnbsmpthreecthree,fitnbsmpthreermse,fitnbsmpthreemse,fitnbsmpthreemaxae,scamptwothreermse,scamptwothreemse,scamptwothreemaxae
CCSD,4.925731799416142,1.4749400263575576,20.337805143340834,1.0,4.925731799416142,1.4749400263575576,20.337805143340834,4.925731799416142,1.4749400263575576,20.337805143340834,1.0,4.925731799416142,1.4749400263575576,20.337805143340834,1.0,1.0,4.925731799416142,1.4749400263575576,20.337805143340834,4.925731799416142,1.4749400263575576,20.337805143340834,1.0,4.925731799416142,1.4749400263575576,20.337805143340834,4.925731799416142,1.4749400263575576,20.337805143340834,1.0,4.925731799416142,1.4749400263575576,20.337805143340834,4.925731799416142,1.4749400263575576,20.337805143340834
$\kappa$-OOMP2,12.433685864201806,-3.523254887319555,56.809971092651026,0.8464673087119536,6.2465337768437506,-0.2288679756713835,32.34937833939076,3.219962445794366,0.3244076065798256,14.628050551701973,0.8146729937323157,1.580059500357701,-0.388668164542999,5.829690064734737,0.999706851421268,0.8134879848281474,1.5800272972635467,-0.38693752115436286,5.838660535913095,119.2875263867597,47.60421419474615,767.9946607048855,-0.11674400625509639,58.60107670472627,14.470157206555985,223.52545787099083,124.56614948293459,51.451876688645534,772.0550831911908,-2.3082601845291157,114.05413694539827,38.722808056572205,758.6221491473804,1.5800272972635467,-0.38693752115436286,5.838660535913095
HF,11.962588500086019,-4.983376646619883,51.13784821562325,0.9035027003907019,10.35620511321548,-2.684333599672804,43.41806735472642,9.167628964430959,-0.8630887404273635,38.56817778616742,0.715722412916153,8.544090697322503,-2.0343942444905285,39.51664048356325,1.0641005962608758,1.0043602927325923,8.280544314637824,-2.3723163863437065,37.563885331012806,11.962588500086019,-4.983376646619883,51.13784821562325,0.9035027003907019,10.35620511321548,-2.684333599672804,43.41806735472642,9.167628964430959,-0.8630887404273635,38.56817778616742,0.715722412916153,8.544090697322503,-2.0343942444905285,39.51664048356325,8.280544314637824,-2.3723163863437065,37.563885331012806
HFLYP,9.235731481810651,-3.8693073092378762,34.95250230358613,0.9042812345199057,7.033584466386489,-1.7252782618360045,29.7988183399779,6.57660584462643,0.018895104591715974,25.739057485479705,0.6465651156913,5.033763426091592,-1.3553312657089542,24.244327149427434,1.035142516600937,0.802792532726159,4.89974667766717,-1.53505364045936,23.54285422043371,9.335283798403417,-3.9440919177448537,35.52289687767684,0.9029978422168554,7.089260515363604,-1.7640615703385716,29.956657885869667,6.549804753584234,-0.055889503915260846,26.111496501657903,0.6542433779176992,5.0750610750276985,-1.4002612364932292,24.64923871426311,4.89974667766717,-1.53505364045936,23.54285422043371
Slater,15.260228320430992,-5.518176136133596,72.91970217030772,0.815678644718363,6.820372353579381,-0.8745113020788665,33.344405033494766,3.43724982148834,-0.6408258528067862,14.04210774784713,0.8702933020912215,2.6063589811677854,-1.2734508526015518,8.417499686436148,0.9781772177304242,0.7875515073107686,2.494936219923938,-1.1272234748024554,8.888516636818315,12.088279619746032,-3.639595614609154,53.38637630332038,0.8515930091111165,5.911905364131692,-0.17952693773879108,33.218940422852185,5.873592141208544,1.2377546687176573,28.690765506842475,0.6870274482019416,2.2920816732681235,-0.28872209546811806,10.674347030100941,2.494936219923938,-1.1272234748024554,8.888516636818315
SPW92,15.373683905950479,-5.063822432858306,72.85916010197221,0.8158034083993938,6.743301663584762,-0.5870304493552916,34.19632681961934,3.0958949835574097,-0.3909481724116171,12.556611967249099,0.8752837226645509,2.2102834785326753,-0.9737316546311675,7.866482140973352,0.9757028800547718,0.7828784034311225,2.0356165829775317,-0.8150025975815046,7.470636939040394,11.480229089814358,-3.2932163091860165,51.08576258417799,0.8610772841541888,5.834214605808526,-0.16275719034517255,32.05206597902457,6.335991757179205,1.3796579512606721,30.470389878748293,0.6516802052668782,1.867720660507188,-0.2479966519518067,10.546573757098702,2.0356165829775317,-0.8150025975815046,7.470636939040394
PBE,14.981198917865683,-4.935714280538554,70.9378028450585,0.820682808994674,6.8190609093099726,-0.5745483804077248,35.028144976019405,3.0428919518852093,-0.3051573992811024,12.602470324973526,0.8733298104260431,2.1529795523664386,-0.8917109172629746,7.383528029702884,0.9827805719056077,0.8069102673222418,2.065440269187329,-0.7804773993785364,7.297689748350558,11.199381129109788,-3.313559442507111,49.67563985885659,0.865055937243961,5.868825129117929,-0.2504907022984059,30.453483405562768,6.239780694502457,1.3169974387503403,30.130615906288302,0.6500615798730367,1.9091285379018414,-0.303412320584931,9.303768429279327,2.065440269187329,-0.7804773993785364,7.297689748350558
BLYP,15.279296967199642,-4.876730683465962,71.35091712008352,0.81736662955277,6.768810956048645,-0.46891220547414536,35.39254994765644,2.978269551677964,-0.2395704028629668,11.903751603635202,0.8764567973487131,2.0778312072557426,-0.8124600351360006,7.10843990153728,0.9771300148317378,0.7891399707993368,1.9103672189874745,-0.6653998111957142,7.135718923396645,11.277469657813592,-3.307689892456274,50.862581038935005,0.8650075109007275,5.950035228390771,-0.2614829166277738,30.234035112882818,6.393225710832255,1.3294703881467222,29.908269010767274,0.6446710363322812,1.7921422626961832,-0.3182469687210486,8.882610492783373,1.9103672189874745,-0.6653998111957142,7.135718923396645
B97M-V,12.704954287363808,-4.193232146614787,59.46233868409315,0.8410834375576256,6.049360172718326,-0.5895897992652823,30.119212780080122,3.320288662399289,0.11684720429508964,14.422968060305424,0.8188868200975173,1.8136537090250906,-0.663764972580226,6.424422900406756,0.9873679990419192,0.7684979053106392,1.7604428181578975,-0.5944979355801266,6.478495144914561,10.66298179598106,-3.604699918884492,47.47997557671362,0.8643533803263206,5.230888450737767,-0.6085665777157411,27.685172401707018,5.216199537938986,0.7053794320253829,24.10139422362661,0.6833780031369736,1.8897855418588616,-0.6592864986977985,6.239342131541784,1.7604428181578975,-0.5944979355801266,6.478495144914561
SCAN,13.891600140173885,-4.869438852003252,64.7067451516036,0.8331183896356256,6.770929775748126,-0.8074948641624947,33.73181321468305,3.2718094816874768,-0.30289909967490297,13.354401809491641,0.8618961283074362,2.429554060134439,-0.9335559197094493,10.724882999828546,0.9905124020456499,0.8238834657339318,2.4065127861828475,-0.8762115282500803,10.559652740782166,10.931207506147175,-3.7198944416911113,48.6591893600079,0.8704825170035271,6.176171706441478,-0.7162896203725866,29.046538894839426,5.794911115810369,0.8466453106372371,26.12594019003064,0.6705794750415287,2.5007766707691568,-0.657666611818495,11.911179094301833,2.4065127861828475,-0.8762115282500803,10.559652740782166
revM06-L,12.84218179319068,-4.413677801863087,59.457764143456885,0.8435588643996219,6.265002418151078,-0.6824325059824956,32.12921992297765,3.027856148814613,-0.006329291616754101,12.869745233329324,0.8448507292497919,1.9325716561885362,-0.6901261989234886,5.9130219981023435,0.9859957978346802,0.7872667698020502,1.864780973718095,-0.6099074360495057,5.916843729025967,10.285677943718467,-3.2258675707219657,47.50746021988383,0.8798489516434539,6.13389031663538,-0.5028863138879216,33.4304342835281,5.325644629391395,1.1814809395243668,23.477740513443013,0.6714546909966413,1.9972859303122041,-0.2665327386600068,10.598836239238022,1.864780973718095,-0.6099074360495057,5.916843729025967
TPSS,14.552539777053635,-4.885399387594321,69.57511685083254,0.826272078070642,6.850295960406964,-0.6357000148445864,35.531948289408795,3.005786087890889,-0.2825859932014559,12.508312599504428,0.8721346748630083,2.1404156599488506,-0.8711262244204003,7.631747701329184,0.9879284062650978,0.8246924857402808,2.097306623360167,-0.794200738772876,7.661769934709073,10.803877390112124,-3.302891816125609,48.04536162495933,0.8708525227641053,5.868655742266321,-0.34808766015082687,29.458013109420197,6.170964226107494,1.299921578267256,30.129872747355105,0.6443515252948688,1.904190840959026,-0.33706198480091365,8.172915973978558,2.097306623360167,-0.794200738772876,7.661769934709073
B3LYP,13.532904719333304,-4.40513427104064,62.888217794516386,0.8332480584970132,6.3557019173525315,-0.5452283325799906,32.6749740861649,3.1952532527598048,-0.059343837776408455,13.43100183310588,0.8398125607846926,1.9193273149084342,-0.7554848786473873,6.620947356115285,0.985889733872835,0.7844474257318831,1.8554189042971962,-0.6694714669211554,6.568754681453097,11.038297900050082,-3.5325079699409434,50.35715592296665,0.864406277384789,5.829948249448086,-0.5121624242155739,29.368983086830237,5.3515946019895235,0.8132824633232876,23.926217430675997,0.6831697279390064,1.7606118265253192,-0.5635955019678821,6.645796867634424,1.8554189042971962,-0.6694714669211554,6.568754681453097
PBE0,12.872590919941857,-4.421469341061599,60.54098272948711,0.8408661269036983,6.326255574876727,-0.7138504759976622,31.74449285175278,3.3559803685290404,-0.11551696102006338,14.454318409953306,0.8275072826927052,2.060871991873483,-0.8582623876492409,7.56753573254617,0.9934895720697015,0.801409125486027,2.04876280926125,-0.818955036808202,7.813338717223464,10.817929987971196,-3.605212401693566,48.93246450944937,0.8664850094751531,5.7928834586750675,-0.6034637793033096,28.459004508044465,5.093914751798148,0.7007399783479701,23.05711872067963,0.6931328984067137,1.9355733682417045,-0.6206151481140889,7.787290150085472,2.04876280926125,-0.818955036808202,7.813338717223464
MN15,12.96008685031245,-4.229062742057382,60.82677542840899,0.8335709551580148,5.919655248295818,-0.5410829523956822,29.8892622507189,3.2542956789647155,0.030159447057866336,13.828334256826091,0.82810795985263,1.872036561657268,-0.7019669444701013,6.461896705052947,0.9814117416972095,0.7553223224612651,1.7654485174978305,-0.60007115350045,6.445025398057532,11.61276460607001,-4.042330516663668,52.223461815289276,0.8484607686244003,5.3176390398036615,-0.7125983922334139,30.80584989173292,4.57508967876564,0.21689167245158159,20.15579822874181,0.7375304016822691,2.099797867184729,-0.9010246646714636,8.357527717852992,1.7654485174978305,-0.60007115350045,6.445025398057532
CAM-B3LYP,12.940523085451725,-4.181815551092118,59.76760926245283,0.8380490981514073,6.1015751507716605,-0.5150401050968533,31.235799394191183,3.2895221004376447,0.0012389638799704897,14.026991056048187,0.8260349322658218,1.8844521463513053,-0.7264663981529088,6.42075564173388,0.9864503502327602,0.7722200502921736,1.826224197716342,-0.6447955913903272,6.369893506917024,10.975220177196785,-3.6107660329710214,50.44268263865349,0.8630743452661601,5.707083990951708,-0.5887855015556389,28.593914934788085,4.997660942338786,0.5722884820010691,21.7003412257381,0.6987372223413488,1.7818480374883012,-0.6879101402769852,6.319005855155341,1.826224197716342,-0.6447955913903272,6.369893506917024
$\omega$B97X-V,12.176746965823062,-3.831197676912631,56.605409433834446,0.8463915656003662,5.971491276500297,-0.4724634606389148,30.47181870245534,3.464607010703511,0.14607452674127924,15.133954582578824,0.8022634143481732,1.7859367099409598,-0.6403776990171625,6.419017559549565,0.9924336715779628,0.771966369316569,1.7672352997999714,-0.5954352969469946,6.398238428306927,10.773860693082407,-3.5514769939954274,49.55141635747646,0.8639614786364143,5.585407322196227,-0.6149715280441025,27.791050519993398,4.721840752992639,0.42579520965848305,20.380327286779416,0.7080021741079975,1.754373762510256,-0.7355596267891524,6.484255131705332,1.7672352997999714,-0.5954352969469946,6.398238428306927
$\omega$B97M-V,12.249891292257756,-3.8361934403085476,56.68702952940034,0.8411715835870585,5.679158658870137,-0.4570978907836279,28.13984507417244,3.5257221724461703,0.14405287804927427,15.351260949281276,0.8012190763875461,1.8448363510284191,-0.6471441613189628,6.41068682849841,0.9842918427445446,0.7390373974350783,1.7704340410552115,-0.5604494384808808,6.378921259334457,11.3006524581234,-3.9662423370440627,51.082611547712595,0.8502886415942045,5.004432023218946,-0.7616436045177779,22.33157120343984,4.5873165200903685,0.014003981313758445,19.750468014579955,0.7334478345717252,2.1933661429513043,-1.0469392937824373,10.134601070884228,1.7704340410552115,-0.5604494384808808,6.378921259334457
\end{filecontents*}
\csvreader[late after line=\\, head to column names, filter not strcmp={\Name}{CCSD}]{newTables/W4.csv}{}{\Name & \scamptwoctwo & \scampthreecthree}
        \hline
    \end{tabular}
    \caption{Scaling coefficients fit over the nonMR portion of the W4-11 dataset for the scaled MP2 (c$_2$) and scaled MP3 (c$_3$) models.}
    \label{tab:fit_coeffs}
\end{table}

\subsubsection{Scaling Parameters}
Scaling parameters for the sMP2 (model \ref{scamp2}) and sMP3 (model \ref{scamp3}) approaches were obtained via least-squares fitting on CCSD(T) values of the non-multireference (nonMR) subset of the W4-11 dataset \cite{karton2011w4}. The resulting scaling parameters, for orbitals resulting from various functionals, are enumerated in Table \ref{tab:fit_coeffs}. Fitted $c_2$ parameters mostly range between 0.8-0.85, which is consistent with the standard expectation that MP2 overcorrelates and a damping factor is useful. 

The inclusion of MP3 energy should ideally ameliorate the MP2 overcorrelation effect, although the MP3 portion of the energy has to be optimally scaled down in practice (similar to MP2.5 and MP2.8:$\kappa-$OOMP2) to prevent overcorrection into the undercorrelation regime. 
It is thus no longer necessary to scale the second-order term once third-order effects are included. This was empirically observed by us on attempting to simultaneously fit both $c_2$ and $c_3$ for model 5, which yielded $c_2$ very close to 1 (and performance similar to sMP3, like the behavior observed in Ref \citenum{bertels2019third}). Table \ref{tab:fit_coeffs} shows that the optimal $c_3$ coefficients for sMP3 are typically between 0.8-0.9 (with the exception of HF/HFLYP) and roughly decreases with the amount of HF exchange present in the functional. The optimal HF $c_3$ for the training set was found to be 0.7 (and not 0.5 as in MP2.5), and HFLYP has an even smaller scaling factor at $0.65$. It is nonetheless worth noting that most functionals predict $c_3$ similar to $\kappa-$OOMP2 (i.e. close to 0.8), with range sepearated hybrid functionals being closest in magnitude. 

\begin{table}[htb!]
    \centering
    
    \begin{tabular}{| c | S S S | S S S | S S S | S S S |}
        \hline
        \textbf{Name} & \textbf{RMSE} & \textbf{MSE} & \textbf{MAX} & \textbf{RMSE} & \textbf{MSE} & \textbf{MAX} & \textbf{RMSE} & \textbf{MSE} & \textbf{MAX} & \textbf{RMSE} & \textbf{MSE} & \textbf{MAX}
        \csvreader[before line=\ifthenelse{\equal{\Name}{CCSD}}{
                                    \\\hline
                                } { \ifthenelse{\equal{\Name}{$\kappa$-OOMP2}}{
                                            \\\hline
                                            & \multicolumn{3}{c|}{\textbf{MP2}} & \multicolumn{3}{c|}{\textbf{sMP2}} & \multicolumn{3}{c|}{\textbf{MP3}} & \multicolumn{3}{c|}{\textbf{sMP3}}\\\hline
                                        }{
                                            \\
                                        }
                                },
                   head to column names]{newTables/W4.csv}{}
                   {
                        \ifthenelse{\equal{\Name}{CCSD}}{
                            \Name & \mptwormse & \mptwomse & \mptwomaxae &   &  &  &  &  &  &  &  & 
                        }
                        {
                            \Name & \mptwormse & \mptwomse & \mptwomaxae &  \scamptwormse & \scamptwomse & \scamptwomaxae & \mpthreermse & \mpthreemse & \mpthreemaxae & \scampthreermse & \scampthreemse & \scampthreemaxae
                        }
                    }
                    \\\hline

    \end{tabular}
    \caption{Root mean square errors (RMSE), mean signed errors (MSE), and maximum absolute errors (MAX) in kcal/mol for MP:DFT on the training set. CCSD values (HF orbitals, and no extra processing) are also supplied as a reference.}
    \label{tab:w411_data} 
\end{table}

\subsubsection{Training Set Results}
Table \ref{tab:w411_data} shows that MP2:HF (i.e. standard MP2 with HF orbitals) has an RMSE of 12 kcal/mol vs CCSD(T). MP2:DFT does not improve this picture--in fact predictions are often significantly degraded when orbitals obtained from local functionals like PBE or TPSS are used. This increased MP2:DFT RMSE (vs MP2:HF) almost completely stems from inclusion of non-Brillouin singles (as shown in the Supporting Information), indicating that the orbital energy differences are not too different from HF values. Scaling alters the picture, with sMP2:HF still having quite high RMSE of 10.4 kcal/mol, but sMP2:DFT reducing it to 6-7 kcal/mol, marking a significant improvement. MSE values show that systematic error is greatly eliminated by scaling (indicating effective overcorrelation in the unscaled case), and the maximum absolute error is also quite reduced.  We also observe that orbitals from hybrid functionals like $\omega$B97M-V seem to yield lower error than ones from local functionals, though the overall RMSE variation is small (roughly 1 kcal/mol, comparable to the anticipated accuracy of CCSD(T) itself over a single-reference thermochemistry dataset of this nature), indicating somewhat functional agnostic behavior. However these sMP2:DFT results are not competitive with the best double hybrid\cite{mardirossian2018survival,santra2019minimally} or hybrid DFT functionals\cite{mardirossian2017thirty,najibi2018nonlocal}, for thermochemistry (as shown in Section \ref{sec:cbs}). It is also worth noting that (s)MP2:$\kappa$OOMP2 behaves very similarly to (s)MP2:DFT approaches, yielding performance close to that of range separated hybrid functionals.

Table \ref{tab:w411_data} further shows that MP3:HF (standard MP3) has an RMSE of 9.2 kcal/mol and sMP3:HF barely improves upon it, yielding an RMSE of 8.5 kcal/mol. Incredibly, every single density functional (save HFLYP) tested improves upon HF by essentially a factor of 3, in both scaled and unscaled regimes. There is also strikingly low variation between different density functionals (a spread of 0.5-0.6 kcal/mol). In fact, even rung 1 SPW92 LSDA yields about the same accuracy as hybrid functionals, indicating a remarkable level of functional agnosticity. It is also worth noting that MP3:DFT fares fairly decently (RMSE of $\sim$ 3.5 kcal/mol, which is lower than CCSD) and sMP3:DFT only improves performance by about 1 kcal/mol. In fact, scaling appears to add systematic error by increasing the magnitude of the MSE. However, sMP3 does appear to reduce the worst case error, which likely contributes to the lower RMSE. The typical $3-3.5$ kcal/mol RMSE of MP3:DFT is quite close to the RMSE of 3.2 kcal/mol obtained from MP3:$\kappa-$OOMP2, which indicates that most of the improvement from explicitly optimizing orbitals in presence of MP2 correlation is captured by using DFT rather than HF orbitals. The sMP3 models perform similarly as well.

\begin{figure}[thb!]
    \centering
    \includegraphics{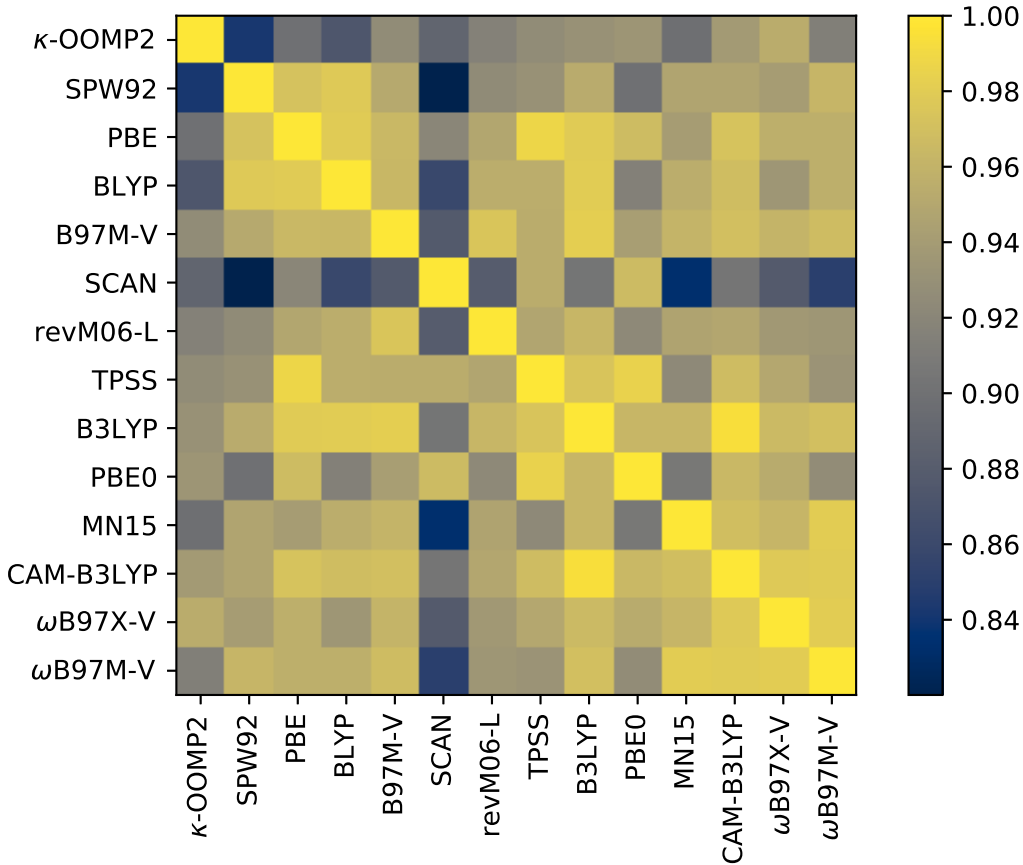}
    \caption{Correlation coefficients computed between sMP3:DFT errors for the non multireference portion of the W4-11 dataset.}
    \label{fig:w4corr}
\end{figure}

This remarkably consistent performance by orbitals generated from many different functionals across the first four rungs of Jacob's ladder raises the question as to whether the accuracy gains are systematic or whether different functionals are improving performance for different species to different extents, resulting in similar final RMSEs over the whole dataset. Fig \ref{fig:w4corr} shows the correlation coefficient ($r$) between sMP3 errors predicted by a characteristic subset of the examined functionals against each other. It is evident that there is quite strong correlation ($r>0.9$) between errors predicted by most DFT methods. SCAN is somewhat of an outlier, by virtue of having somewhat lower $r$ vs other approaches (and perhaps not so coincidentally, the largest sMP3 RMSE and MAX). Correlation between sMP3:DFT and sMP3:$\kappa-$OOMP2 errors is also quite strong ($r>0.8$ for even local functionals like SPW92). Overall,  Fig \ref{fig:w4corr} indicates that the near functional agnosticity of sMP3:DFT is generally a consequence of consistent improvement in predictions over different initial orbital choices (with sMP3:$\kappa-$OOMP2 acting similarly as well). The Supporting Information shows that the correlation between sMP3:DFT and CCSD is much lower, showing that the two approaches are improving different aspects over the dataset. Correlation is also low with sMP3:HF or sMP3:HFLYP.

\begin{table}[htb!]
    \centering
    \begin{tabular}{| c | S S S S | S S S S |}
        \hline
        \textbf{Name} & \multicolumn{4}{c|}{\textbf{NSC}} & \multicolumn{4}{c |}{\textbf{SC}}
        \begin{filecontents*}{newTables/W4_sp_nsp.csv}
Name,Rmptwormse,Rmptwomse,Rmptwomaxae,scaRmptwoctwo,scaRmptwormse,scaRmptwomse,scaRmptwomaxae,Rmpthreermse,Rmpthreemse,Rmpthreemaxae,scaRmpthreecthree,scaRmpthreermse,scaRmpthreemse,scaRmpthreemaxae,Umptwormse,Umptwomse,Umptwomaxae,scaUmptwoctwo,scaUmptwormse,scaUmptwomse,scaUmptwomaxae,Umpthreermse,Umpthreemse,Umpthreemaxae,scaUmpthreecthree,scaUmpthreermse,scaUmpthreemse,scaUmpthreemaxae
CCSD,4.916901277123885,1.6825634847812467,19.294408278544108,1.0,4.916901277123885,1.6825634847812467,19.294408278544108,4.916901277123885,1.6825634847812467,19.294408278544108,1.0,4.916901277123885,1.6825634847812467,19.294408278544108,4.934570193505972,1.2667584403574614,20.337805143340834,1.0,4.934570193505972,1.2667584403574614,20.337805143340834,4.934570193505972,1.2667584403574614,20.337805143340834,1.0,4.934570193505972,1.2667584403574614,20.337805143340834
$\kappa$-OOMP2,12.221290414197865,-3.67273462357334,56.809971092651026,0.8558176566941991,6.26088100626849,0.26719812039818325,26.751722565058586,3.0467571899186354,0.598253823839327,13.533794082371418,0.8197602610262567,1.471962859716925,-0.17154801908220485,5.135173008086305,12.643070269745118,-3.3733733238177765,45.27985717288844,0.8354439897300114,6.1503508885969405,-0.8108001420291394,35.01849082738561,3.384745891441024,0.049825243575002884,14.628050551701973,0.8100048012528644,1.6782401105969464,-0.6005660485876972,5.64093748084913
HF,8.101397465049791,-3.3040963358686612,35.421703333097135,0.9089013760062375,5.392524862766539,-0.8529451747243364,21.986618457762727,5.274666267617486,0.6440077761772233,22.37385849535664,0.630998083648245,2.8583727201416926,-0.8128502071239523,7.615222974766416,14.858757061718949,-6.667171151754845,51.13784821562325,0.8962669695440034,13.613654294451186,-4.516262222259678,43.506263215693636,11.849889255686756,-2.374236591754006,38.56817778616742,0.8109099971643926,11.653220680154586,-3.1859875998776404,39.19905697710113
HFLYP,7.998962780387462,-3.0046215181525415,34.95250230358613,0.9073710424520511,5.118428876478287,-0.5485169243425848,22.135044848050047,5.136018368082343,0.8648074165021906,22.078301756371616,0.627708579355819,2.4436258469578034,-0.5757477786621187,6.69372369327892,10.328183079398563,-4.7363175244927955,33.26644837149473,0.9002890841190373,8.527036299849469,-2.9144067344092446,29.65419360990591,7.756942817978751,-0.8292911651464753,25.739057485479705,0.6665570789041452,6.682063530062345,-2.132061447205415,24.328876228248358
Slater,15.073183172526152,-5.528325538983354,72.91970217030772,0.824695911933672,6.624273985703157,-0.14419309121993648,28.214632557590225,3.011909953641561,-0.23209525655542587,13.328993970088362,0.8794117653100386,2.2116997386877277,-0.8707583168249249,7.919131017282379,15.445502185407973,-5.507999449942847,54.17061803799157,0.8048907825674582,6.934878960688656,-1.6724103749067223,36.078030325173906,3.8164309224167794,-1.050655187220112,14.04210774784713,0.8618027414812593,2.94167379397759,-1.6666479446026314,8.540127132668959
SPW92,15.221078527832232,-5.289902182218903,72.85916010197221,0.824772276293707,6.681725099738858,0.04058319930492754,28.465923007661377,2.74499901102927,-0.12876546926139237,11.79410695609341,0.8925199048903382,2.058576907078385,-0.6834849340440327,7.697332426244063,15.525193493292141,-4.837134942235987,55.106121005336384,0.8051962453849866,6.725251518550509,-1.2971464267838906,36.90182523010351,3.411688945466533,-0.653835667774611,12.556611967249099,0.8597886494504109,2.317976728076632,-1.240381708799956,6.78605399862287
PBE,14.772303337978673,-5.01191916081373,70.9378028450585,0.8300862237088994,6.726074301211594,0.100757591253939,28.339391830838736,2.65074005319409,0.033323318688711114,11.560240626897443,0.8865012815901504,1.8726789483159265,-0.5393052368017481,7.364262274530128,15.187771532442294,-4.859304548434683,53.99897550119407,0.8095449894109472,6.826527223764672,-1.3289203964954759,37.82444147091488,3.3908645060250513,-0.6445480116540607,12.602470324973526,0.8614268274012649,2.382499588291391,-1.228600196687009,7.276336828316031
BLYP,15.078828883430146,-5.090517237031475,71.35091712008352,0.8268562377425509,6.728881943646828,0.13814000312841268,28.666595989227677,2.609692768456231,0.011685546621630441,10.997980012682401,0.8927967751370093,1.8935051358070552,-0.5352870456909102,7.355793697953387,15.477697209221413,-4.662369434863985,55.78570070216952,0.8061937425877724,6.720295213824977,-1.1632115449995697,38.227873355517865,3.306844302347633,-0.4915017715666085,11.903751603635202,0.8618744044333162,2.21559895668038,-1.0676053515893815,6.891727254193862
B97M-V,12.504377667503686,-4.234289820057755,59.46233868409315,0.8498140807623447,5.968741211955582,0.0144299963673668,25.061066126853945,3.1084886817866946,0.363896708569711,13.500630231755212,0.8273893951083128,1.7284506243543114,-0.42979904954148285,6.540667356519015,12.902939917574223,-4.152064103081917,46.60619784585018,0.8307441083685355,6.055763595438507,-1.2666219157091148,32.61648361066706,3.5198852216615513,-0.1308664115501622,14.422968060305424,0.8111729785442471,1.8870541704407586,-0.8901771943268524,5.514611520364004
SCAN,13.472583947444221,-4.641708197606979,64.7067451516036,0.8435347314006226,6.44611598255992,-0.019848973129380564,26.870549969444696,2.773638901870157,0.18055698345696486,12.33399687442219,0.8609134852197041,1.7424570005995526,-0.49015507392359187,6.22916670145001,14.299420036004038,-5.097781685578008,50.172764474049956,0.8208025715522863,6.985154687559965,-1.6702064440416444,36.80093162518237,3.7046589159240697,-0.7876547959334694,13.354401809491641,0.8628009867128377,2.9625280863447303,-1.3789999523351661,10.70869659403596
revM06-L,12.804222666291816,-4.289063078697778,59.457764143456885,0.8514000918093145,6.306714233708501,0.05305349736653809,26.311686694050078,2.784898890594761,0.259857112704059,12.088195324012865,0.8542571301694631,1.753458492601856,-0.4031155706209201,6.0171388241915436,12.880130640132029,-4.538627510843357,48.95542838951297,0.8340433332443526,6.160911296679174,-1.4739106102247752,34.58044593779253,3.2532999029787733,-0.2732312507878919,12.869745233329324,0.8358661921304171,2.08774547138772,-0.973326981023473,5.668307582604184
TPSS,14.327636504112474,-4.851179029339509,69.57511685083254,0.8357251288151716,6.750002536488685,0.07731518231244291,27.96578782711493,2.5869483541501905,0.09639137149524013,11.28569916426437,0.8832076576564316,1.7770598379989906,-0.48144696452795754,7.129421997128073,14.774610670695978,-4.9197117360594955,52.73692940040404,0.8150596901681325,6.865449434856489,-1.4230050395210028,38.32506167325361,3.373943557484518,-0.6625821142548634,12.508312599504428,0.862064654449092,2.4388216600257917,-1.2497907596934923,7.7797308167035615
B3LYP,13.338956830847463,-4.519472270721664,62.888217794516386,0.8423539591359839,6.29408159789914,0.04121037793321239,26.920521353865087,2.9127693283870277,0.22892685763064208,12.538897388171389,0.850223550018118,1.7489378985738258,-0.48227150691102927,6.486707660090309,13.724622435820548,-4.290488911145419,49.6139928099139,0.8224368990442521,6.337503720925432,-1.2098134311730575,35.36748524847765,3.455387063570427,-0.34838945440767155,13.43100183310588,0.8304045523236513,2.0641070244685413,-1.0169515765578006,6.539662026184246
PBE0,12.65761685845398,-4.327611999611418,60.54098272948711,0.8497535647106677,6.189290155710593,-0.024191723306446356,26.047905477309815,3.04601030793173,0.30057174718658425,13.441575359770205,0.8323218852209264,1.7009379610973618,-0.47547337832765413,6.15403502054524,13.08459716731064,-4.5155789871930985,46.67922734325773,0.8302315648856743,6.387528081840973,-1.469813557635307,34.3645819624244,3.64037967097691,-0.5327241872595249,14.454318409953306,0.8230233377171391,2.3652997034894256,-1.2375965361090404,7.6576501310273715
MN15,13.041353365514807,-4.368837759520193,60.82677542840899,0.8401791155948889,6.0025264786123955,0.07059253797904859,26.85782644391341,3.031370796984373,0.2541728232707097,12.879823027873844,0.8402105134323906,1.7753667409483294,-0.48453566415007243,6.005894871099473,12.87808698909518,-4.0889119853003155,48.31322940382175,0.825570503472568,5.7916601971046795,-1.2062366630473444,31.81820432217558,3.4634431675251314,-0.19445611565017276,13.828334256826091,0.8163093949026717,1.9466588381451984,-0.9098310708710492,5.755572565812313
CAM-B3LYP,12.83018219186371,-4.308851297425603,59.76760926245283,0.8462972725027551,6.077907984025838,0.05153435253749951,26.31691112000232,3.0441672390160552,0.2895479281532284,13.223572458069167,0.8350517461436316,1.7203139289477922,-0.4689499946404766,5.9594118979055715,13.050223919895632,-4.054438310816879,47.94860902428062,0.8281720547084567,6.058689600599917,-1.1508543954537058,33.683024367122805,3.5183994291074545,-0.2878450244907959,14.026991056048187,0.817604825147365,2.026786254285438,-0.9748534655490034,6.347510849389739
$\omega$B97X-V,12.105942668678688,-3.9658275569901766,56.605409433834446,0.854137899314963,5.997588574453579,0.06664275915290664,25.47953296142599,3.2406436102826617,0.44096611902565747,14.287432391412667,0.8117499945824524,1.6639126803589086,-0.38861281435833755,6.075109237119372,12.247330618003415,-3.6962058885553066,44.88554934258447,0.8371059854876217,5.887076813760231,-1.0785304608627222,32.7344836094703,3.6754958004233704,-0.1496097848771968,15.133954582578824,0.7933415753093666,1.8903611722080143,-0.8825437486772536,5.330186732177996
$\omega$B97M-V,12.204254481053491,-4.056255220898204,56.68702952940034,0.8483125689237145,5.671736124008163,0.05450998622337021,23.83523332856718,3.3097005570933358,0.3801053333346017,14.44273459381794,0.8115371687515655,1.7572309157933614,-0.45598373715498713,6.295813245802786,12.295480711658763,-3.615540095792574,44.959043196645844,0.8326479959169252,5.636286742853243,-1.0325674689196422,30.166536339554668,3.7297835391942047,-0.0926341268470351,15.351260949281276,0.7914872823260908,1.9167397618107114,-0.8272048245415067,5.5890929946385395
\end{filecontents*}
\csvreader[before line=\ifthenelse{\equal{\Name}{CCSD}}{
                                    \\\hline
                                } { \ifthenelse{\equal{\Name}{$\kappa$-OOMP2}}{
                                            \\\hline
                                             & \textbf{MP2} & \textbf{sMP2} & \textbf{MP3} & \textbf{sMP3} & \textbf{MP2} & \textbf{sMP2} & \textbf{MP3} & \textbf{sMP3}\\\hline
                                        }{
                                            \\
                                        }
                                }, 
                   head to column names]{newTables/W4_sp_nsp.csv}{}
                   {
                        \ifthenelse{\equal{\Name}{CCSD}}{
                            \Name & \Rmptwormse & & &  & \Umptwormse & & &
                        }{
                            \Name & \Rmptwormse &   \scaRmptwormse & \Rmpthreermse& \scaRmpthreermse & \Umptwormse &   \scaUmptwormse & \Umpthreermse& \scaUmpthreermse 
                        }
                    }
                    \\\hline
    \end{tabular}
    \caption{Root mean square errors in kcal/mol for MP:DFT on the non-spin contaminated (NSC) and spin contaminated (SC) subsets of the W4-11 dataset. CCSD values (HF orbitals, and no extra processing) are also supplied as a reference.}
    \label{tab:w411_spin_data}
\end{table}

\subsubsection{Role of spin-contamination}\label{sec:spincont}
A natural question to consider is whether the significant accuracy gains stem mostly from DFT functionals lowering spin-contamination, or whether other factors are also at play. This can be addressed by dividing the nonMR W4-11 energies into non spin-contaminated (NSC) and spin-contaminated (SC) subsets. An energy is classified as NSC if all of the participating species have HF $\langle S^2\rangle$ values that deviate by no more than 10\% from the exact, spin-pure $\langle S^2\rangle$ value. The remaining energies are classified as SC, with the NSC and SC subsets having 373 and 372 reaction energies respectively. Table \ref{tab:w411_spin_data} presents the RMSEs predicted with various approaches for these subsets. The performance of CCSD is nearly the same across both subsets, likely as a consequence of the removal of the explicitly multireference species from W4-11. However, there is a significant degradation in performance of MP:HF on moving from the NSC to SC subset, leading to an increase in RMSE by $\sim$7-9 kcal/mol. 

Use of DFT orbitals however greatly ameliorates this issue, with MP:DFT methods (save MP:HFLYP) yielding fairly similar RMSEs over both subsets (although errors are slightly larger for the SC subset). This leads to significantly better sMP2, MP3 and sMP3 results over the SC subset relative to HF, highlighting the importance of reducing spin-contamination. However, it is also worth noting that MP3:DFT has $\sim$ 3 kcal/mol RMSE over the NSC subset vs 5 kcal/mol RMSE for MP3:HF, showing that errors are reduced even in the absence of spin-contamination. Similarly, sMP3:DFT with meta-GGA and hybrid functionals reduces the sMP3:HF RMSE by 1 kcal/mol over the NSC subset. $\kappa-$OOMP2 orbitals yield similar behavior across both subsets, reproducing CCSD(T) to a slightly better extent than DFT methods. It thus appears that while reduction of spin-contamination is an important reason for improved performance of MP:DFT (or MP:$\kappa-$OOMP2) over MP:HF, it is not the sole reason as prediction quality is also improved for non spin-contaminated species as well, albeit to a smaller extent. 

\begin{table}[hbt!]
    \centering
    \begin{tabular}{| c | S S S | S S S | S S S | S S S |}
        \hline
        \textbf{Name} & \textbf{RMSE} & \textbf{MSE} & \textbf{MAX} & \textbf{RMSE} & \textbf{MSE} & \textbf{MAX} & \textbf{RMSE} & \textbf{MSE} & \textbf{MAX} & \textbf{RMSE} & \textbf{MSE} & \textbf{MAX}
        \begin{filecontents*}{newTables/BH76RC.csv}
Name,mptwormse,mptwomse,mptwomaxae,scamptwoctwo,scamptwormse,scamptwomse,scamptwomaxae,mpthreermse,mpthreemse,mpthreemaxae,scampthreecthree,scampthreermse,scampthreemse,scampthreemaxae,nbsmptwormse,nbsmptwomse,nbsmptwomaxae,fitnbsmptwoctwo,fitnbsmptwormse,fitnbsmptwomse,fitnbsmptwomaxae,nbsmpthreermse,nbsmpthreemse,nbsmpthreemaxae,fitnbsmpthreecthree,fitnbsmpthreermse,fitnbsmpthreemse,fitnbsmpthreemaxae,scamptwothreermse,scamptwothreemse,scamptwothreemaxae
CCSD,1.904917738072994,-0.6448948666666666,7.175495999999995,1.0,1.904917738072994,-0.6448948666666666,7.175495999999995,1.904917738072994,-0.6448948666666666,7.175495999999995,1.0,1.904917738072994,-0.6448948666666666,7.175495999999995,1.904917738072994,-0.6448948666666666,7.175495999999995,1.0,1.904917738072994,-0.6448948666666666,7.175495999999995,1.904917738072994,-0.6448948666666666,7.175495999999995,1.0,1.904917738072994,-0.6448948666666666,7.175495999999995,1.904917738072994,-0.6448948666666666,7.175495999999995
$\kappa$-OOMP2,6.240851645081003,1.031777681429136,24.04443596841722,0.846467308711953,3.8106999780403834,0.7901217429815512,14.592463338284979,1.5742495325654184,-0.43724682461529035,6.213232249887,0.814672993732315,0.8210616492718681,-0.16499691077621173,1.465756488985198,42.365942246548755,22.310622408721777,144.9495073858536,1.0,42.365942246548755,22.310622408721777,144.9495073858536,37.15951140384587,20.84159790267735,134.79203175302433,1.0,37.15951140384587,20.84159790267735,134.79203175302433,0.8207598110779266,-0.16371751180876928,1.4682430123737125
HF,6.341137237523657,-0.09878007892209108,21.224608478739796,0.903503,5.649094808729891,-0.11197498432896587,21.633996443742376,4.511950039467275,-0.978633038153881,21.615421083950054,0.715722,4.333381382016966,-0.7285101986093859,21.50432165816609,6.341137237523657,-0.09878007892209108,21.224608478739796,0.903502697,5.649092978838916,-0.11197502576088986,21.633997729218123,4.511950039467275,-0.978633038153881,21.615421083950054,0.715722413,4.333381165869811,-0.7285105619886582,21.5043218195717,4.270430713837872,-0.9737040890062845,21.345176904743
SPW92,7.331482819301718,1.3879659087802665,28.743621201139646,0.815803,4.001137011739261,0.9880628406548466,15.833203555246847,1.6311806273815213,-0.4410682498368866,6.044373343038714,0.875284,0.9856384948077295,-0.2129584257107896,2.1388130675114527,5.713626653310761,0.7909480521804999,20.74425186804475,0.861077284,3.630054800266154,0.5722775941563374,12.118415170716311,3.3743007785136014,-1.0380861064366536,14.043742676133618,0.651680205,1.4243906635936319,-0.4009973032591292,3.6390331145304025,0.8645845415942746,-0.0966949835075288,1.7148516538096743
PBE,7.248678824950072,1.4320205964632828,28.582320012661725,0.820683,4.033771856034956,1.0069194687404677,16.088568051086213,1.5879016271819917,-0.4048925651550929,5.812383265743023,0.87333,0.9389924673556194,-0.17221077497289297,1.967815465231375,5.607588442448375,0.7793352325732026,20.697597515130326,0.865055937,3.563408538830838,0.5475036348806372,12.35948683230874,3.3315758332242495,-1.057577929045173,13.697105763274422,0.65006158,1.348254186340422,-0.41477143959123386,3.9474023643475853,0.852996803969023,-0.0910235402523892,1.6573412359796933
BLYP,7.398451028347054,1.3983137413588937,28.93599663524695,0.817367,4.0793800018564745,1.0023724600385548,16.09652928107647,1.5722482376587723,-0.4453648039558232,5.779376292349667,0.876457,0.9438997381716977,-0.2175912254320073,2.0994761411742853,5.6811146103712575,0.8644249614247633,20.68212031940505,0.865007495,3.638785476834749,0.6438374147329581,12.30608705535262,3.3850210660910323,-0.9792535838899535,14.033252608191574,0.644671153,1.3603183842412705,-0.3241414121446373,3.816701099536459,0.8397102757340594,-0.1061880214751098,1.5233028003021207
B97M-V,6.330920140596608,1.21213997867696,25.200812063648073,0.841083,3.6836769316643747,0.8725839923760819,14.899270955682113,1.6930871702857453,-0.44985191165157,6.275886513517378,0.818887,0.7855382661654466,-0.14884357441849883,1.8252364139673105,5.01338289854639,0.6324986202149598,19.01416882968907,0.864353423,3.059812757271542,0.4212906247830833,11.060292762318959,3.040609591922141,-1.02949327011357,12.462529747476381,0.683377911,1.0776238598112589,-0.503269925896692,2.885044718642547,0.7475185814936515,-0.09208810305614713,1.610110512683228
SCAN,6.84560604165962,1.3981026891667259,27.07597838307997,0.833118,4.018520454001867,0.9989882785561294,15.861978019919505,1.5823320069999418,-0.33786832886711515,5.7352498594569425,0.861896,0.9477431632260214,-0.09812378739257022,2.238786350759746,5.479178017379612,0.7883908191373726,20.351659835553775,0.870482395,3.580102004373626,0.5576053867627683,12.519359210043085,2.983545482651509,-0.947580198896468,12.459568406983138,0.670581665,1.228892171473332,-0.37571951652750535,2.868322499280012,0.9145343694794918,-0.0548257888749023,2.077699862432354
revM06-L,6.471235163344275,1.3478850264232134,25.73542463128271,0.843559,3.915058469516667,0.9854156784499282,15.552919661867541,1.638554483261364,-0.2885678751043902,5.658329806239607,0.844851,0.9918474372962581,-0.03467384388528421,2.178527006750727,5.407477019045236,0.7722837068634968,19.658529019120614,0.879849018,3.675888419508578,0.5630563623297268,12.568230688042398,2.7565828822556853,-0.8641691946641069,11.735225418401697,0.671454696,1.2698730408169703,-0.32652027865003874,2.4098470752107097,0.9600474526427517,0.027112790256137456,1.9348366680745954
TPSS,7.124259119986581,1.3877615500199818,28.354025963374184,0.826272,4.051454874501174,0.991821242708792,16.28599676141134,1.529081187917959,-0.37713909710024635,5.594894921428889,0.872135,0.908793942268984,-0.15147007585621838,2.126114505910266,5.510117022460289,0.8168714864560944,20.51407792726708,0.870852401,3.5777388903970055,0.5962626366897574,12.555340256850364,3.2074942090747287,-0.9480291606641338,13.434842957535992,0.644351728,1.2639229935921061,-0.3203452952641424,3.479635134898384,0.8527309102608549,-0.09525096885775694,1.9213198750626006
B3LYP,6.535227117693653,1.1121357518700608,25.18211857053855,0.833248,3.822412949443976,0.8346984908478896,14.648661335225086,1.6377504827349483,-0.48448120498986325,6.022561130195285,0.839813,0.8922353367122065,-0.22872392452134244,1.78207903188477,5.5541259613663,0.8147001086886869,19.976715740695248,0.864406277,3.5882698228399685,0.6294335138395782,12.117296996734808,2.668044915208025,-0.7819168481712369,11.227963960038593,0.683169728,1.1339030102652412,-0.27606026344949497,2.2952783059940813,0.8499452531701678,-0.16380144898669693,1.5483711609087578
PBE0,6.232067704633374,1.0964580670245712,24.044206300108677,0.840866,3.744329201062081,0.8204966252266955,14.30719873235818,1.675694023324205,-0.44568818026635765,6.1178609220318805,0.827507,0.8997829559192414,-0.1796787476324034,1.9729698955056776,5.429486735263593,0.7609013755033006,19.681513167764976,0.866483847,3.5184888024571865,0.5741672353866013,12.094486963929391,2.5054300404446153,-0.781244871787628,10.480554054375588,0.693137289,1.0761731663863878,-0.3080176935854568,2.1172229453948646,0.8811301770911558,-0.1507210992856525,1.8670332429630676
MN15,6.102007544809368,0.9253001763634747,23.015273517723642,0.833571,3.485837071176659,0.650952990051064,13.052241126924152,1.766315815918188,-0.5736324544612793,6.596003954752476,0.828108,0.8942889214281976,-0.315977926683551,1.918276363357073,5.204997135459281,0.4641838849882144,18.79136378987034,0.848460825,3.10332692084293,0.2842587716244343,10.359777123521006,2.6847842544770986,-1.0347487458365403,10.819913682605787,0.737529995,1.138914417237965,-0.6413238907293032,3.047841536348585,0.8090498883202994,-0.2375177034017168,1.6298058217390903
CAM-B3LYP,6.203340315534188,0.9750175052645218,23.002892765159224,0.838049,3.6882603249811363,0.7441439461508811,13.427595497917096,1.7011380798633444,-0.5154952098017329,6.352361035330148,0.826035,0.9194967004226737,-0.256198165325232,1.6708816388330163,5.496555929950536,0.7762209284147716,19.32691680525422,0.863074345,3.5494812169531116,0.6082432675263518,11.734569884468314,2.4352342555990574,-0.7142917866514827,10.02833699523515,0.698737222,1.0844022628953662,-0.2652557854663002,2.1913763513432087,0.8843736729694126,-0.19530278675891982,1.5778278510318149
$\omega$B97X-V,5.858922771512293,0.9050241055661307,21.557708162721433,0.846392,3.575675991328608,0.7086098676129751,12.886952112908979,1.7510675022898603,-0.5263862560005717,6.659420014304999,0.802263,0.86094069707047,-0.24334346533545675,1.5634239647587478,5.39542021349106,0.7622468050019138,18.99558439765663,0.863961489,3.5175584535054676,0.607721390845694,11.665125973785777,2.287682225490635,-0.6691635565647888,9.22154377936981,0.708002087,1.0583060551319265,-0.2511947183407358,2.0633822960121293,0.8432402186383529,-0.20965045743983904,1.4370194528941838
$\omega$B97M-V,5.741848228278041,0.820746718483919,21.149676710553265,0.841172,3.350512557505703,0.6251978050426151,12.168781772212263,1.8460818124943281,-0.5965327679043827,6.969961668228194,0.801219,0.8644492276386174,-0.31480453432062944,1.6270663263056986,5.119762244514549,0.6283033418627928,18.207762096617166,0.85028865,3.075193529731695,0.47278979871694454,10.182804170916718,2.48730445446929,-0.7889761445255092,9.9118762821643,0.733447797,1.0224300278413265,-0.4111971751619985,2.416524724736746,0.8052307157171695,-0.2460149390144057,1.327511743807376
\end{filecontents*}
\csvreader[before line=\ifthenelse{\equal{\Name}{CCSD}}{
                                    \\\hline
                                } { \ifthenelse{\equal{\Name}{$\kappa$-OOMP2}}{
                                            \\\hline
                                            & \multicolumn{3}{c|}{\textbf{MP2}} & \multicolumn{3}{c|}{\textbf{sMP2}} & \multicolumn{3}{c|}{\textbf{MP3}} & \multicolumn{3}{c|}{\textbf{sMP3}}\\\hline
                                        }{
                                            \\
                                        }
                                },
                   head to column names]{newTables/BH76RC.csv}{}{\ifthenelse{\equal{\Name}{CCSD}}{\Name & \mptwormse & \mptwomse & \mptwomaxae &   &  &  &  &  &  &  &  & }{\Name & \mptwormse & \mptwomse & \mptwomaxae &  \scamptwormse & \scamptwomse & \scamptwomaxae & \mpthreermse & \mpthreemse & \mpthreemaxae & \scampthreermse & \scampthreemse & \scampthreemaxae}}\\
        \hline
    \end{tabular}
    \caption{Root mean square errors (RMSE), mean signed errors (MSE), and maximum absolute errors (MAX) in kcal/mol for MP:DFT on the BH76RC dataset. }
    \label{tab:bh76rc_data}
\end{table}

\subsection{Test Sets}

\subsubsection{Thermochemistry}

The behavior of MP:DFT (as well as the transferability of the fit coefficients) for other thermochemistry was tested using the BH76RC \cite{Goerigk2010, Zhao2005, Zhao2006} and RSE43 \cite{AndreasGansauer2006,goerigk2017look} datasets. The BH76RC dataset consists of the overall reaction energies obtained by taking the difference of the forward and reverse barrier heights in the HTBH38 and NHTBH38 datasets. Table \ref{tab:bh76rc_data} shows that the RMSEs for this dataset present a very similar picture as W4-11. MP2:DFT tends to fare worse than MP2:HF, but scaling leads to a significant improvement in performance (lowering RMSE by $\sim 2$ kcal/mol). MP3:DFT however marks a significant improvement over MP3:HF (by almost a factor of 3), with the resulting RMSE of $\sim$ 1.7 kcal/mol being competitive with CCSD. sMP3:DFT fares even better, giving RMSEs of less than 1 kcal/mol, compared to 4.3 kcal/mol with sMP3:HF (or 1.9 kcal/mol from CCSD). There is also fairly low variation in predictions from orbitals generated from different functionals for the sMP2,and (s)MP3 approaches.

\begin{table}[htb!]
    \centering
    
    \begin{tabular}{| c | S S S | S S S | S S S | S S S |}
        \hline
        \textbf{Name} & \textbf{RMSE} & \textbf{MSE} & \textbf{MAX} & \textbf{RMSE} & \textbf{MSE} & \textbf{MAX} & \textbf{RMSE} & \textbf{MSE} & \textbf{MAX} & \textbf{RMSE} & \textbf{MSE} & \textbf{MAX}
        \begin{filecontents*}{newTables/RSE43.csv}
Name,mptwormse,mptwomse,mptwomaxae,scamptwoctwo,scamptwormse,scamptwomse,scamptwomaxae,mpthreermse,mpthreemse,mpthreemaxae,scampthreecthree,scampthreermse,scampthreemse,scampthreemaxae,nbsmptwormse,nbsmptwomse,nbsmptwomaxae,fitnbsmptwoctwo,fitnbsmptwormse,fitnbsmptwomse,fitnbsmptwomaxae,nbsmpthreermse,nbsmpthreemse,nbsmpthreemaxae,fitnbsmpthreecthree,fitnbsmpthreermse,fitnbsmpthreemse,fitnbsmpthreemaxae,scamptwothreermse,scamptwothreemse,scamptwothreemaxae
CCSD,0.44647941824195486,0.31593176744186047,0.9732230000000008,1.0,0.44647941824195486,0.31593176744186047,0.9732230000000008,0.44647941824195486,0.31593176744186047,0.9732230000000008,1.0,0.44647941824195486,0.31593176744186047,0.9732230000000008,0.44647941824195486,0.31593176744186047,0.9732230000000008,1.0,0.44647941824195486,0.31593176744186047,0.9732230000000008,0.44647941824195486,0.31593176744186047,0.9732230000000008,1.0,0.44647941824195486,0.31593176744186047,0.9732230000000008,0.44647941824195486,0.31593176744186047,0.9732230000000008
$\kappa$-OOMP2,1.2980725798130017,-1.0234819489808737,3.0413251627843576,0.846467308711953,0.5430358654621168,-0.3659550816427972,1.4429933410883113,0.5208711427079245,-0.4162088008331317,1.550431852373853,0.814672993732315,0.6248101206229472,-0.528752915366105,1.7019428174603017,6.092487034908087,1.8730184566447456,14.879411855985554,1.0,6.092487034908087,1.8730184566447456,14.879411855985554,6.387438094597951,2.4802916047924883,16.583876759728096,1.0,6.387438094597951,2.4802916047924883,16.583876759728096,0.6242209935759244,-0.5282170846186092,1.7008392700300938
HF,4.09861100475382,2.2526001100706186,16.444956542696843,0.903503,3.687373616064881,2.1740779272367585,14.525657646422307,2.4311252410795405,1.5623085427494081,9.346737887039065,0.715722,2.8762573996240004,1.758543248924347,11.364605290032149,4.09861100475382,2.2526001100706186,16.444956542696843,0.903502697,3.687372358761413,2.1740776806775766,14.525651619835308,2.4311252410795405,1.5623085427494081,9.346737887039065,0.715722413,2.876256729467804,1.7585429638339298,11.364602358467842,2.664646693120676,1.6114595650279473,10.590740984312689
SPW92,2.5968322446912353,-2.3197217448552734,5.8191129780137665,0.815803,0.8964152870234434,-0.6965142211244805,2.489793322024284,0.9750179788332042,-0.8509906698087725,2.1578291889348247,0.875284,1.1496430088846878,-1.0341649345642718,2.5991925658285076,0.9654946565990703,-0.3837349080128018,2.576816002551677,0.861077284,0.8111096714673508,0.5715474530756592,1.8645071415953451,1.2522354546824113,1.0849961670336994,3.460996412169579,0.651680205,0.7847232552921125,0.5734080600633725,2.178046856104025,1.0681554854453976,-0.9557709502680204,2.499965056423575
PBE,2.3277994685697694,-2.0549714657184,5.208377293581938,0.820683,0.8891152677560792,-0.7124415648665208,2.3976857602670294,1.0392071412514465,-0.9097441915483527,2.381925203008338,0.87333,1.177273469705816,-1.0548101303674724,2.7082496232255946,0.9428310322033442,-0.5030495159717133,2.4195057713880956,0.865055937,0.5656057375814526,0.2978416298080903,1.2268041210736842,0.7851595615629063,0.6421777581983338,2.3865171439616466,0.65006158,0.4593184951176283,0.24141873533436065,1.3835179113741878,1.1204012981215354,-1.0019590552398383,2.628495658532976
BLYP,2.3904007604814974,-2.1139157210018324,5.2796685910717205,0.817367,0.882937056222172,-0.7042734234609115,2.3637843538465977,1.0110276421889517,-0.8775513745926404,2.3285994881546106,0.876457,1.1558454135693943,-1.0302955350410712,2.680127617405218,0.9357989816458552,-0.471730728571314,2.469603328843867,0.865007495,0.6007243606917326,0.3485184848548484,1.3828048048025325,0.9059013338404409,0.7646336178378779,2.645986530631535,0.644671153,0.5270906442687718,0.3253177001563912,1.5453407515203053,1.0818426218702977,-0.961730382149268,2.5739686697662614
B97M-V,1.8680795113512265,-1.62717975514576,3.9965444629532105,0.841083,0.8180776645829302,-0.6667565823840336,2.0121365341207835,0.9429048371120126,-0.8122060897032412,2.2773633783590643,0.818887,1.072893880367046,-0.959808415172532,2.572009231400978,0.8913493494040221,-0.46618173215499714,2.3593358786098193,0.864353423,0.4898343178065046,0.19612001855497724,1.1236226280577046,0.4608507009912027,0.3487919332875219,1.3688101838292168,0.683377911,0.3030055952103259,0.09075326885512443,0.7634562177644035,1.0336710556651123,-0.92453197278032,2.5019292130958677
SCAN,1.641733500368785,-1.2386026872361438,3.4616411468432844,0.833118,0.7453723473154934,-0.421286198709438,2.0099829603664023,0.821045666577424,-0.7308947868761033,2.0747291244628134,0.861896,0.8819535202038562,-0.8010112787474263,2.100389595483387,1.0818618876637975,-0.3509475218944798,3.7862043191987578,0.870482395,0.7156359106618497,0.16840720811271623,3.2725603011837414,0.3576778034837635,0.15676037846556057,1.2402871162890705,0.670581665,0.3818709465638833,-0.010487912737389773,1.4534541994040016,0.85335488913762,-0.7738426229627756,2.0288036512557652
revM06-L,1.9113281062832272,-1.6219311531216059,4.116170504119069,0.843559,0.9001245156326845,-0.7347729236357754,2.0054598475069185,1.1308210656212196,-0.9764882377679917,2.607044056742849,0.844851,1.2221499651986887,-1.0766280606421894,2.7795728262262624,0.9520121751953201,-0.5125727632916016,2.338046945517469,0.879849018,0.5540709870914827,0.035498607786730835,1.2352625277139229,0.3633379700218954,0.13287015206201216,0.9869601014145435,0.671454696,0.3578838529839601,-0.07918708677748727,0.732460260674471,1.1734899196564366,-1.0343802337776113,2.687727990608842
TPSS,2.0468948425555276,-1.777760609028715,4.5591439320117,0.826272,0.8170628559479595,-0.6495044464933597,2.1730494053991785,0.9970891460192052,-0.8551341263237447,2.3974946537097392,0.872135,1.104289441903123,-0.973105761534816,2.6738939386798197,0.9104103396766401,-0.5379655206005369,2.3406822323444523,0.870852401,0.45217009297332156,0.14065188649508492,1.0294494705934518,0.5222133303343655,0.3846609621044331,1.6498935532628016,0.644351728,0.30617771426968504,0.05653044782897267,0.8348162845916431,1.065734356293317,-0.9384777386842608,2.6106442836322286
B3LYP,1.7651481598017542,-1.52185601781934,3.608838027304723,0.833248,0.7238174045803716,-0.5773894625127584,1.6784244983100174,0.8597317243936463,-0.7243154279744665,2.159900921039642,0.839813,0.9672392912958218,-0.852071062439947,2.3920018092809263,0.9129053210747794,-0.5444143109419792,2.3509657243624957,0.864406277,0.4268937146727647,0.09103993743036083,1.013280011187275,0.37149469213085695,0.2531262789028945,1.136730372798181,0.683169728,0.24862606487262343,0.0004412768913026797,0.6138415215644208,0.9261884661341693,-0.8163100739868678,2.3088784844721433
PBE0,1.5177939633394621,-1.269515239912742,3.1067095062790706,0.840866,0.6389985069393078,-0.48325353470006,1.3480203288242838,0.7939553317450544,-0.6645396503017879,2.06016464059389,0.827507,0.8735462904095579,-0.7688937046805502,2.2051403505411873,0.8827130181841223,-0.49181098301616116,2.246909922953325,0.866483847,0.4386083886715547,0.06403998087946793,1.2175705694564325,0.2703512027574487,0.1131646065947932,0.8236808163747507,0.693137289,0.23963281523687946,-0.0724798429220475,0.6305122525127951,0.8545953028874347,-0.7525172407925568,2.16355911816676
MN15,1.5768277826017252,-1.3226668574715732,3.3057928313918516,0.833571,0.6166100415958428,-0.46928719510444944,1.3017011394045106,0.7843338597535191,-0.6550079345239234,2.0107410793280867,0.828108,0.8765286587642487,-0.7697731621072408,2.1204916053430773,0.837686930153171,-0.448646565216108,2.293343957421051,0.848460825,0.41054867897921754,0.19593586658714102,1.1520448390563107,0.3572015769494509,0.21901235773154168,0.8721124002203531,0.737529995,0.2215965746468136,0.043771916887177506,0.5356140918556573,0.8227693942906915,-0.7230577430878126,1.9977528823346518
CAM-B3LYP,1.411720431357364,-1.156441766412587,2.9489849388560287,0.838049,0.5799862913596289,-0.41591944667284797,1.3069791180063692,0.6799433977832796,-0.561777754934693,1.825037018288278,0.826035,0.7540822792558327,-0.6652284796914447,1.9139985283607386,0.8978131658271715,-0.5000320820288171,2.2764856770866295,0.863074345,0.43856011269028,0.03618230967161452,1.3674486312233984,0.257991669678372,0.09463192944907685,0.6726154292740489,0.698737222,0.23457767279957908,-0.08451820262537744,0.6537942120210438,0.7189171832477823,-0.6352728336575584,1.8255380251434024
$\omega$B97X-V,1.3457337324967484,-1.1023990860788344,2.7551725091402233,0.846392,0.5755938205734209,-0.4126965254422273,1.2985543923675076,0.6278209320503214,-0.5222559247169761,1.646178128009001,0.802263,0.7104672280306755,-0.636971693015186,1.7359475496069343,0.8833050116421354,-0.4860007268683993,2.202177379099723,0.863961489,0.4366358028543483,0.04096061115569762,1.4428918034092835,0.25257083905818184,0.09414243449345894,0.6172135381919255,0.708002087,0.22808688299494764,-0.07525815786542589,0.6346719526812734,0.6915755056450733,-0.6205768203097577,1.6874512055894098
$\omega$B97M-V,1.4373224699641696,-1.2025449985790686,2.8243323424778026,0.841172,0.5912816296244111,-0.4484252106579628,1.3282638519829257,0.6612758760507573,-0.5444808341855794,1.7349079006025652,0.801219,0.7646706933318612,-0.6752914868478814,1.9081773871636651,0.9290723826212556,-0.5785958571360686,2.381729604011255,0.85028865,0.3834469844928991,0.03882555320935346,1.064488983741688,0.25904571107054264,0.07946830725742049,0.6953771427929558,0.733447797,0.2507735702981402,-0.09594014547701839,0.7181442538072886,0.7261029907389815,-0.6416292092168956,1.8166998047515475
\end{filecontents*}
\csvreader[before line=\ifthenelse{\equal{\Name}{CCSD}}{
                                    \\\hline
                                } { \ifthenelse{\equal{\Name}{$\kappa$-OOMP2}}{
                                            \\\hline
                                            & \multicolumn{3}{c|}{\textbf{MP2}} & \multicolumn{3}{c|}{\textbf{sMP2}} & \multicolumn{3}{c|}{\textbf{MP3}} & \multicolumn{3}{c|}{\textbf{sMP3}}\\\hline
                                        }{
                                            \\
                                        }
                                }, 
                   head to column names]{newTables/RSE43.csv}{}{\ifthenelse{\equal{\Name}{CCSD}}{\Name & \mptwormse & \mptwomse & \mptwomaxae &   &  &  &  &  &  &  &  & }{\Name & \mptwormse & \mptwomse & \mptwomaxae &  \scamptwormse & \scamptwomse & \scamptwomaxae & \mpthreermse & \mpthreemse & \mpthreemaxae & \scampthreermse & \scampthreemse & \scampthreemaxae}}\\
        \hline
    \end{tabular}
    \caption{Root mean square errors (RMSE), mean signed errors (MSE), and maximum absolute errors (MAX) in kcal/mol for MP:DFT on the RSE43 dataset. }
    \label{tab:rse43_data}
\end{table}

The RSE43 dataset consists of reaction energies for hydrogen abstraction of hydrocarbons (by a methyl radical). Table \ref{tab:rse43_data} shows somewhat different behavior than the datasets considered previously. MP2:DFT improves significantly over MP2:HF, with several methods reducing RMSE by nearly a factor of 3. This is likely on account of the DFT approaches significantly reducing spin-contamination effects in the reference, which degrades the performance of standard MP2\cite{gill1988does}. The role of spin-contamination in degrading MP2:HF predictions can be gauged by exclusion of reactions involving heavily spin-contaminated species from the dataset. Nine radicals are found to have UHF $\langle S^2 \rangle \ge 0.9$ vs the ideal value of 0.75 (with the benzyl radical being an extreme case with UHF $\langle S^2\rangle\sim 1.3$) and three singlet organic molecules are found to have UHF $\langle S^2 \rangle \ge 0$; discarding reactions involving these species (9 reactions in total) lowers MP2:HF RMSE to 1 kcal/mol. The effect on MP2:DFT RMSEs is much more muted (as can be seen from the supporting information), due to significantly lower spin-contamination with the tested DFT functionals (with benzyl radical again being the worst case, with a SCAN $\langle S^2\rangle$ of 0.86). Scaling further reduces RMSE (over the full dataset) by a roughly a factor of 2, with sMP2:DFT approaches having RMSE below 1 kcal/mol and being fairly competitive with CCSD (the best functionals yielding an RMSE of 0.6 kcal/mol and CCSD 0.4 kcal/mol). 

Moving on to third-order, the performance of MP3:HF is also negatively affected by the spin-contamination in HF orbitals, resulting in a fairly large RMSE of 2.4 kcal/mol (vs 0.9 kcal/mol when the nine most spin-contaminated species are removed). MP3:DFT brings down the RMSE by a factor of 2-3, although sMP3:DFT offers no further improvement in performance. Indeed, (s)MP3 models do not improve upon sMP2 for this dataset, indicating that the excellent performance of sMP2:DFT may be somewhat fortuitous. 
The (s)MP3:DFT RMSEs show somewhat stronger functional dependence for this dataset (relative to preceding ones), with hybrid functionals reproducing CCSD(T) values somewhat better than local functionals. The overall effect is small in absolute terms ($\sim$ 0.5 kcal/mol spread of RMSE, comparable to the W4-11 spread) but is significant in percentage terms-with $\omega$B97X-V having essentially half the RMSE of revM06-L. This argues for use of hybrid functionals for orbital generation (over local functionals). It is also worth noting that use of $\kappa-$OOMP2 orbitals reproduces the CCSD(T) benchmark for RSE43 to a slightly (but perceptibly) better extent than the tested functionals.

\begin{table}[hbt!]
    \centering
    
    \begin{tabular}{| c | S S S | S S S | S S S | S S S |}
        \hline
        \textbf{Name} & \textbf{RMSE} & \textbf{MSE} & \textbf{MAX} & \textbf{RMSE} & \textbf{MSE} & \textbf{MAX} & \textbf{RMSE} & \textbf{MSE} & \textbf{MAX} & \textbf{RMSE} & \textbf{MSE} & \textbf{MAX}
        \begin{filecontents*}{newTables/HTBH38.csv}
Name,mptwormse,mptwomse,mptwomaxae,scamptwoctwo,scamptwormse,scamptwomse,scamptwomaxae,mpthreermse,mpthreemse,mpthreemaxae,scampthreecthree,scampthreermse,scampthreemse,scampthreemaxae,nbsmptwormse,nbsmptwomse,nbsmptwomaxae,fitnbsmptwoctwo,fitnbsmptwormse,fitnbsmptwomse,fitnbsmptwomaxae,nbsmpthreermse,nbsmpthreemse,nbsmpthreemaxae,fitnbsmpthreecthree,fitnbsmpthreermse,fitnbsmpthreemse,fitnbsmpthreemaxae,scamptwothreermse,scamptwothreemse,scamptwothreemaxae
CCSD,2.205894441653741,1.877403447368421,4.146062000000001,1.0,2.205894441653741,1.877403447368421,4.146062000000001,2.205894441653741,1.877403447368421,4.146062000000001,1.0,2.205894441653741,1.877403447368421,4.146062000000001,2.205894441653741,1.877403447368421,4.146062000000001,1.0,2.205894441653741,1.877403447368421,4.146062000000001,2.205894441653741,1.877403447368421,4.146062000000001,1.0,2.205894441653741,1.877403447368421,4.146062000000001,2.205894441653741,1.877403447368421,4.146062000000001
$\kappa$-OOMP2,4.030625708901079,-1.9862156709746182,8.732473851169958,0.846467308711953,2.24894126756796,0.8295867846729965,5.530201432376295,0.7304804066860118,0.34645111209738305,1.7549363514181238,0.814672993732315,0.6736632356875908,-0.0858550394294223,1.3769527251816998,33.74497299188975,-31.128033530736758,65.96564667589752,1.0,33.74497299188975,-31.128033530736758,65.96564667589752,30.673606551451805,-28.795366747664758,57.00814395358772,1.0,30.673606551451805,-28.795366747664758,57.00814395358772,0.6737955614970798,-0.0832428926511322,1.3734419363982209
HF,4.047034862932203,2.9888551651246233,12.147754649649833,0.903503,4.564966153944054,3.998769216304346,10.332911039509256,3.8847997846696267,3.5084200669380663,7.214337064752952,0.715722,3.7050274300993937,3.3607191957803444,6.322828006219861,4.047034862932203,2.9888551651246233,12.147754649649833,0.903502697,4.56496824039454,3.9987723874283962,10.332905340911225,3.8847997846696267,3.5084200669380663,7.214337064752952,0.715722413,3.7050275628702565,3.360719410360649,6.322828825283338,3.2331058208423795,2.8398199482422384,5.832832296649935
SPW92,7.637045232445383,-4.954783908026767,24.166407622999074,0.815803,2.928396159612298,0.2206414323284192,8.551913069938161,1.4181489094211461,-0.5575869191920951,5.053355720448778,0.875284,1.9465450252898993,-1.1059877388516006,7.437059101527243,3.913960456245642,-0.49261130493265665,10.745716750278262,0.861077284,3.963557565847146,2.790834960431703,11.275906301019504,4.858540701997636,3.9045856839020163,14.430707928998928,0.651680205,3.132411502240973,2.372954930176506,7.801080743166857,1.7917177956788413,-0.8296367441615262,7.085357915317809
PBE,6.822771107678517,-4.462098354220034,20.30546397270706,0.820683,2.6398332232870216,0.2046263120767638,7.160399886958908,1.5032504263476594,-0.6861998427612042,5.203794728098501,0.87333,1.9394683118835685,-1.1644929072076937,7.116723171313065,3.662644582772917,-0.5248670873070221,9.867333605382797,0.865055937,3.565447304401185,2.4557459734543667,9.91120923649126,3.962337826462891,3.2510314241518063,11.355158747429499,0.65006158,2.54008130784522,1.929699464971552,6.172427401174097,1.8043470643277044,-0.9671636361808873,6.8157082711685035
BLYP,7.208980671777054,-4.674646551359178,22.14558391440265,0.817367,2.784841044037003,0.15082529739448777,7.876025464076628,1.620670225596129,-0.8204094860786872,5.770673808561142,0.876457,2.1019171985109817,-1.2965734958346342,7.793679327767119,3.7011261064907726,-0.5022835651729324,9.818575716830715,0.865007495,3.575667662990198,2.5012083343981826,9.82281707163866,4.096408993491605,3.3519535001075575,12.260110058582057,0.644671153,2.56497441279342,1.9824318876367777,6.441632130944747,1.924453782973875,-1.0288499461387173,7.39797814797981
B97M-V,4.79296253442114,-3.18666335809901,9.720259380170546,0.841083,2.0500377023198175,0.36162085151791207,3.9875679799760846,1.1356358769663535,-0.4904896936961104,3.0901561727324864,0.818887,1.3800628152294192,-0.9788017945771129,4.181313592447124,3.3174203898514927,-0.3503154006459493,8.339895876229306,0.864353423,3.203324430091559,2.293648045591944,7.623270886355037,2.540120090164203,2.34585826375695,4.830655774994041,0.683377911,1.7497722127396234,1.4921901258269188,3.459841238074091,1.2658712518847415,-0.8326131512058738,3.995264660311933
SCAN,5.039402439138954,-3.2357490936876676,10.738286056160618,0.833118,2.210073373140237,0.4429644796221741,4.54426850718621,1.069287994623081,-0.4324482487462075,3.419761550154883,0.861896,1.3153905321519253,-0.8195953086360032,4.336452195755382,3.514821513702998,-0.5231432441213577,9.161641376547763,0.870482395,3.1779345646731327,1.9805868340142438,8.655710575883923,2.5501154334277314,2.2801576008201025,5.252302774168584,0.670581665,1.817645871975348,1.3566989039753938,4.674617565128912,1.2445315626496922,-0.7170056011979133,4.190870810292807
revM06-L,4.693830401454076,-3.000569289056919,9.635258499276391,0.843559,2.15431296953703,0.42350457272580494,4.41100164253108,1.135034637461551,-0.47887648371920905,2.9503534427236264,0.844851,1.328866865996454,-0.8701146007745489,3.7380740634329976,3.510250809460881,-0.10190504433725862,9.099209062864873,0.879849018,3.3619933405791618,2.1796003101625407,8.266826160902625,2.599715099485693,2.4197877610004515,4.952469265342954,0.671454696,1.993388353736943,1.5912974316761606,4.109992090610405,1.2126020925426553,-0.7088137627902277,3.526721745104057
TPSS,5.756172064914224,-3.670116272904941,15.054550236858766,0.826272,2.3720293035406765,0.3990547766776835,4.877970486823459,1.308094165758115,-0.5386455600281611,4.402007653547464,0.872135,1.6196681681477783,-0.9390510627301506,5.7640950109625635,3.459967530245315,-0.49038365623357977,9.273915788835595,0.870852401,3.1989433899890187,2.1239411973477047,8.922137405625627,3.055327231337828,2.6410870566431988,7.710518704438535,0.644351728,1.9724752251800266,1.5273849087899647,5.096355460177449,1.5272095032007478,-0.8048591337005557,5.530714539939694
B3LYP,4.943919441983658,-3.2032606408802526,10.060243756411047,0.833248,2.1437860846075685,0.417107875212557,4.708632839913534,1.0317073516057376,-0.4007418703156315,2.7938087725114773,0.839813,1.279488898615711,-0.8496689446160662,3.9206391053033105,3.5245917320251614,-0.7466075184674728,8.978025905303575,0.864406277,2.984641228831793,1.8641739421588104,8.344980065291132,2.2505896540248873,2.0559112520971476,4.5531451280709145,0.683169728,1.5140575091797543,1.1679884677340535,4.157023103219167,1.1722703536283012,-0.6984896481926761,3.656549161104137
PBE0,4.413949452809781,-2.734582755749375,8.98976509133457,0.840866,2.0681861707370084,0.5179967037690091,4.663332822762904,0.9002892255876508,-0.16866476742649586,2.2396964012596534,0.827507,1.0213545305151122,-0.611267658986274,2.8785977449715316,3.4295755089981483,-0.7639484258398693,8.749444582404095,0.866483847,2.8484786821723924,1.7019099638289845,7.963649527646339,1.9778233593430803,1.8019695624830092,4.208975580584669,0.693137289,1.3667206794222215,1.0145850123825846,3.72867450234331,0.9780599088980726,-0.5451734752811103,2.792269918338686
MN15,4.629068180795828,-3.0378248286626968,9.78580418346939,0.833571,2.0187839366824636,0.5033453506726437,4.563156461295087,0.9518883965980058,-0.4160998228213388,2.647765503970934,0.828108,1.1387990780817792,-0.8667533775254216,2.6445042500260927,3.4120215817980006,-0.48511443179262376,8.89783176112779,0.848460825,3.187251525581234,2.352403914484927,7.9382493442759525,2.267272466926845,2.1366105740487344,4.21747951980594,0.737529995,1.6065713862344333,1.4484863986569283,3.2854118179517613,0.9800559597316625,-0.6620746378833801,2.3120596924065877
CAM-B3LYP,4.3514151225131865,-2.6282113932829296,8.98709190299632,0.838049,2.08411307406818,0.5721269100919638,4.838222597913827,0.8092887357130429,-0.12918322221316672,2.128928512590722,0.826035,0.9202004060958646,-0.5639266579933182,2.3766051142254483,3.478053486008198,-0.9420422835573191,8.676571522969674,0.863074345,2.713463876176895,1.5328865673661363,7.846281847696723,1.7322576475562832,1.5569858875124438,3.851023111693999,0.698737222,1.16776187525828,0.8041217183957078,3.5078626614661204,0.8447010960262636,-0.430648251442271,2.1397945569326433
$\omega$B97X-V,3.9813456346083953,-2.252705549067738,8.23521061919344,0.846392,2.0761060800180964,0.6594432512008724,4.840078251861058,0.7612991211228084,0.07129682091795984,2.148895848973101,0.802263,0.7401183465889993,-0.3882444357159023,1.8660187002032562,3.405012129550672,-0.8532056197545759,8.60138496336372,0.863961489,2.7050603414133785,1.5354693928823238,7.678108767133821,1.6372566098851031,1.470796750231122,3.676058605011226,0.708002087,1.1430516490002676,0.7921929083882449,3.264414900246045,0.7088427866028512,-0.31521612670450416,1.7954543636540756
$\omega$B97M-V,4.190223682612099,-2.618065728687797,8.610931632957923,0.841172,1.9572705391939882,0.5198454265725237,4.511157043039638,0.828417027561212,-0.17273862190346134,2.5073227146845305,0.801219,0.8952306013724217,-0.6588231895171583,2.051465992247385,3.3688459821013934,-1.0175631469304653,8.066234495006604,0.85028865,2.6620763173475486,1.7006200310388657,7.057553813814749,1.6372137692567568,1.4277639598538705,3.7143456901508407,0.733447797,1.0078135990483346,0.7759566324848892,2.4778189042101317,0.7872062619481318,-0.5005408687102498,1.8593151341472725
\end{filecontents*}
\csvreader[before line=\ifthenelse{\equal{\Name}{CCSD}}{
                                    \\\hline
                                } { \ifthenelse{\equal{\Name}{$\kappa$-OOMP2}}{
                                            \\\hline
                                            & \multicolumn{3}{c|}{\textbf{MP2}} & \multicolumn{3}{c|}{\textbf{sMP2}} & \multicolumn{3}{c|}{\textbf{MP3}} & \multicolumn{3}{c|}{\textbf{sMP3}}\\\hline
                                        }{
                                            \\
                                        }
                                },
                   head to column names]{newTables/HTBH38.csv}{}{\ifthenelse{\equal{\Name}{CCSD}}{\Name & \mptwormse & \mptwomse & \mptwomaxae &   &  &  &  &  &  &  &  & }{\Name & \mptwormse & \mptwomse & \mptwomaxae &  \scamptwormse & \scamptwomse & \scamptwomaxae & \mpthreermse & \mpthreemse & \mpthreemaxae & \scampthreermse & \scampthreemse & \scampthreemaxae}}\\
        \hline
    \end{tabular}
    \caption{Root mean square errors (RMSE), mean signed errors (MSE), and max absolute errors (MAX) in kcal/mol for MP2.n:DFT on the HTBH38 dataset. }
    \label{tab:htbh38_data}
\end{table}

\subsubsection{Kinetics}

The performance of the MP:DFT models in predicting reaction kinetics was tested via the HTBH38 \cite{Zhao2005} and NHTBH38 \cite{Zhao2006} datasets, consisting of 38 hydrogen-transfer and non-hydrogen transfer barrier heights respectively. In the case of HTBH38, MP2:DFT does not represent an improvement over standard MP2, but sMP2:DFT fares quite well, halving the RMSE and yielding performance comparable to CCSD. Inclusion of third-order contributions further improves performance, with MP3:DFT having RMSEs ranging from $0.8-1.6$ kcal/mol vs the CCSD(T) benchmark. These results exhibit greater functional sensitivity than W4-11 or BH76RC (but similar to RSE43), with spread in RMSEs being 0.8 kcal/mol and the ratio of the largest RMSE to smallest being 2. The overall performance of MP3 is generally degraded by the scaling (significantly for local functionals, less so for hybrids), with the range of RMSEs being $0.7-2.1$ kcal/mol. LSDA/GGAs perform the poorest, while hybrids have the smallest RMS deviation from the CCSD(T) benchmark. Meta-GGAs like SCAN/B97M-V  reduce the error significantly relative to LSDA/GGAs, for both the unscaled and scaled models (with TPSS yielding performance between LSDA/GGA and more modern meta-GGAs). However, meta-GGAs still have larger sMP3 RMSE than hybrids in general.

\begin{figure}[htb!]
    \centering
    \includegraphics{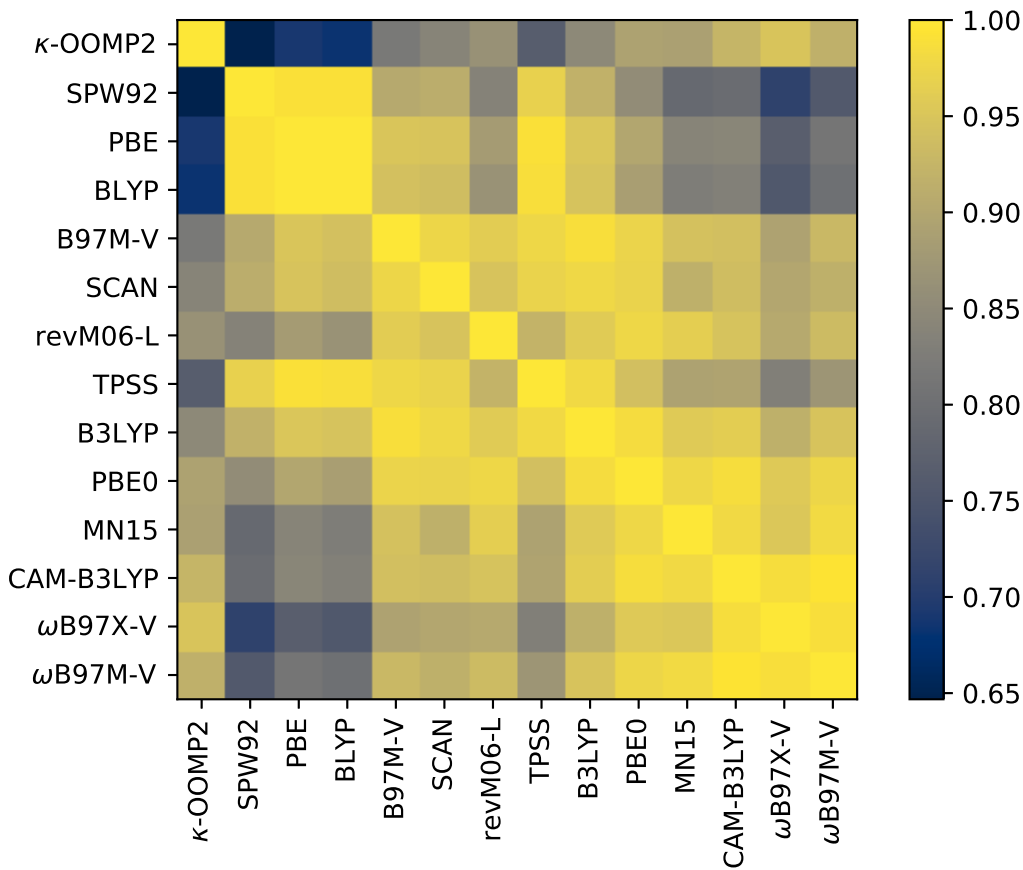}
    \caption{Correlation coefficients computed between sMP3:DFT errors for the HTBH38 dataset.}
    \label{fig:htbh38_corr}
\end{figure}

This significant difference in performance based on functional rung is nicely illustrated by the correlation coefficients computed between each functional for the HTBH38 dataset (plotted in Fig. \ref{fig:htbh38_corr}). There is clearly reduced correlation between LSDA/GGA and hybrids, with meta-GGAs occupying an intermediate spot. Indeed, correlation with $\kappa-$OOMP2 effectively provides a visual representation of Jacob's ladder.
The ability of hybrid functionals to reproduce CCSD(T) values more accurately relative to local functionals may stem from the underlying delocalization error in the reference functional. Transition states have more `fractional charge' character than equilibrium geometries\cite{mori2006many}, making DFT modeling of barrier heights somewhat more challenging than ground state thermochemistry. Indeed, a number of functionals with increased HF exchange contributions were explicitly developed for improving barrier height predictions\cite{boese2004development,lynch2000adiabatic,kang2001prediction}. It thus appears that reference orbitals with reduced delocalization error are superior for improving performance of MP theory. However \textit{all} choices of DFT orbitals perform much better than the overly localized HF orbitals.

\begin{table}[htb!]
    \centering
    
    \begin{tabular}{| c | S S S | S S S | S S S | S S S |}
        \hline
        \textbf{Name} & \textbf{RMSE} & \textbf{MSE} & \textbf{MAX} & \textbf{RMSE} & \textbf{MSE} & \textbf{MAX} & \textbf{RMSE} & \textbf{MSE} & \textbf{MAX} & \textbf{RMSE} & \textbf{MSE} & \textbf{MAX}
        \begin{filecontents*}{newTables/NHTBH38.csv}
Name,mptwormse,mptwomse,mptwomaxae,scamptwoctwo,scamptwormse,scamptwomse,scamptwomaxae,mpthreermse,mpthreemse,mpthreemaxae,scampthreecthree,scampthreermse,scampthreemse,scampthreemaxae,nbsmptwormse,nbsmptwomse,nbsmptwomaxae,fitnbsmptwoctwo,fitnbsmptwormse,fitnbsmptwomse,fitnbsmptwomaxae,nbsmpthreermse,nbsmpthreemse,nbsmpthreemaxae,fitnbsmpthreecthree,fitnbsmpthreermse,fitnbsmpthreemse,fitnbsmpthreemaxae,scamptwothreermse,scamptwothreemse,scamptwothreemaxae
CCSD,2.5342178242335245,2.067267263157896,7.645958000000007,1.0,2.5342178242335245,2.067267263157896,7.645958000000007,2.5342178242335245,2.067267263157896,7.645958000000007,1.0,2.5342178242335245,2.067267263157896,7.645958000000007,2.5342178242335245,2.067267263157896,7.645958000000007,1.0,2.5342178242335245,2.067267263157896,7.645958000000007,2.5342178242335245,2.067267263157896,7.645958000000007,1.0,2.5342178242335245,2.067267263157896,7.645958000000007,2.5342178242335245,2.067267263157896,7.645958000000007
$\kappa$-OOMP2,5.21396101004478,-2.968335022251439,21.796467683200632,0.846467308711953,2.8635236273128655,-0.5569020142988336,10.535899133300688,1.6684295632794872,1.0764299585638248,7.175271441759307,0.814672993732315,0.7852355646410848,0.32682577361296233,1.806025763362129,51.79692682629816,-38.217619854146676,161.0656756760789,1.0,51.79692682629816,-38.217619854146676,161.0656756760789,46.54819795507041,-34.172854873331424,149.14966254658577,1.0,46.54819795507041,-34.172854873331424,149.14966254658577,0.785118694534836,0.3266369827002914,1.7931944655146026
HF,7.039540561895595,4.374088503657099,25.644487975201244,0.903503,7.105204613359416,4.785165619099963,24.958869328181933,6.654041974825059,5.162332282673036,23.292584243200224,0.715722,6.55566957228534,4.938251917661943,23.96117873232602,7.039540561895595,4.374088503657099,25.644487975201244,0.903502697,7.1052050147529595,4.785166909879634,24.958867175343514,6.654041974825059,5.162332282673036,23.292584243200224,0.715722413,6.555669593037584,4.938252243206625,23.96117776098977,6.419645555043353,4.8926988350047305,23.73777250479681
SPW92,8.035797726835414,-5.455571143272164,23.727331701461736,0.815803,4.091931538409948,-1.0792811497555548,8.595474755250535,1.8063741226459205,-0.10585050083595393,5.179566918527115,0.875284,1.8990609931254607,-0.7730462604780292,4.066746270451722,3.6868168867221733,-0.7306869664283849,15.289655363982035,0.861077284,3.229859525277921,1.913549007508538,7.06913136855729,5.540083322531562,4.619033676007826,13.617243256006816,0.651680205,3.090348961392564,2.7556200785271763,6.634766000859955,1.9515204097389998,-0.6901262607594639,3.946991902517441
PBE,7.709003592466638,-5.211656083103892,24.09158185068452,0.820683,3.8926580328800493,-1.1877441543006249,9.544192523074074,1.566046218476519,-0.1632365291611973,4.795531410533627,0.87333,1.660834214470502,-0.8027198340591186,3.7750271284198664,3.6063824478398727,-0.8969567150018349,15.780440506773317,0.865055937,2.8670856373679685,1.5489739298071405,6.405068289824566,5.075035344896939,4.15146283894086,13.10667275444483,0.65006158,2.661259084735673,2.3848268767370495,5.183073009110458,1.6835890220121092,-0.7516378087674184,3.6944336575095864
BLYP,7.652812682190059,-5.172799900149812,24.388274380968795,0.817367,3.8381550138960874,-1.091280402086536,9.36088553554643,1.5107430887090794,-0.16310689901105735,4.518448536503342,0.876457,1.5756528556772378,-0.7820194014507418,3.464322627628251,3.709704866945914,-0.9157912622226533,15.34314172109309,0.865007495,2.9679653549839013,1.5263844863138534,6.849358982813698,4.973317018198228,4.093901738916101,13.563581196379047,0.644671153,2.559082393919439,2.3138133009974973,5.049331280204669,1.604532377448879,-0.7083473451444068,3.362388247488303
B97M-V,6.054131589155762,-3.850851129670933,21.05623924153329,0.841083,3.239983741792805,-0.8195701737966047,8.913793134749923,1.330680548221617,0.20876764969491193,5.08307485961673,0.818887,1.1436094987653571,-0.5264820862923747,2.681660387974164,3.3466343744025715,-0.7890699463669094,14.36838731277679,0.864353423,2.706975780485741,1.3830164595678949,6.149133158638076,4.075906933317172,3.270548832998935,11.770926788373231,0.683377911,2.237656752083371,1.9851838545324916,5.552681158001633,1.1598582920456864,-0.4900916513586084,2.638467988168996
SCAN,6.675298632248508,-4.104802370633054,22.28947012902225,0.833118,3.6259666940358843,-0.8529509478841101,9.444677919264606,1.390395382053282,0.15688025156362576,4.850953005397827,0.861896,1.3509146450945155,-0.4316751652922245,3.0200756916684792,3.525498159763245,-0.5068068154455379,15.206071792671551,0.870482395,2.9379943786121667,1.5509607534215635,6.36474317829251,4.512306071976224,3.754875806751143,11.934351341748524,0.670581665,2.573350590029144,2.3509994130486778,5.562737599814767,1.3705860540199664,-0.40879197724373345,2.9895331328383765
revM06-L,6.178194091320338,-3.8560312164654413,21.631740295593488,0.843559,3.3375017264984512,-0.8136974538207277,9.57903417736074,1.2654411347234689,0.27007775601125167,4.784047995837653,0.844851,1.0817787444269973,-0.3700839249595352,2.4341961696597565,3.331346774999238,-0.4401835243776624,14.287627406464182,0.879849018,2.8782587373947313,1.4859948568883439,6.655010028084387,4.412606266454738,3.685925448099031,12.12816088496696,0.671454696,2.5388837990101405,2.330311721399548,5.107572000999994,1.1059968077338862,-0.33534370122576346,2.388153358788818
TPSS,7.120683154746573,-4.640128316930526,23.932545969727983,0.826272,3.6991210807405364,-1.021569378591068,9.951356364384054,1.3688488094301337,-0.005567957282310912,4.663032885193559,0.872135,1.366344778460286,-0.5981660176687297,3.26669659737623,3.6255864689785056,-0.8445893599984075,15.499976927999285,0.870852401,2.8787101022604236,1.3552251987376502,6.359789082702061,4.632078792066096,3.789970999649805,13.095601926922257,0.644351728,2.3781285664965583,2.1416976162612213,4.834431275521553,1.384597136419568,-0.5665972759206453,3.2247626581205733
B3LYP,6.070350070930172,-3.907409011244141,20.70103777206336,0.833248,3.2643826277643395,-0.8044834286698146,8.3578242689577,1.33215627419381,0.26967218703815804,4.865850069068728,0.839813,1.09166562698009,-0.39944191887108876,2.435913901622367,3.578720329958033,-1.0386905214862945,14.768229458712355,0.864406277,2.725390557113371,1.0954608064949822,6.581450747584004,3.8150034679134377,3.1383906767960053,10.798658382419731,0.683169728,2.0057311544496264,1.8149649045781375,4.699150399896851,1.114657646548734,-0.3681509470101812,2.3940210613191795
PBE0,5.736886926725508,-3.5809576923021504,19.53284191731146,0.840866,3.2053340062191573,-0.7702952287623941,8.217388357813604,1.3681985027857944,0.4019187179830254,5.251774379170271,0.827507,1.0702726138743084,-0.28509958265629515,2.4490732715111534,3.476867617708761,-0.9933599995702732,14.763801253209252,0.866483847,2.6199529166897366,1.0193478828153428,6.187801191452127,3.641346116381643,2.989516410714902,10.020815043272478,0.693137289,1.9594009048178531,1.7673201578768452,4.743986739989069,1.0846001147028432,-0.27406352667357925,2.431976062319647
MN15,5.2940374717933345,-3.3759643732605147,19.03501392725063,0.833571,2.7321998019612797,-0.5823608030276294,7.332849200868111,1.3795087921316347,0.570151569014563,5.365290757254769,0.828108,0.7494391902819908,-0.10815419253498494,1.6725014230270645,3.3782281301268853,-1.2918892272426472,14.338440951760731,0.848460825,2.461879502466642,0.9359612991376023,5.96483404109205,3.4069531111434226,2.654226715032431,10.06186373274467,0.737529995,1.904010116077657,1.6184896439329117,5.332987986610462,0.7393691914936193,-0.08336624234668234,1.6275553610703675
CAM-B3LYP,5.235594648219917,-3.1592765983188285,18.490110211993823,0.838049,2.9241585876533227,-0.5170730515435352,7.328852377771213,1.3718995740371644,0.6179766720527194,5.314073472672817,0.826035,0.7986457891462708,-0.039133193127466714,1.6667860987322118,3.508009508852482,-1.1235825002353466,14.298994355484126,0.863074345,2.5954311385307673,0.8315979700741714,6.298277897512005,3.3023962425582885,2.653670770136201,9.505189329182514,0.698737222,1.7318158664143333,1.5157249566944837,4.607268096423837,0.8255256092015519,-0.021340088413939837,1.630988237826557
$\omega$B97X-V,4.783752646298354,-2.783237206314558,17.526701919406364,0.846392,2.7364019403066018,-0.4001332761715696,7.278209640536431,1.452645181558733,0.8006823576140766,5.778197954341692,0.802263,0.6763612155493363,0.09200885480152035,1.3299845147752158,3.4010465827384055,-1.041302588796141,14.380672746722766,0.863961489,2.5418494311549966,0.8322547325642549,5.932754780915239,3.1362123538203215,2.5426169751324945,8.92422712702529,0.708002087,1.683529587537786,1.4961199421054623,4.211382266710416,0.6901869665793235,0.10080721677107733,1.309311998382828
$\omega$B97M-V,4.9083697402223585,-3.0532028072230215,17.69850133343563,0.841172,2.657874378967204,-0.5101644755735228,6.988561368195164,1.4468583170428952,0.6632190801330048,5.800292546245103,0.801219,0.7069628898690766,-0.07553497905751336,1.4386905656889883,3.3509170865388778,-1.3163056420525636,14.49739657433463,0.85028865,2.3157523269401867,0.8207303060216217,5.30136049852652,3.1139216452872516,2.400116245303463,9.001397305346103,0.733447797,1.6676114308024022,1.4094958039512961,4.84022774110079,0.7214677577042967,-0.05512446206242425,1.3973560973774752
\end{filecontents*}
\csvreader[before line=\ifthenelse{\equal{\Name}{CCSD}}{
                                    \\\hline
                                } { \ifthenelse{\equal{\Name}{$\kappa$-OOMP2}}{
                                            \\\hline
                                            & \multicolumn{3}{c|}{\textbf{MP2}} & \multicolumn{3}{c|}{\textbf{sMP2}} & \multicolumn{3}{c|}{\textbf{MP3}} & \multicolumn{3}{c|}{\textbf{sMP3}}\\\hline
                                        }{
                                            \\
                                        }
                                },
                   head to column names]{newTables/NHTBH38.csv}{}{\ifthenelse{\equal{\Name}{CCSD}}{\Name & \mptwormse & \mptwomse & \mptwomaxae &   &  &  &  &  &  &  &  & }{\Name & \mptwormse & \mptwomse & \mptwomaxae &  \scamptwormse & \scamptwomse & \scamptwomaxae & \mpthreermse & \mpthreemse & \mpthreemaxae & \scampthreermse & \scampthreemse & \scampthreemaxae}}\\
        \hline
    \end{tabular}
    \caption{Root mean square errors (RMSE), mean signed errors (MSE), and maximum absolute errors (MAX) in kcal/mol for MP:DFT on the NHTBH38 dataset. }
    \label{tab:nhtbh38_data}
\end{table}

The NHTBH38 dataset presents a similar picture (except that sMP2:DFT is perceptibly worse than CCSD in this case). The sMP3 results exhibit a similar level of functional dependence as well, further highlighting the delocalization driven challenges associated with barrier heights (however unscaled MP3:DFT values show significantly less variation).  Interestingly, sMP3 performs better than MP3 for hybrid functionals, while little to no benefit is obtained from scaling for local functionals.

\begin{table}[htb!]
    \centering
    \begin{tabular}{| c | S[round-precision=2] S[round-precision=2] S[round-precision=2] | S[round-precision=2] S[round-precision=2] S[round-precision=2] | S[round-precision=2] S[round-precision=2] S[round-precision=2] | S[round-precision=2] S[round-precision=2] S[round-precision=2] |}
        \hline
        \textbf{Name} & \textbf{RMSE} & \textbf{MSE} & \textbf{MAX} & \textbf{RMSE} & \textbf{MSE} & \textbf{MAX} & \textbf{RMSE} & \textbf{MSE} & \textbf{MAX} & \textbf{RMSE} & \textbf{MSE} & \textbf{MAX}
        \begin{filecontents*}{newTables/TA13.csv}
Name,mptwormse,mptwomse,mptwomaxae,scamptwoctwo,scamptwormse,scamptwomse,scamptwomaxae,mpthreermse,mpthreemse,mpthreemaxae,scampthreecthree,scampthreermse,scampthreemse,scampthreemaxae,nbsmptwormse,nbsmptwomse,nbsmptwomaxae,fitnbsmptwoctwo,fitnbsmptwormse,fitnbsmptwomse,fitnbsmptwomaxae,nbsmpthreermse,nbsmpthreemse,nbsmpthreemaxae,fitnbsmpthreecthree,fitnbsmpthreermse,fitnbsmpthreemse,fitnbsmpthreemaxae,scamptwothreermse,scamptwothreemse,scamptwothreemaxae
CCSD,0.7220632643691166,0.5392092307692307,1.4703170000000014,1.0,0.7220632643691166,0.5392092307692307,1.4703170000000014,0.7220632643691166,0.5392092307692307,1.4703170000000014,1.0,0.7220632643691166,0.5392092307692307,1.4703170000000014,0.7220632643691166,0.5392092307692307,1.4703170000000014,1.0,0.7220632643691166,0.5392092307692307,1.4703170000000014,0.7220632643691166,0.5392092307692307,1.4703170000000014,1.0,0.7220632643691166,0.5392092307692307,1.4703170000000014,0.7220632643691166,0.5392092307692307,1.4703170000000014
$\kappa$-OOMP2,1.5682582515295387,-0.5291858297609863,4.9531083030780465,0.846467308711953,0.7753956544729143,0.13819277397261007,1.737675088058655,0.8082012319428163,-0.4419586998423667,2.4632377611356837,0.814672993732315,0.8186571474352359,-0.4581242426955071,2.0839867206382596,21.466009715344942,9.68084686761308,65.18090730535134,1.0,21.466009715344942,9.68084686761308,65.18090730535134,20.636536640367687,9.7680739975317,63.17813075567918,1.0,20.636536640367687,9.7680739975317,63.17813075567918,0.8174489536415623,-0.4569533423753136,2.0805016444720934
HF,1.790733220451875,0.16126985305491576,5.1610197592999185,0.903503,1.6419768899085663,0.44597829074038814,3.929041771482712,1.4352414328652758,0.394337315036104,4.001169022972413,0.715722,1.4898534262880392,0.32808136307901614,3.8367372347560105,1.790733220451875,0.16126985305491576,5.1610197592999185,0.903502697,1.6419765882439399,0.44597918472317083,3.929043361235667,1.4352414328652758,0.394337315036104,4.001169022972413,0.715722413,1.4898533162030845,0.32808145933587746,3.836737473643046,1.4673441792913235,0.20622744831853984,3.667371608800293
SPW92,3.288857602773682,-2.1012172253351324,9.214547022435521,0.815803,0.8163378659911841,0.3797874287339314,1.688106838582236,0.8031547835517248,-0.4886277026685575,2.407058460805626,0.875284,0.8823043791476759,-0.6897434175774424,2.1844772391707856,2.7217934876497276,2.2794815730226907,4.498514053480163,0.861077284,3.9827346000356747,3.5420947380807672,6.951986684345263,4.812556408622308,3.892071095689266,9.198793502261292,0.651680205,3.9132251051720313,3.330374243734896,6.824680562898138,0.6962933162938353,-0.5114927826492349,1.736401521426835
PBE,3.1708553493168985,-1.7997339051884302,9.36755254190695,0.820683,0.7854645074507973,0.20424807712475643,1.5602382186675676,0.771702587083355,-0.5635488983742144,2.1713284298725917,0.87333,0.9347336045082136,-0.720136453187371,1.9546464638686025,1.890862633794019,1.4411844367435385,3.7197682936252647,0.865055937,3.029579487731147,2.511927823731878,5.9086616913952845,3.747389383579812,2.677369443557755,8.10453853899902,0.65006158,2.913453320670077,2.2447808154454987,5.768674476454166,0.8083434526958525,-0.6098105407208964,1.747876901102991
BLYP,3.227596976579776,-1.7356232414303236,9.58158958671196,0.817367,0.8902390016536917,0.27271347723747746,1.9228170280212566,0.7813698979815967,-0.5612971064540679,2.18377193333329,0.876457,0.9297655279797697,-0.7063768801474399,1.9467795953374107,1.8582766688129195,1.3750588223285225,3.6100915175935953,0.865007495,2.941661729340193,2.439594699757834,5.810138572486176,3.663881887212518,2.549384957304777,8.00723109486555,0.644671153,2.8027094846234077,2.132113005761698,5.665980663231521,0.762819716683346,-0.557423940609397,1.7064670518007619
B97M-V,2.79532057199434,-1.4676474537921906,8.307947119596609,0.841083,1.0558316073466887,-0.09174534666991985,2.7680740817453646,0.8662478327564881,-0.614313411045389,2.153106669265483,0.818887,1.1533476949448114,-0.7688632995293905,3.267828287746304,1.156425367928939,0.6796128741830783,2.164995759187576,0.864353423,1.9276805543801512,1.5627713328213368,3.1373452982710788,2.27730441829299,1.5329469169298804,5.147253531344219,0.683377911,1.7167007644865462,1.262762509700573,3.1984950904986773,1.0870018250239193,-0.7024941942648366,3.137610923294968
SCAN,2.9209998536279955,-1.6826402430817358,8.675968597913783,0.833118,0.8744868308818219,-0.1448068603447911,2.4872549722196027,0.8696011399410805,-0.6601954210458693,1.9557174382212565,0.861896,1.097948796298547,-0.8013991407483103,2.8838110043794334,1.092510799477911,0.7138287822848953,1.8786521811673076,0.870482395,1.9452949681803928,1.59696080948263,3.116593619445558,2.5282663197289077,1.7362736043207616,5.861476933293144,0.670581665,1.8762865536195716,1.3994615334163352,3.647702985485413,1.0441926847501064,-0.7528325201536851,2.78741142101529
revM06-L,2.782173609441714,-1.5036640569003406,8.33445937484209,0.843559,0.976709548126144,-0.2004983091515719,2.8834724082486503,0.9043973655639403,-0.6096346074155737,2.2592372664286344,0.844851,1.1520553933624624,-0.7483423824736855,3.2018019013268733,0.9347340068484971,0.6159899973263597,1.727702381172449,0.879849018,1.6585268916422173,1.3621785448371475,2.718584680903845,2.230117493076021,1.510019446811126,5.332052045309666,0.671454696,1.6358634730323236,1.216290269545201,3.3360663508334465,1.081027829101039,-0.6831697606066744,3.0636859856650247
TPSS,3.051876342102327,-1.7031182752415908,9.104161914230296,0.826272,0.8040448144649304,0.06794168495244514,1.9331469144410747,0.8101805586512091,-0.598015071544207,1.9949607503287972,0.872135,0.9865597472398339,-0.7393190926849731,2.284248268276861,1.5612791601112885,1.1042002197022394,2.649081124984015,0.870852401,2.5581180558288987,2.0582294396040925,4.366317267099346,3.2626028074397664,2.2093034233996236,7.31099329311288,0.644351728,2.4702537604842756,1.816275378622985,4.5298980459635745,0.9051808974741817,-0.6686812042942157,2.1569435994171835
B3LYP,2.7691646875314184,-1.4562610616380505,8.26660682667147,0.833248,0.9650786839326005,-0.08149969493679397,2.4777981402896785,0.8331142161730778,-0.6372483823138132,1.735085273828986,0.839813,1.0643524554015285,-0.7684435663767246,2.7813501168141643,1.0130818038439844,0.4185084983322263,1.996932231091904,0.864406277,1.5809345209204992,1.2821832301756155,2.94228189850282,1.950649991853708,1.2375211776564639,4.856122738819515,0.683169728,1.3539300599243729,0.9780331676947173,2.786738988658568,0.9891910524094055,-0.6974612848768644,2.653144588552048
PBE0,2.551947417992336,-1.4154657361180378,7.562102472364906,0.840866,0.9277145014878475,-0.1988138824147048,2.5711825954105265,0.8338647308463687,-0.6343304654728007,1.8396596486771344,0.827507,1.0694232672806745,-0.7690708317122095,2.8267409786635094,0.8451212503986237,0.24075748488726223,1.7653503373136559,0.866483847,1.272836139281158,1.0404166863225517,2.539543901395078,1.5800647501533958,1.0218927555324993,3.9986761807094666,0.693137289,1.0594975481512559,0.7821914687245834,2.2426718622901425,1.0379577364767063,-0.7396849872300575,2.771912151864383
MN15,2.633198224909947,-1.5173495454924375,7.879421223495953,0.833571,0.9962197211754088,-0.23997992615278416,2.8284000964153186,0.8850223896978515,-0.6473170055633136,2.144308993079875,0.828108,1.1477788115471288,-0.7968686389168108,3.1301289045905554,1.0643671459648136,0.08726408678703966,2.879536555188146,0.848460825,1.26974882375491,1.0071900536591019,2.7844973253770178,1.3399450843467666,0.9572966267161636,3.482985519462723,0.737529995,0.954037619736971,0.728939181610804,1.9776905836698537,1.0692796447309336,-0.7175289068597336,2.9834300429927563
CAM-B3LYP,2.4850793815124805,-1.2468608256278477,7.608243209149318,0.838049,0.9910291162788777,-0.13679784758160363,2.6344390896719467,0.8164852503937903,-0.6193692027992401,1.8246559084896994,0.826035,1.034686929427173,-0.7285307829646186,2.8307976732489504,0.9496952637256083,0.030555688178821476,2.219188261539518,0.863074345,1.0459923258687782,0.7941760051903499,2.499891089378661,1.2124440131639105,0.6580473110074292,3.564399039120101,0.698737222,0.697633041752243,0.46900744154235474,1.8220194621178631,0.9750255914269728,-0.6694232923373272,2.7258967515632113
$\omega$B97X-V,2.3330845859385594,-1.138335557572213,7.157315241732103,0.846392,1.013587391631756,-0.15786225252236125,2.7023592355273927,0.770105793159341,-0.5730054590006334,1.7932114324795734,0.802263,1.0027341584535003,-0.6847921367018821,2.853893227409741,0.9317460651931746,-0.007615689116381646,2.145619231864103,0.863961489,0.9742059413711925,0.706890870588055,2.5266619072960808,1.0851604880403654,0.5577144094551979,3.218484577388427,0.708002087,0.6122286163887848,0.3926392005162126,1.6521774599713384,0.971866764584026,-0.6536264877610165,2.7969795215129842
$\omega$B97M-V,2.3069476963614144,-1.1369709668730417,7.0634336202547665,0.841172,0.9987640494036135,-0.10224332152651791,2.5572523282484525,0.7663170537388659,-0.576065772109613,1.8291049472977434,0.801219,1.003423970145247,-0.687563067629882,2.869590035236813,1.0124255844636554,-0.010210168546026691,2.2240621866888652,0.85028865,1.0884785893637532,0.7964357326531165,2.5983512713236223,1.0161648485556942,0.550695026217402,3.010266486268158,0.733447797,0.6247751525541052,0.4011845108790663,1.6150446472653965,0.9401110963752745,-0.6201072914432372,2.749412261732896
\end{filecontents*}
\csvreader[before line=\ifthenelse{\equal{\Name}{CCSD}}{
                                    \\\hline
                                } { \ifthenelse{\equal{\Name}{$\kappa$-OOMP2}}{
                                            \\\hline
                                            & \multicolumn{3}{c|}{\textbf{MP2}} & \multicolumn{3}{c|}{\textbf{sMP2}} & \multicolumn{3}{c|}{\textbf{MP3}} & \multicolumn{3}{c|}{\textbf{sMP3}}\\\hline
                                        }{
                                            \\
                                        }
                                },
                   head to column names]{newTables/TA13.csv}{}{\ifthenelse{\equal{\Name}{CCSD}}{\Name & \mptwormse & \mptwomse & \mptwomaxae &   &  &  &  &  &  &  &  & }{\Name & \mptwormse & \mptwomse & \mptwomaxae &  \scamptwormse & \scamptwomse & \scamptwomaxae & \mpthreermse & \mpthreemse & \mpthreemaxae & \scampthreermse & \scampthreemse & \scampthreemaxae}}\\
        \hline
    \end{tabular}
    \caption{Root mean square errors (RMSE), mean signed errors (MSE), and maximum absolute errors (MAX) in kcal/mol for MP:DFT on the TA13 dataset. }
    \label{tab:ta13_data}
\end{table}

\subsubsection{Non-covalent Interactions}
The applicability of MP:DFT for predicting non-covalent interactions was tested via the TA13 \cite{Tentscher2013}  and A24 \cite{Rezac2013} datasets.The results were counterpoise corrected to account for the small basis set size (as was the CCSD(T) benchmark). TA13 consists of nonbonded interaction energies for small radicals with closed-shell species. MP2:DFT again performs worse than MP2:HF, but scaling greatly improves performance, resulting in RMSE $\sim 1$ kcal/mol for sMP2:DFT. No systematic behavior with rungs of Jacob's ladder is seen, with the SPW92 LSDA functional being one of the best performers. Interestingly, third-order contributions do not perceptibly improve predictions as both MP3 and sMP3 approaches have quite similar RMSE as sMP2. sMP3:DFT in fact slightly worsens predictions relative to MP3:DFT, with an average RMSE increase of 0.21 kcal/mol. Interestingly, (s)MP3:$\kappa-$OOMP2 does not show a similar degradation, leading to perceptibly better performance than tested DFT methods at the sMP3 level.

\begin{table}[htb!]
    \centering
    \begin{tabular}{| c | S[round-precision=2] S[round-precision=2] S[round-precision=2] | S[round-precision=2] S[round-precision=2] S[round-precision=2] | S[round-precision=2] S[round-precision=2] S[round-precision=2] | S[round-precision=2] S[round-precision=2] S[round-precision=2] |}
        \hline
        \textbf{Name} & \textbf{RMSE} & \textbf{MSE} & \textbf{MAX} & \textbf{RMSE} & \textbf{MSE} & \textbf{MAX} & \textbf{RMSE} & \textbf{MSE} & \textbf{MAX} & \textbf{RMSE} & \textbf{MSE} & \textbf{MAX}
        \begin{filecontents*}{newTables/A24.csv}
Name,mptwormse,mptwomse,mptwomaxae,scamptwoctwo,scamptwormse,scamptwomse,scamptwomaxae,mpthreermse,mpthreemse,mpthreemaxae,scampthreecthree,scampthreermse,scampthreemse,scampthreemaxae,nbsmptwormse,nbsmptwomse,nbsmptwomaxae,fitnbsmptwoctwo,fitnbsmptwormse,fitnbsmptwomse,fitnbsmptwomaxae,nbsmpthreermse,nbsmpthreemse,nbsmpthreemaxae,fitnbsmpthreecthree,fitnbsmpthreermse,fitnbsmpthreemse,fitnbsmpthreemaxae,scamptwothreermse,scamptwothreemse,scamptwothreemaxae
CCSD,0.24744173901706973,0.22579704166666661,0.4288280000000002,1.0,0.24744173901706973,0.22579704166666661,0.4288280000000002,0.24744173901706973,0.22579704166666661,0.4288280000000002,1.0,0.24744173901706973,0.22579704166666661,0.4288280000000002,0.24744173901706973,0.22579704166666661,0.4288280000000002,1.0,0.24744173901706973,0.22579704166666661,0.4288280000000002,0.24744173901706973,0.22579704166666661,0.4288280000000002,1.0,0.24744173901706973,0.22579704166666661,0.4288280000000002,0.24744173901706973,0.22579704166666661,0.4288280000000002
$\kappa$-OOMP2,0.19722322655588298,-0.13824027240968204,0.6003655554031306,0.846467308711953,0.1368728443633877,0.09947451683897303,0.32795308024860925,0.1062883164096344,0.043377827231178556,0.373027892075265,0.814672993732315,0.07648564889440351,0.009719088540711712,0.2430235850809117,2.1488728866316817,0.07197419673542609,4.114533665536581,1.0,2.1488728866316817,0.07197419673542609,4.114533665536581,2.0841331438188915,0.25359229637628666,4.370805951964655,1.0,2.0841331438188915,0.25359229637628666,4.370805951964655,0.07646139007271341,0.009957752338867068,0.24292747399197623
HF,0.5157748455806793,0.07841688393583353,2.4051154170423272,0.903503,0.5302701373071844,0.20359480490947243,2.431646598307655,0.48856869727356056,0.1874504946135591,2.2025429616572754,0.715722,0.4884238810515634,0.1564546378373166,2.260129854129227,0.5157748455806793,0.07841688393583353,2.4051154170423272,0.903502697,0.5302702392375992,0.20359519796739187,2.4316466816154048,0.48856869727356056,0.1874504946135591,2.2025429616572754,0.715722413,0.48842387658990666,0.15645468286819778,2.260129770466803,0.4713023263421776,0.10477272704426364,2.184035620128318
SPW92,0.5020105249398112,-0.3471837743771727,1.8743782528457409,0.815803,0.20366543215851057,0.13010022963040094,0.46999672198909503,0.1690434635221773,0.025542714127630945,0.4768869066398933,0.875284,0.17002720453033834,-0.02094224261273417,0.3985308206306728,0.3900855477439524,0.24134871988651505,0.9984586813186327,0.861077284,0.6448308129423246,0.5195592516792035,1.6195062234313031,0.7905008458513796,0.6140752083913187,2.606629254156542,0.651680205,0.6348069837026827,0.48424719432425567,2.0193148795759157,0.15643470644479585,0.007573021925112698,0.36965607202193107
PBE,0.43110622757365763,-0.30181225905568926,1.5969376986692367,0.820683,0.16403934421150035,0.09436205539248428,0.429269548913326,0.1269363876065675,-0.006585223741558899,0.3613813250079547,0.87333,0.13407712148611226,-0.04398163230479981,0.3660902701842703,0.21654530149757362,0.061753701423885886,0.49692895291768413,0.865055937,0.40120997901818917,0.3108315252756409,1.0529453656487635,0.5053463520873295,0.35698073673801617,1.8598870180655118,0.65006158,0.37687959069335714,0.25366945445890526,1.3666933749007377,0.12222351090111924,-0.025547785484884677,0.34757491444675015
BLYP,0.43165929676147835,-0.31040625627397783,1.543825336919915,0.817367,0.1514275971597274,0.08044777632767049,0.4263783083511301,0.11310709658772657,-0.0501798427433394,0.2509146406482756,0.876457,0.13312710910911826,-0.082328994550155,0.40031374051586166,0.18262713077999557,0.004002719925297248,0.3975079793041556,0.865007495,0.32847523246519583,0.25045817408622506,0.8535077669237303,0.4070337252448477,0.2642291334559357,1.5774767808599943,0.644671153,0.2893637073308233,0.1717631819771487,1.113879984803392,0.11127578431098165,-0.05610695338047928,0.3743050721313308
B97M-V,0.347388033726026,-0.2619042554284255,1.1801022952561517,0.841083,0.13487600750408316,0.05229618461287314,0.4306137294090684,0.1223049705795815,0.011917663506269967,0.385979677818973,0.818887,0.11855254789994812,-0.03767504569774953,0.3003963058393273,0.13748022290379855,-0.034877674962833734,0.33188252332319257,0.864353423,0.23874772411446962,0.20251860746091144,0.6128649020807351,0.2994273064732113,0.23894424397186173,0.9879486416045538,0.683377911,0.19493882032691423,0.15224617598476975,0.6478109326471921,0.1118407777289124,-0.02649748291079344,0.29495233929110753
SCAN,0.37905998631976545,-0.27223409941900245,1.3391319011088036,0.833118,0.152760057673398,0.0632816830966725,0.45843538774249204,0.14281336558086177,0.0230695442879818,0.43474967728570335,0.861896,0.1370205697965494,-0.01771307012252755,0.32905557507485295,0.15754533523559838,-0.034825230900402696,0.3784024010363316,0.870482395,0.2556510793392231,0.19482091454273376,0.6640514513496758,0.3552562697439872,0.26047841280658157,1.2563328603946005,0.670581665,0.24124876641890394,0.16319997817719362,0.8595369788749521,0.13297790455319625,-0.009862850734769077,0.3241307768903927
revM06-L,0.30917191139191863,-0.22604708385668604,1.0994987641938527,0.843559,0.1350475789910592,0.05842615478968349,0.39991071304763093,0.12509571365565203,0.0322235932865278,0.4125546839696501,0.844851,0.1095575632822894,-0.007846844001564652,0.2977897881079845,0.13673345809313608,-0.07780301639619856,0.34250016385916426,0.879849018,0.16134885823820608,0.12286856883812995,0.40717311527021094,0.2572322662692608,0.18046766074701528,0.9131973849846964,0.671454696,0.15349975057967039,0.09561404261071225,0.5720250110564917,0.10591560799489441,0.002745853660820636,0.2910459550717712
TPSS,0.3777483586602196,-0.2548188944365246,1.4000693001158098,0.826272,0.16642473166099508,0.09468496078441076,0.42512989700218196,0.1211740213536591,0.009629870278327854,0.35489288196397983,0.872135,0.12120762086858844,-0.02418387102193682,0.3084411394231159,0.16421987651768677,0.008842785736025439,0.3491569675893831,0.870852401,0.2989075811153954,0.23460906519159389,0.7211484150530529,0.3865112212165424,0.2732915504508779,1.4177830194853627,0.644351728,0.2687602561986212,0.179240804247506,0.9726275458294551,0.11551187561914719,-0.012443812535487932,0.30007775341902754
B3LYP,0.33095179219503645,-0.2447824108988316,1.0996493663573972,0.833248,0.13966145192654844,0.06825996447026798,0.40288394433085895,0.10286957042492642,-0.01602008302253992,0.2944386648868238,0.839813,0.10970918488410598,-0.05266483403805946,0.3270711701753868,0.16223454787318642,-0.0750111687177314,0.3985429298919074,0.864406277,0.19199981694470447,0.1565180741262505,0.4969269801655132,0.200935488387427,0.15375115915856027,0.6339809803417609,0.683169728,0.11295718282232801,0.08127232859416164,0.34251590414311206,0.10046857346594028,-0.038841883750440793,0.3190466352709924
PBE0,0.31152035445958604,-0.22673028428234498,1.0648979179334654,0.840866,0.14229138767230942,0.06488331088488604,0.4232153921276023,0.12182023873220672,0.016924400940361444,0.3721493794382462,0.827507,0.11441027544061687,-0.025104326677758804,0.30466866336402987,0.16372619298090968,-0.08014855614299904,0.37380677612064783,0.866483847,0.18331294864191103,0.14494921899780697,0.5189907896608545,0.2038821943626692,0.1635061290797074,0.5958828381802017,0.693137289,0.11532682770949029,0.08873759182441605,0.31396859450771775,0.11163745058241442,-0.019533629555043227,0.30239431217023593
MN15,0.33097397473002343,-0.2560150834602029,1.0326125798918548,0.833571,0.1597727150449725,0.07971119593964195,0.3738617688566057,0.1330739704342777,0.024133694967137003,0.4253872429979517,0.828108,0.12212846967106883,-0.024021638854295316,0.29404475762551785,0.15503950624809537,-0.002538504481273939,0.3343853863715536,0.848460825,0.3056320307817319,0.2647398782553988,0.8156994983717383,0.3080369955805885,0.27761027394606597,0.6032197361678966,0.737529995,0.22588650463913507,0.20407962267149818,0.49105198425012975,0.11618433477517903,-0.006916197556496172,0.289861066526405
CAM-B3LYP,0.2857227212373428,-0.2168647969565629,0.8912787278141439,0.838049,0.13050674929751965,0.06869155810152569,0.33727460436399426,0.104762235916029,-0.007976202024148461,0.3151778373451013,0.826035,0.10287823818726985,-0.044315506441565934,0.27380096512037166,0.18692022368076147,-0.11406264086822555,0.4413281056816434,0.863074345,0.14933189353405885,0.11329210086076542,0.3589262228252603,0.12340014394870284,0.09482595406418888,0.3215496892329017,0.698737222,0.050848751001086885,0.03189559566233299,0.10740526607558731,0.09500614481634335,-0.03166512206971563,0.26767673982956364
$\omega$B97X-V,0.25826336064462085,-0.1942551247600529,0.8014006137402809,0.846392,0.1319481701510656,0.06892816985551453,0.3125517716957136,0.12032465616174134,0.022490405770346184,0.38182556822520564,0.802263,0.10416612929365075,-0.020368205200143315,0.24047040107494322,0.1789803402058152,-0.10930818819634047,0.4397144035333813,0.863961489,0.14767785370094377,0.11221647887416537,0.37055804527811786,0.12997419737130578,0.1074373423340586,0.3414419127757058,0.708002087,0.05594641602944802,0.044148099767104265,0.13270296255575298,0.10143228396981624,-0.013971786323807314,0.2381148474909942
$\omega$B97M-V,0.28220606942475107,-0.21632409336714983,0.8885490636656845,0.841172,0.13346096368281268,0.06516589862718754,0.3493184325495866,0.12212275632581915,0.006033982243459329,0.37998696685913425,0.801219,0.11337701824072496,-0.03816657838449316,0.2877966968239072,0.17864113652750047,-0.10859185006941226,0.4497395500736845,0.85028865,0.17541321309567653,0.1406120136229534,0.48223889686735166,0.1412012644741515,0.11376622554119686,0.36874043680663426,0.733447797,0.06933917452834967,0.054496190632348396,0.17421084107996876,0.10588920284235749,-0.02415404612254199,0.28109102661654983
\end{filecontents*}
\csvreader[before line=\ifthenelse{\equal{\Name}{CCSD}}{
                                    \\\hline
                                } { \ifthenelse{\equal{\Name}{$\kappa$-OOMP2}}{
                                            \\\hline
                                            & \multicolumn{3}{c|}{\textbf{MP2}} & \multicolumn{3}{c|}{\textbf{sMP2}} & \multicolumn{3}{c|}{\textbf{MP3}} & \multicolumn{3}{c|}{\textbf{sMP3}}\\\hline
                                        }{
                                            \\
                                        }
                                }, 
                   head to column names]{newTables/A24.csv}{}{\ifthenelse{\equal{\Name}{CCSD}}{\Name & \mptwormse & \mptwomse & \mptwomaxae &   &  &  &  &  &  &  &  & }{\Name & \mptwormse & \mptwomse & \mptwomaxae &  \scamptwormse & \scamptwomse & \scamptwomaxae & \mpthreermse & \mpthreemse & \mpthreemaxae & \scampthreermse & \scampthreemse & \scampthreemaxae}}\\
        \hline
    \end{tabular}
    \caption{Root mean square errors (RMSE), mean signed errors (MSE), and maximum absolute errors (MAX) in kcal/mol for MP:DFT on the A24 dataset.}
    \label{tab:a24_data}
\end{table}

The A24 set consists of interaction energies of closed shell molecules. MP2:DFT is able to slightly reduce the RMSE relative to MP2:HF (from 0.5 kcal/mol to 0.3-0.4 kcal/mol). A much more significant reduction in error is acheived by sMP2:DFT, which lowers RMSE to 0.1-0.2 kcal/mol. Inclusion of third-order contributions does not significantly improve the already excellent sMP2:DFT predictions, with both MP3:DFT and sMP3:DFT yielding RMSE on the scale of 0.1 kcal/mol as well. However, scaling does improve $\kappa-$OOMP2 performance, allowing it to reproduce the benchmark values with lower RMSE than tested DFT methods at the sMP3 level.

\subsubsection{Dipoles}
Having examined the performance of MP:DFT on some representative datasets of ground state energetics, we investigated its performance at predicting molecular dipole moments (and thereby densities, indirectly). The dataset from Ref \citenum{hait2018accurate} was employed, with the 152 species within being separated into 81 not spin-polarized (NSP) and 71 spin-polarized (SP) cases, depending on whether unrestricted HF broke spin-symmetry or not. For consistency of assessment, the regularized error expression $\dfrac{\mu-\mu_\textrm{ref}}{\textrm{max}\left(\mu_\textrm{ref}, 1\,D\right)}\times 100\%$ (vs CCSD(T) reference value $\mu_\textrm{ref}$) from Ref \citenum{hait2018accurate} was employed in order to not weigh ionic species with large dipoles or nearly non-polar molecules with small dipoles too heavily.  

\begin{table}[htb!]
    \centering
    \begin{tabular}{| c | S S | S S | S S | S S |}
        \hline
        \textbf{Name} & \textbf{RMSRE} & \textbf{MSRE}  & \textbf{RMSRE} & \textbf{MSRE} & \textbf{RMSRE} & \textbf{MSRE} & \textbf{RMSRE} & \textbf{MSRE}
        \begin{filecontents*}{newTables/DIPoles_nsp.csv}
Name,mptwormse,mptwomse,mptwomaxae,scamptwoctwo,scamptwormse,scamptwomse,scamptwomaxae,mpthreermse,mpthreemse,mpthreemaxae,scampthreecthree,scampthreermse,scampthreemse,scampthreemaxae,scamptwothreermse,scamptwothreemse,scamptwothreemaxae
CCSD,2.8479226965330278,1.5806887004514425,10.15952980688497,1.0,2.8479226965330278,1.5806887004514425,10.15952980688497,2.8479226965330278,1.5806887004514425,10.15952980688497,1.0,2.8479226965330278,1.5806887004514425,10.15952980688497,2.8479226965330278,1.5806887004514425,10.15952980688497
$\kappa$-OOMP2,5.730296943052797,-1.1173945603942321,22.42141061803299,0.846467308711953,3.734302068271177,0.535557387320307,18.577145163049526,2.6024718028997387,1.4743214163214842,9.12343216314515,0.814672993732315,1.7407504075627176,0.9829302226657978,7.512428117726833,1.7403240964355078,0.9830147611459082,7.5267743766817325
HF,3.6513886025979962,1.3194961062616386,14.905237871933346,0.903503,3.4998440420493453,2.017330130103324,13.50021642453375,3.942247037977466,2.479854525060682,11.700336802897455,0.715722,2.941387953394839,2.146895690620962,10.177994561862613,3.174731064191195,2.0177644126551573,10.253553129204391
SPW92,8.52727176307637,-3.165258587257412,29.525591334677546,0.815803,4.782032863937996,-0.4168028747585744,22.875299682654703,2.763782150118793,0.9201964864749286,8.713940097699458,0.875284,2.4428730159750582,0.3998647123828415,10.103525499898335,2.406115937870856,0.384248212271724,11.27673871344087
PBE,8.376017257621719,-3.1158511534462887,29.207873056597855,0.820683,4.745175937025614,-0.43239432707124303,22.710981145870747,2.655692201249088,0.8916424336212436,8.63535069525211,0.87333,2.3305221333569204,0.3739476273192888,9.940154563391712,2.3019633157935733,0.36503139952791286,10.781669211654787
BLYP,8.310937949193793,-3.060868463397385,28.775776179626867,0.817367,4.619567476677657,-0.32396261411152966,22.337925027112128,2.668274329285823,0.948022676193807,9.065524194769235,0.876457,2.2927003190695436,0.4426272405054389,9.606148573647644,2.2402852867756167,0.434795804330958,10.709749856088413
B97M-V,6.578472350367887,-2.1054146087729024,23.781244846769255,0.841083,4.0085656193448695,-0.024399390085670537,19.254842390709527,2.535397412015975,1.158821214317614,8.94970825801741,0.818887,1.8101915336017547,0.5579005806367078,7.962582460814859,1.802808913163614,0.5588486087141172,8.576165623328736
SCAN,7.152389200785363,-2.297169066430194,26.170486302985406,0.833118,4.242187863699993,-0.08706058833268311,20.751692911992958,2.6200122443465133,1.1301122031112703,9.321973789158623,0.861896,2.038760429245101,0.6475157509145802,8.086085195735283,2.008352286941845,0.6426838379418205,8.575595300487471
revM06-L,6.36334213624912,-1.4753141499671938,23.743118659178965,0.843559,3.9973791535141325,0.35307298590101455,19.53138651657208,2.6433257245701602,1.3978281249614182,9.280610944527806,0.844851,1.9741067333581412,0.9434074393100917,7.358480719132309,1.9472454130562253,0.9414436786222863,8.098219243564165
TPSS,7.8811335515138214,-2.7180208882785477,27.619281648139193,0.826272,4.5618083219483,-0.2593315052024588,21.640831729874844,2.5697464131846046,1.0016844297084901,8.763575492446803,0.872135,2.164364604283142,0.5167491766533961,8.832360589037924,2.1393269522032194,0.5108665705043972,9.438914308971501
B3LYP,6.556183260939019,-1.8611192867428994,24.620021100177237,0.833248,3.970117581335228,0.17219686310859778,19.957767161791374,2.64322060649344,1.1984031237885042,9.421244578653297,0.839813,1.9854721824059116,0.6992500590591059,8.055590848469674,1.9524172284430341,0.701826373674532,8.753122665258218
PBE0,6.21110027313993,-1.6644953636869875,23.98458454401785,0.840866,3.9024421188583935,0.1985794614639158,19.696306186886904,2.6264114296124603,1.2180921485940648,9.18547641833536,0.827507,1.9209253487059765,0.7122073946502169,7.957341031542553,1.9074673402228404,0.7131621503345827,8.287381038978925
MN15,6.002465252578471,-1.0326717340012712,23.590613872697375,0.833571,3.7954477750553806,0.7191173444318052,19.024532937381057,2.712275937609025,1.4841180011349655,9.851418078170424,0.828108,1.9137898894304433,1.0432092613373964,7.04656843196455,1.8857793435608645,1.0557067374377027,7.990723167581438
CAM-B3LYP,5.8237997308081075,-1.16587964682162,23.476235291866278,0.838049,3.747246404527617,0.5568940861134604,19.296047913482262,2.7233111495107862,1.4081658148075114,9.652876452156963,0.826035,1.9543519511144263,0.9525241407333148,7.729450181228071,1.9283285406057677,0.9581286110850288,8.405585798162107
$\omega$B97X-V,5.474121971531611,-0.84880741657301,22.929759852124388,0.846392,3.7162443517164756,0.692978731096573,19.099030837032448,2.7279552400409646,1.4953015917822703,9.520515391660027,0.802263,1.8813418787962555,1.0243335097329453,7.850253332648604,1.8769673163454141,1.029291223962817,8.231038690353875
$\omega$B97M-V,5.556271213034631,-1.0325402454629315,23.27289559461766,0.841172,3.669207038904275,0.6181379258337548,19.177959877327766,2.7200992721995663,1.422691752647134,9.446062276026977,0.801219,1.8721638101764253,0.9271903160695146,8.192647361067118,1.866235412049715,0.9378700403828768,8.958027169886973
\end{filecontents*}
\csvreader[before line=\ifthenelse{\equal{\Name}{CCSD}}{
                                    \\\hline
                                } { \ifthenelse{\equal{\Name}{$\kappa$-OOMP2}}{
                                            \\\hline
                                            & \multicolumn{2}{c|}{\textbf{MP2}} & \multicolumn{2}{c|}{\textbf{sMP2}} & \multicolumn{2}{c|}{\textbf{MP3}} & \multicolumn{2}{c|}{\textbf{sMP3}}\\\hline
                                        }{
                                            \\
                                        }
                                },
                   head to column names]{newTables/DIPoles_nsp.csv}{}{\ifthenelse{\equal{\Name}{CCSD}}{\Name & \mptwormse & \mptwomse & & & & & & }{\Name & \mptwormse & \mptwomse &  \scamptwormse & \scamptwomse & \mpthreermse & \mpthreemse & \scampthreermse & \scampthreemse }}\\
        \hline
    \end{tabular}
    \caption{Root mean square regularized errors (RMSRE) and mean signed regularized errors (MSRE), in \%  for MP:DFT on the non spin-polarized (NSP) subset of the dipole dataset.}
    \label{tab:dipole_cs_data}
\end{table}

Table \ref{tab:dipole_cs_data} shows that MP2:HF is already quite effective at predicting dipole moments for NSP (i.e. unambiguously closed-shell) species, with a RMS regularized error (RMSRE) of 3.7\%, vs 2.8\% for CCSD. Similar to many other datasets, MP2:DFT actually degrades performance, with LSDA/GGAs predicting RMSRE $> 8\%$. Rung 3 meta-GGAs have lower RMSRE, and hybrid functionals yield better results still, suggesting that the error is influenced by delocalization error in the reference functional. Nonetheless, neither $\kappa-$OOMP2 nor range separated hybrid functionals with low delocalization error (like $\omega$B97X-V) are able to attain MP2:HF level results.  sMP2:DFT leads to some improvement, with $\kappa$OOMP2 and range seperated hybrids attaining (s)MP2:HF level accuracy, but the error still appears to depend on delocalization error in the underlying functional. In contrast, MP3:DFT RMSEs are virtually functional agnostic and represent a significant improvement over MP3:HF, attaining close to (or slightly better than) CCSD level accuracy for all functionals tested. Scaling the third-order term leads to further improvement, with mGGA and hybrid functionals leading to $\sim 2\%$ RMSE, comparable to the best double hybrid density functionals (and much better than predictions from the best hybrid functionals)\cite{hait2018accurate}. $\kappa-$OOMP2 orbitals lead to slightly better performance, with an RMSE of 1.7\%. It is quite interesting that $c_3$ scaling parameters trained on W4-11 energies prove to be quite effective for a molecular property like dipole moments (i.e. derivative of the  energy vs an applied field).

\begin{table}[htb!]
    \centering
    \begin{tabular}{| c | S S | S S | S S | S S |}
        \hline
        \textbf{Name} & \textbf{RMSRE} & \textbf{MSRE}  & \textbf{RMSRE} & \textbf{MSRE} & \textbf{RMSRE} & \textbf{MSRE} & \textbf{RMSRE} & \textbf{MSRE} 
        \begin{filecontents*}{newTables/DIPoles_sp.csv}
Name,mptwormse,mptwomse,mptwomaxae,scamptwoctwo,scamptwormse,scamptwomse,scamptwomaxae,mpthreermse,mpthreemse,mpthreemaxae,scampthreecthree,scampthreermse,scampthreemse,scampthreemaxae,scamptwothreermse,scamptwothreemse,scamptwothreemaxae
CCSD,4.6155214370235225,2.0316681778433145,11.6,1.0,4.6155214370235225,2.0316681778433145,11.6,4.6155214370235225,2.0316681778433145,11.6,1.0,4.6155214370235225,2.0316681778433145,11.6,4.6155214370235225,2.0316681778433145,11.6
$\kappa$-OOMP2,14.477193857118358,-2.761706314367478,57.539757861753124,0.846467308711953,8.913934883851649,-0.5557683562587532,44.487362464517894,9.425485208895978,1.5655355120696461,50.72787794071447,0.814672993732315,8.505058212972566,0.7528860732644029,51.99030325317949,8.499754263635726,0.7519967536433538,51.97345367084427
HF,51.02815206384378,14.356304069664652,325.34630499545955,0.903503,45.581297544887015,13.551486008732061,285.51007914112233,29.00508093225374,9.991474167654218,133.92022903637414,0.715722,34.57910048645775,11.19598337129133,186.40075758922544,31.8898632052287,10.46701149195604,157.57955074210207
SPW92,14.567528207153986,-4.793122965714588,32.75773209929623,0.815803,7.629839458475956,-1.0222607933149295,26.081924536746627,6.395495075323831,1.2913255526542846,23.517291299066194,0.875284,5.3153142719434525,0.5218328592332013,24.359301589691402,4.987667805496108,0.45798076358411266,24.365355330333742
PBE,14.089074895896193,-4.334376638360805,30.844380581342456,0.820683,7.475532793823842,-1.0654899889470768,23.600803238228725,6.674904975083718,1.48672044465264,24.849054680987102,0.87333,5.641180694994635,0.7396079478921309,24.290115258310678,5.385382032025509,0.6675110767530025,23.482304143478917
BLYP,14.433718857384733,-4.432506208986833,32.503557491280624,0.817367,7.651908761334373,-0.9887231471395967,26.718148878533665,6.931568547158973,1.4310277790052701,28.028293011982186,0.876457,5.969182541825496,0.6969915100647991,27.54748512237823,5.644008294101807,0.6171453881202855,26.4861701698912
B97M-V,13.485995232039288,-2.808501157165525,46.03801360942663,0.841083,7.9254910615619805,-0.575708307631491,32.367045619929705,7.301798658010297,0.4871914093893369,31.8919258171178,0.818887,5.516803520188916,-0.12098454892941933,31.72581325880735,5.331565540724501,-0.10849687225619083,31.652152701953774
SCAN,13.086307144323278,-3.309207113898983,37.23615924710305,0.833118,6.978415533279713,-0.7741961961545499,22.75748623686425,6.215291237832565,0.8970854280023821,21.290536262429708,0.861896,4.795104215677954,0.3071472062414787,20.242389994525656,4.63023532861971,0.2919336090041876,20.26583129453844
revM06-L,14.01544459486613,-1.9444262474626213,60.27179252518955,0.843559,8.30803553394407,0.028468344261671972,34.4274159548855,8.00400968832027,0.5303251481807822,47.21865561236448,0.844851,5.933175928640198,0.13550277167058222,30.541620074076352,5.577873633280447,0.17091755610411596,26.66537987779017
TPSS,13.27690062485427,-3.850444357048761,30.159225175599406,0.826272,7.110841543189877,-0.8839485251895591,20.280672908472603,6.130656893832365,1.520962018287577,22.327284568020026,0.872135,5.02739494388421,0.8246816958914025,20.402449351072978,4.847870977201431,0.7765505503834766,19.903143888578377
B3LYP,12.825710567011258,-2.463245450120459,34.09886872659351,0.833248,7.0220092562778955,-0.22850409021398616,24.701834212032576,7.164922796757995,1.2075135821884853,29.05149615321594,0.839813,5.7105992962941166,0.6091372183777883,28.308075157782035,5.4795436412611425,0.5960467158080294,27.560729403626503
PBE0,11.855672772972689,-2.16941054051832,37.01077919459579,0.840866,6.55345803506415,-0.1982568843151379,21.73703160969843,6.484263824621362,1.3365818797415774,23.031041959438838,0.827507,4.934560534826164,0.7214157094351527,22.26990563597178,4.8358772078987755,0.711091387477216,21.96685134863814
MN15,13.378243471872986,-2.0308369934669352,45.57522168951567,0.833571,8.077251655631285,-0.09836742642732282,38.41819801985824,6.717012231471231,0.845907206916554,31.741033594993052,0.828108,5.444196283311119,0.3401305790109287,33.44933542288231,5.3200820885615565,0.34827571145391645,33.80848006377726
CAM-B3LYP,11.827261519327815,-1.6845139241558322,36.983733578045815,0.838049,6.562334112058231,0.22461498074164638,23.418883887312937,6.813583433984877,1.440143487277657,28.247550826569846,0.826035,5.331740682736295,0.8861893550861819,27.43541882576813,5.131727334763821,0.8788723635951947,26.719857106448604
$\omega$B97X-V,10.86096021238551,-1.2907265662122447,31.845068073407063,0.846392,6.287455545478201,0.4775962908911848,21.29329750275348,5.861888503930092,1.3217295717085942,21.49788592098549,0.802263,4.002670847652124,0.7945026671482681,17.176429649375674,3.921632314220238,0.8031040265283848,17.26368971231547
$\omega$B97M-V,11.530054104783211,-1.5399759514208486,37.51563053271027,0.841172,6.577279605499169,0.26259007622767894,23.120773055311822,6.018922162739167,1.225257058442508,21.263490645879738,0.801219,4.143260175077687,0.6646385368094662,18.56978850907524,3.9887968805871177,0.6723286137474316,18.807643926509705
\end{filecontents*}
\csvreader[before line=\ifthenelse{\equal{\Name}{CCSD}}{
                                    \\\hline
                                } { \ifthenelse{\equal{\Name}{$\kappa$-OOMP2}}{
                                            \\\hline
                                            & \multicolumn{2}{c|}{\textbf{MP2}} & \multicolumn{2}{c|}{\textbf{sMP2}} & \multicolumn{2}{c|}{\textbf{MP3}} & \multicolumn{2}{c|}{\textbf{sMP3}}\\\hline
                                        }{
                                            \\
                                        }
                                },
                   head to column names]{newTables/DIPoles_sp.csv}{}{\ifthenelse{\equal{\Name}{CCSD}}{\Name & \mptwormse & \mptwomse & & & & & &}{\Name & \mptwormse & \mptwomse &  \scamptwormse & \scamptwomse & \mpthreermse & \mpthreemse & \scampthreermse & \scampthreemse }}\\
        \hline
    \end{tabular}
   \caption{Root mean square regularized errors (RMSRE) and mean signed regularized errors (MSRE) in \%  for MP:DFT on the spin-polarized (SP) subset of the dipole dataset.}
    \label{tab:dipole_os_data}
\end{table}

A somewhat different picture is obtained for dipole moments of SP species. Unrestricted MP2 (UMP2) is known to be problematic for prediction of molecular properties, as the one particle density matrix could be non $N$-representable (i.e. occupation numbers greater than 2 or less than 0) in regions where the orbital rotation hessian is singular (or nearly so)\cite{kurlancheek2009violations}. The consequences of this behavior are clearly seen for the SP subset of species, with MP2:HF having an RMSRE of 50\%. Use of DFT orbitals significantly reduces spin-contamination, permitting a dramatic reduction of RMSRE to 10-15\% for MP2:DFT (which is still significantly larger than the corresponding NSP RMSRE). sMP2:DFT leads to further improvements, leading to 6\%-8\% RMSRE vs 46\% RMSRE with sMP2:HF. Third-order contributions also help reduce error (with the MP3 RMSRE going down relative to MP2) for each orbital choice, and scaling provides some further improvement. However, sMP3:DFT SP RMSREs are still fairly large relative to the corresponding NSP values, and a strong level of functional dependence persists. It is nonetheless encouraging that sMP3:$\omega$B97X-V and sMP3:$\omega$B97M-V marginally outperform CCSD even for the SP dataset. Interestingly, $\kappa-$OOMP2 orbitals yield much worse results for this dataset, with sMP3:$\kappa-$OOMP2 having over double the RMSRE of the best sMP3:DFT methods. The most significant instance of MP:$\kappa-$OOMP2 failure is prediction of near zero dipole moments for the very challenging NaLi molecule\cite{hait2018accurate} (vs a CCSD(T) benchmark of 0.59 D). The second most significant case of failure is BH, where (s)MP3:$\kappa$OOMP3 predicts a dipole of $\sim$ 2 D vs 1.4 D from CCSD(T). It is also worth noting that there exists a slightly higher energy spin unpolarized (i.e. restricted) $\kappa-$OOMP2 solution for both NaLi and BH that leads to much more accurate dipole moments with (s)MP3. It thus appears that sMP3:DFT can be effective for predicting dipole moments for both NSP and SP species (though with much lower error for NSP species), although sMP2:HF is the best $O(N^5)$ scaling approach. It will be interesting to see whether similar behavior holds for other properties like static polarizabilities, where MP2:HF is excellent for NSP species, but more problematic for SP cases\cite{hait2018accuratepolar}.

\begin{figure}[htb!]
    \centering
    \begin{minipage}{0.48\textwidth}
    \centering
    \includegraphics[width=\textwidth]{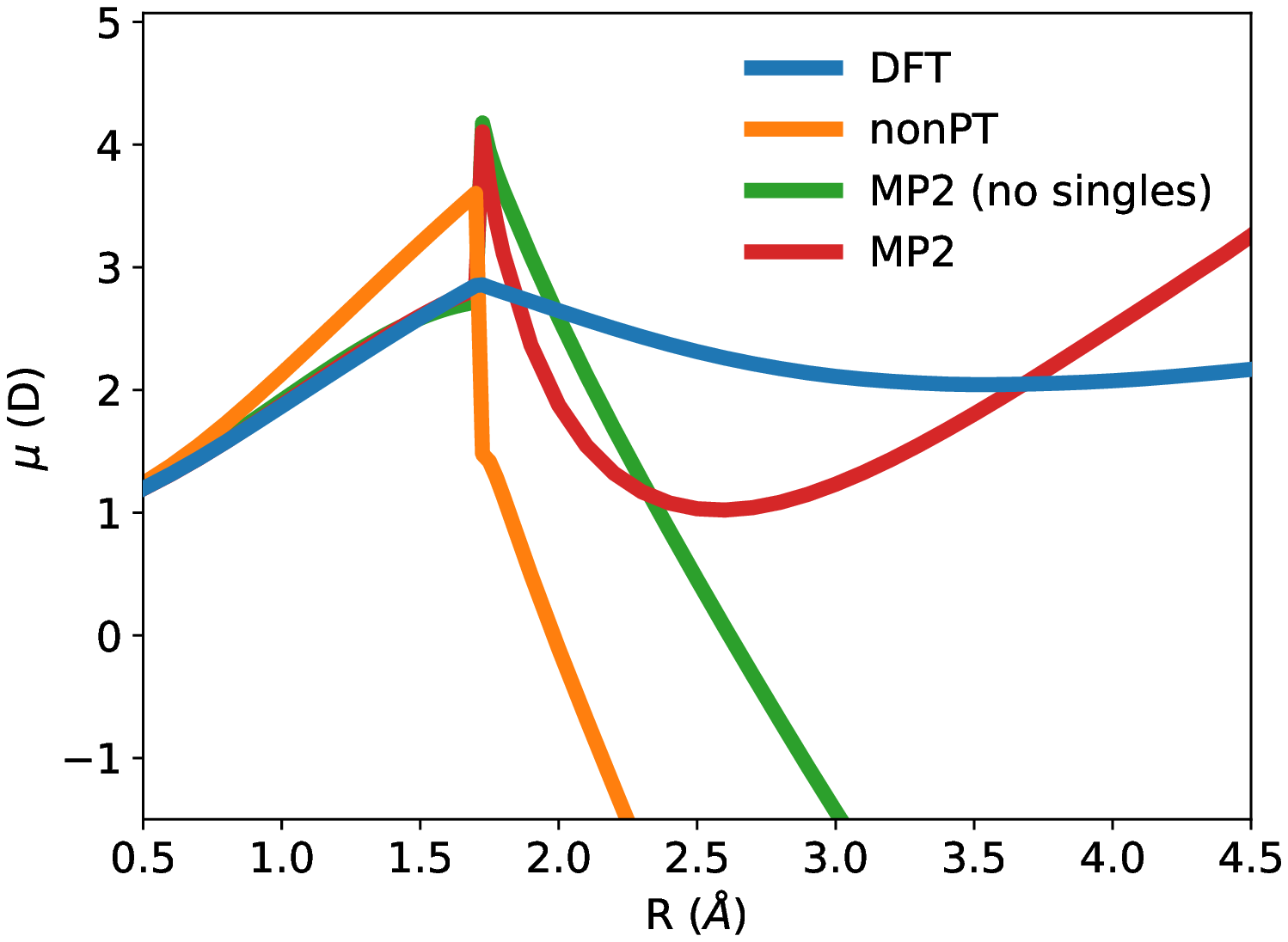}
    \label{stretch:pbe_ns}
    \end{minipage}
    \begin{minipage}{0.48\textwidth}
     \centering
    \includegraphics[width=\textwidth]{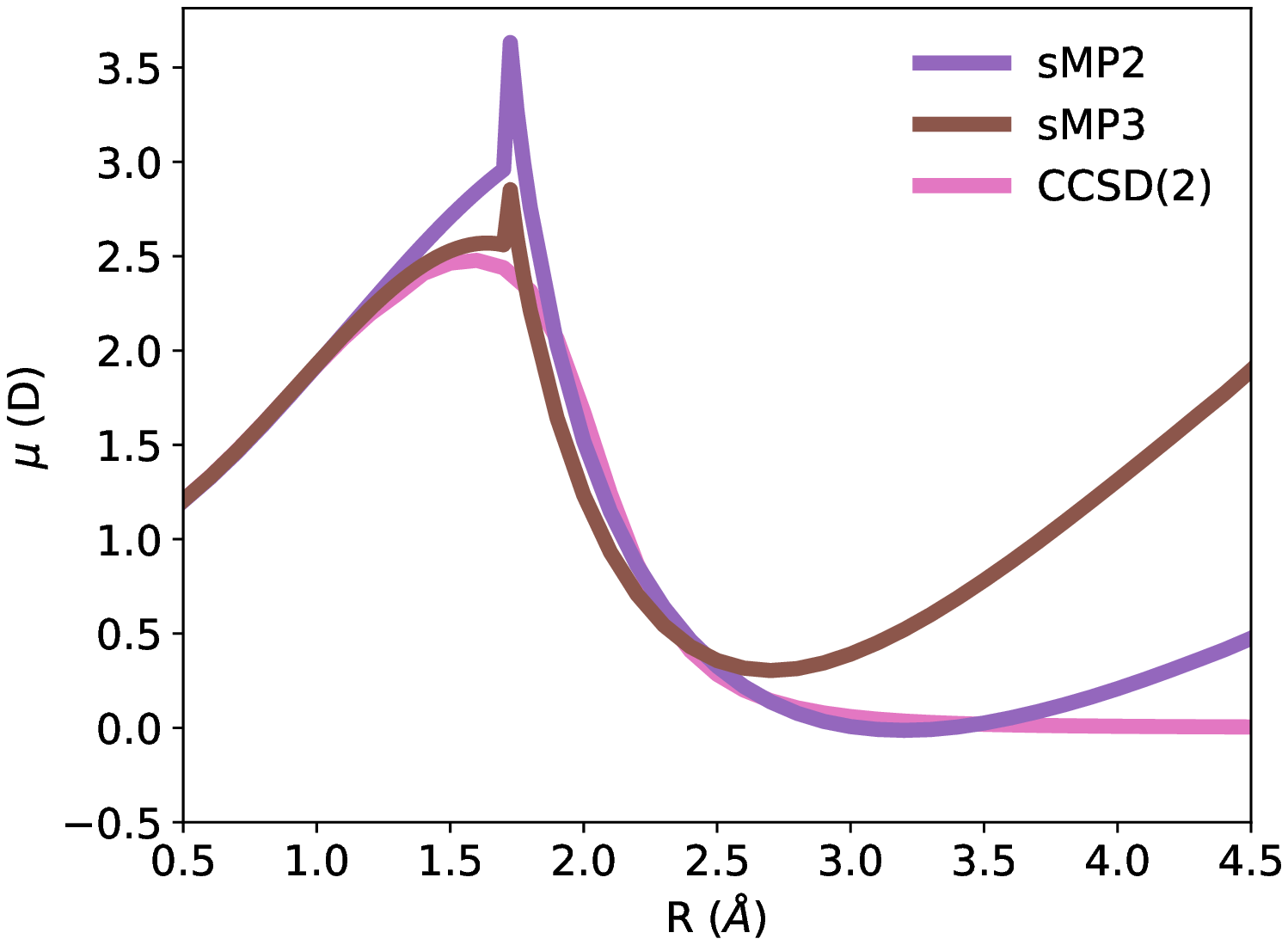}
    \label{stretch:pbe_scaled}
    \end{minipage}
    \begin{minipage}{\textwidth}
    \vspace*{-25pt}
    \subcaption{MP:PBE}
    \label{fig:fh_pbe}
    \end{minipage}
    \begin{minipage}{0.48\textwidth}
    \centering
    \includegraphics[width=\textwidth]{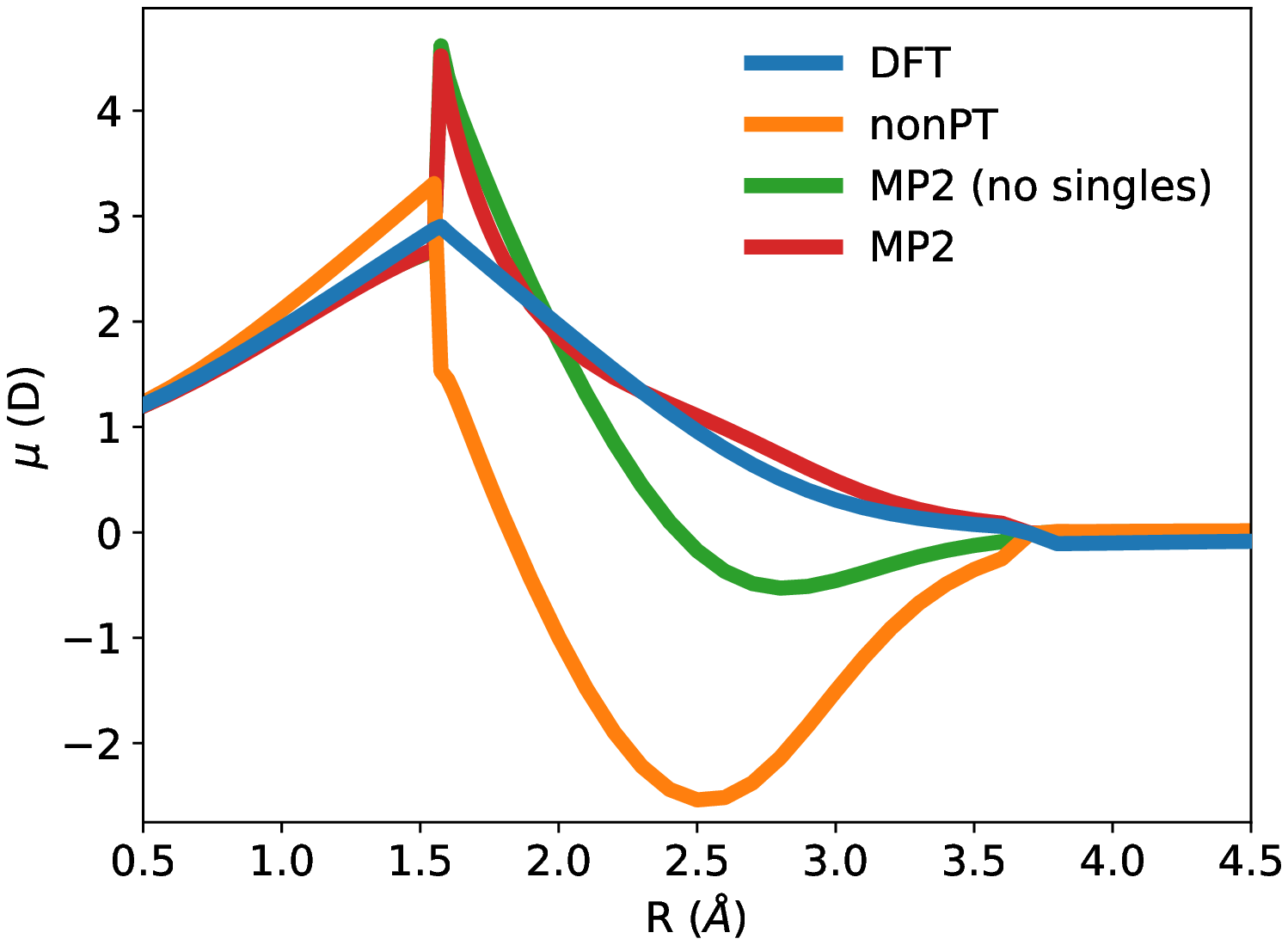}
    \label{stretch:lrcwpbeh_nbs}
    \end{minipage}
    \begin{minipage}{0.48\textwidth}
    \centering
    \includegraphics[width=\textwidth]{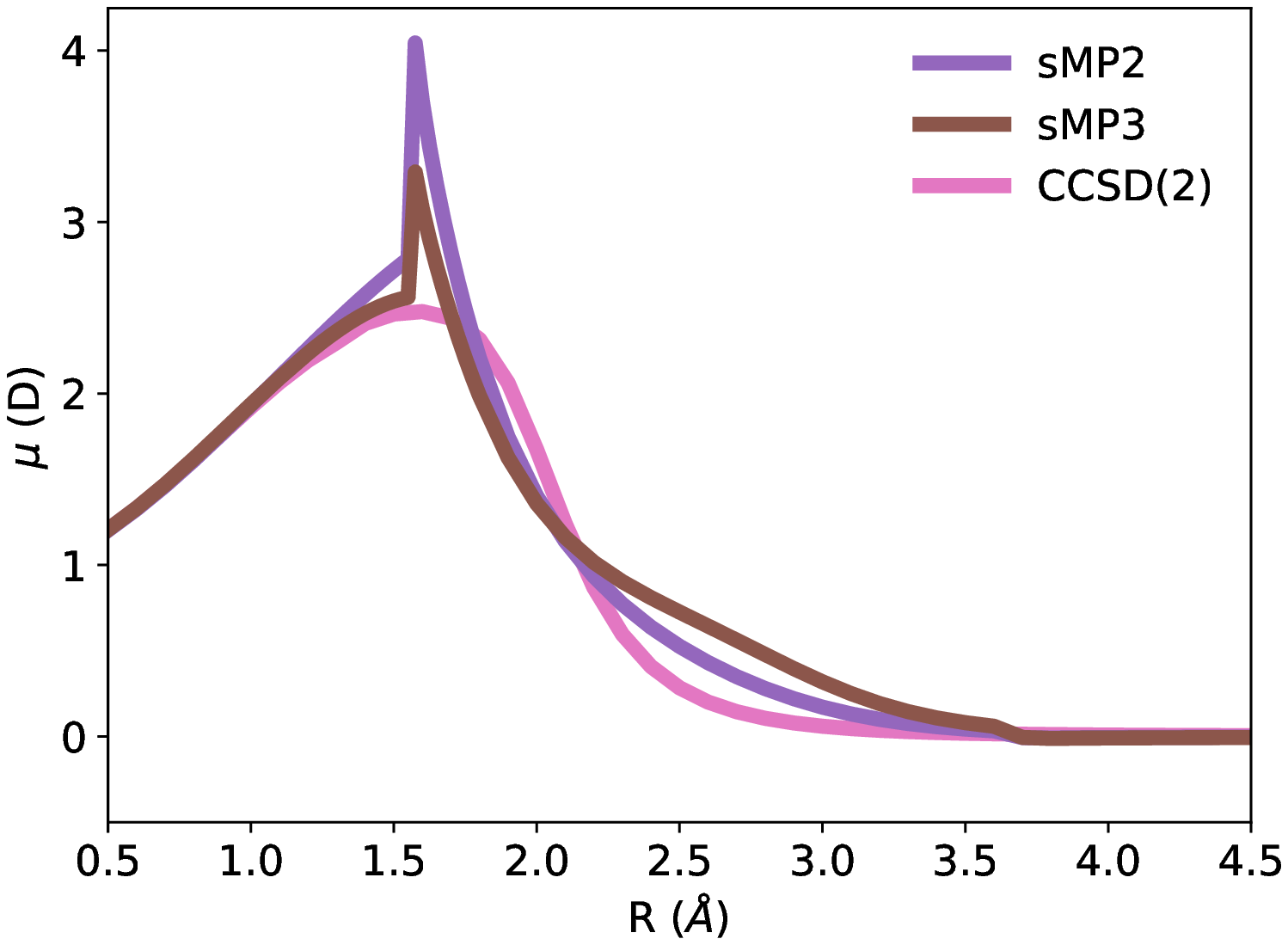}
    \label{stretch:lrcwpbeh_scaled}
    \end{minipage}
    \begin{minipage}{\textwidth}
    \vspace*{-25pt}
    \subcaption{MP:LRC-$\omega$PBEh}
    \vspace*{-15pt}
    \label{fig:fh_lrcwpbeh}
    \end{minipage}
    \caption{Dipole moments $(\mu)$ as a function of internuclear distance ($R$) for the FH molecule, for various MP:PBE and MP:LRC-$\omega$PBEh approaches. A CCSD(2)\cite{gwaltney2001second} benchmark is also supplied. MP2 (no singles) does not contain non-Brillouin singles, while the non-perturbative (nonPT) approach does not contain any perturbative correction whatsoever (i.e. is obtained from the HF energy computed from DFT densities). MP3:DFT was very similar to sMP3:DFT and was thus not shown separately. A positive value of $\mu$ indicates a polarity of H$^+$F$^-$.}
    \label{stretch:all}
\end{figure}

\subsection{Non-equilibrium configurations}\label{sec:stretch}
We next examine bond dissociation problems where MP theory is particularly challenged. Orbital degeneracies lead to divergent behavior when restricted orbitals are employed, resulting in catastrophic failure. UMP2 has its own challenges, with a derivative discontinuity in the energy at the Coulson-Fischer (CF) point as well as a discontinuity in first-order (and higher) properties, due to non $N$-representability of the one particle density matrix\cite{kurlancheek2009violations}. Furthermore, xDH functionals are known to yield unphysical behavior (such as inversion of dipole moments) for highly stretched bonds, due to non-Hellman-Feynman terms coming from the non-perturbative component of the energy (as the orbitals are obtained from another method and are thus not self-consistent)\cite{hait2018communication}. The use of non self-consistent orbitals in MP:DFT is thus similarly expected to pose a challenge, making bond dissociations an interesting regime for characterizing the limitations of MP:DFT.  

A plot of the MP:DFT dipole moment as a function of bond stretch for the hydrogen fluoride (FH) molecule is shown in Fig \ref{stretch:all}. The behavior of MP2 without non-Brillouin singles and the contribution from the non-perturbative (nonPT) terms (i.e. HF:DFT) is also plotted. It can be seen that the MP2 CF point discontinuity is unsurprisingly retained by MP:DFT, though the precise location of the CF point is functional sensitive.  It can also be seen that the nonPT models predict unphysically large negative dipole moments at large stretches for all functionals, spuriously indicating a polarity of H$^-$F$^+$. Inclusion of MP2 doubles (i.e. MP2 without non-Brillouin singles) partially amerliorates this, but the qualitatively problematic behavior persists. This is exactly the behavior seen for xDH functionals, where the non-Hellman-Feynman terms leads to an overcorrection of the delocalization error in the underlying functional\cite{hait2018communication}. Orbitals from functionals with less delocalization error have smaller failures, as can be seen from the relative ``success" of the range-separated hybrid LRC-$\omega$PBEh\cite{lrcwpbeh} functional, compared to the local PBE functional. 

Interestingly however, the unphysical dipole inversion is eliminated upon inclusion of non-Brillouin singles in MP2, yielding the correct H$^+$F$^-$ polarity at all distances. While exponentiated single excitations are equivalent to orbital rotation\cite{thouless1960stability}, it is a little surprising that inclusion of merely the first-order term in the power series of the orbital rotation operator is adequate to address the dipole inversion problem. This does not however mean that MP:DFT is qualitatively successful in the non-equilibrium regime, as the behavior of the models is very functional sensitive. MP:PBE predicts existence of partial charges in the dissociation limit (as can be seen from the asymptotically divergent dipoles in Fig \ref{fig:fh_pbe}), similar to the behavior predicted by the baseline functional on account of delocalization error\cite{ruzsinszky2006spurious,dutoi2006self,hait2018accurate}. Indeed, the dissociation limit partial charges appear to be larger for MP:PBE than PBE itself, based on asymptotic $\mu$ divergence rates. The use of local functionals for generating MP:DFT reference orbitals for cases with catastrophic delocalization error is thus clearly problematic, and cannot be recommended.  

MP:LRC-$\omega$PBEh does however go to the correct dissociation limit of neutral atoms, similar to the underlying functional. A discontinuity at the CF point is nonetheless retained (in contrast to the derivative discontinuity predicted by the reference DFT functional itself), although reasonable agreement with a CCSD(2)\cite{gwaltney2001second} benchmark is observed at other regimes.  Overall however, MP:DFT appears to be ill-suited for non-equilibrium problems like stretched bonds, due to limitations of both standard MP theory and delocalization error from DFT functionals. Even dissociation of nonpolar bonds could prove problematic, as many exchange-correlation functionals yield unphysical behavior for highly stretched nonpolar bonds (relative to unrestricted HF) due to incomplete spin-localization\cite{hait2019well}, making MP:DFT on such references likely much less optimal than MP:HF.

\section{Discussion}
The good performance of MP:DFT relative to MP:HF naturally raises questions about the factors that contribute to improved accuracy. A related, but equally intriguing issue is the relatively low impact the choice of the reference functional has on MP3:DFT errors, even in the absence of functional specific scaling factors. Traditional wisdom\cite{cremer2011moller} would indicate that the main issue with MP:HF is spin-contamination in HF orbitals, leading to unreasonably slow convergence of the MP series. Indeed, we do observe significant reduction of error for datasets like RSE43 or SP dipole moments, where several species are heavily spin-contaminated at the HF level (but not with the functionals studied). However, this cannot be the sole factor as 
Sec \ref{sec:spincont} demonstrated that significant reduction in error is possible even for systems without heavy spin-contamination. Similarly, the A24 results in Table \ref{tab:a24_data} show improved performance vs MP:HF for unambiguously closed-shell species.

Another potential factor behind improved performance could be the higher quality of DFT densities over HF\cite{hait2018accurate} for molecules at equilibrium. This general idea has precedence in quantum chemistry, with the density-corrected DFT (DC-DFT)\cite{kim2013understanding,kim2014ions} approach evaluating DFT energies with HF densities for problems where self-consistent DFT densities are qualitatively problematic (possibly due to delocalization error\cite{kim2013understanding} or incomplete spin localization\cite{hait2019well}). However, improved reference densities are unlikely to be responsible for the success of MP:DFT for many of the datasets studied. The unscaled MP3:PBE and MP3:PBE0 errors are quite similar for the non-barrier height datasets, despite PBE0 predicting significantly more accurate densities than PBE\cite{brorsen2017accuracy,hait2018accurate}. Similar parallels can be drawn between many other functional pairs, indicating that accurate reference densities are not the most critical factor. This is not to insinuate that reference density quality does not matter at all---the barrier height datasets demonstrate a dependence on delocalization error in the reference functional, and Fig \ref{stretch:all} clearly shows that a catastrophically poor reference is unsalvageable with MP theory. However, even MP3:SPW92 exceeds MP3:HF in accuracy for the two barrier height datasets, showing that factors other than density quality or delocalization error are playing an important role in reducing error. 

A third possible contributor is the role of non-Brillouin singles in MP:DFT. 
MP2 non-Brillouin singles are a major contributor to the worse performance of MP2:DFT relative to MP2:HF (as shown in the Supporting Information), via extra correlation. However, exclusion of singles leads to significantly poorer MP3:DFT performance and considerably greater variation over different functionals. It thus appears that a subtle partial cancellation between MP2 singles and MP3 doubles is responsible for a significant reduction of error at the MP3:DFT level for many functionals, especially LSDA/GGAs. This cancellation also appears to be responsible for the low functional dependence of (s)MP3:DFT. However it is also worth noting that hybrid functionals (especially range separated ones) continue to have lower MP3:DFT errors than MP3:HF even in the absence of MP2 singles, and thus presence of singles cannot be the main factor behind improved performance of MP:DFT for these functionals. Indeed, it does not appear that there is a single obvious factor that is responsible for the greater accuracy of MP:DFT over MP:HF. It is almost easier to conclude that HF orbitals are uniquely poor references for quantitative accuracy with low order MP theory, and almost any DFT approach (or $\kappa-$OOMP2)  can do significantly better (especially if a close eye is kept on delocalization error or other qualitative failures). 

\section{Behavior at the Complete Basis Set Limit}\label{sec:cbs}

We next gauge the practical utility of MP:$\omega$B97M-V by comparison to various density functionals and MP:HF. Comparison to DFT is only meaningful at the CBS limit, due to different basis set convergence rates of MP and DFT. The datasets considered are the nonMR section of W4-11, 24 diverse barrier heights (DBH24\cite{karton2008highly}, a subset of HTBH38 and NHTBH38) and TA13. These datasets also contain post CCSD(T) corrections to benchmark values (at least to the CCSDT(Q) level), although such effects could be small in practice (especially for TA13). The $c_2$ coefficient for sMP2 and $c_3$ coefficient for sMP3 were reparameterized by fitting to the original benchmark values\cite{Karton2011} of the nonMR W4-11 subset, though the resulting values of 0.8425 and 0.7679 are not too different from the values found from fitting to CCSD(T) for aug-cc-pVTZ (0.8412 and 0.8012 respectively, given in Table \ref{tab:fit_coeffs}).  The corresponding $c_2$ and $c_3$ values for MP:HF are 0.9007 and 0.7060 respectively, again quite close to the values in Table \ref{tab:fit_coeffs} (0.9035 and 0.7157).

\begin{table}[htb!]
\begin{tabular}{|l|rrr|rrr|rrr|}
\hline
\textbf{}           & \multicolumn{3}{c}{\textbf{nonMR W4-11}}          & \multicolumn{3}{c}{\textbf{DBH24}}          & \multicolumn{3}{c}{\textbf{TA13}}\vline           \\\hline
\textbf{Method}     & \textbf{RMSE} & \textbf{MSE} & \textbf{MAX} & \textbf{RMSE} & \textbf{MSE} & \textbf{MAX} & \textbf{RMSE} & \textbf{MSE} & \textbf{MAX} \\ \hline
MP2:$\omega$B97M-V  & 12.2 & -4.1 & 53.5 & 5.0 & -2.5 & 17.1 & 2.42 & -1.36 & 7.22 \\
sMP2:$\omega$B97M-V & 5.3  & -0.8 & 25.4 & 2.7 & 0.0  & 6.5  & 0.99 & -0.14 & 2.55 \\
MP3:$\omega$B97M-V & 4.2  & 0.1  & 17.7 & 1.8 & 0.7  & 6.9  & 0.79 & -0.58 & 1.74 \\
sMP3:$\omega$B97M-V & 2.2  & -0.9 & 8.4  & 0.8 & 0.0  & 1.4  & 1.10 & -0.76 & 3.01\\ \hline
MP2:HF & 12.8 & -5.4 & 58.2 & 7.2 & 4.8 & 19.0 & 1.97 & 0.07 & 5.81 \\
sMP2:HF & 11.2 & -3.0 & 50.1 & 7.4 & 5.4 & 18.1 & 1.79 & 0.47 & 4.24 \\
MP3:HF & 10.2 & -1.0 & 43.2 & 6.5 & 5.1 & 17.4 & 1.51 & 0.44 & 4.23 \\
sMP3:HF & 9.6  & -2.3 & 44.2 & 6.4 & 5.0 & 14.2 & 1.59 & 0.33 & 4.03\\ \hline
BLYP-D3(BJ)         & 6.7           & 0.6          & 31.8         & 9.6           & -8.3         & 21.3         & 6.10          & -3.99        & 15.62        \\
PBE-D3(BJ)          & 10.0          & -2.2         & 42.4         & 10.4          & -8.4         & 29.7         & 6.54          & -4.66        & 15.98        \\
SCAN-D3(BJ)         & 5.5           & -0.7         & 17.5         & 8.0           & -6.9         & 17.1         & 4.89          & -3.57        & 11.12        \\
B97M-V              & 3.8           & -1.2         & 26.7         & 5.0           & -3.7         & 12.1         & 4.12          & -2.49        & 9.02         \\
B3LYP-D3(BJ)        & 3.7           & 0.3          & 17.5         & 5.4           & -4.6         & 9.9          & 3.85          & -2.73        & 8.99         \\
PBE0-D3(BJ)         & 4.8           & -0.9         & 20.3         & 4.7           & -3.5         & 13.3         & 3.31          & -2.45        & 8.57         \\
$\omega$B97X-V      & 3.6           & 0.1          & 11.9         & 1.8           & -0.1         & 4.2          & 2.88          & -1.29        & 8.76         \\
$\omega$B97M-V      & 2.5           & -0.2         & 11.3         & 1.5           & -0.7         & 3.9          & 2.75          & -0.98        & 7.92         \\ 
$\omega$B97M(2)     &     1.8      &     -0.6     & 7.4         &0.7            &-0.3          & 1.4          &1.14          &-0.53         &3.37          \\ \hline
\end{tabular}
 \caption{Root mean square errors (RMSE), mean signed errors (MSE), and maximum absolute errors (MAX) in kcal/mol for MP:$\omega$B97M-V, MP:HF and various DFT approaches. $\omega$B97M(2)\cite{mardirossian2018survival}/CBS values were obtained in a similar manner as MP:$\omega$B97M-V/CBS (but without a frozen core correction as the functional was trained without them). Other DFT results and the reference values were obtained from Ref \citenum{mardirossian2017thirty}.}
 \label{tab:cbs}
\end{table}

Table \ref{tab:cbs} presents MP:DFT errors for these datasets, along with those of MP:HF and several classic and recent DFT functionals. DFT performs excellently on the nonMR W4-11 subset (as exemplified by the 2.5 kcal/mol RMSE of $\omega$B97M-V), which is relatively unsurprising\cite{mardirossian2017thirty} due to the lack of multreference species. On the other hand, MP:DFT is not particularly competitive, with sMP3:$\omega$B97M-V only marginally lowering the RMSE of reference functional despite much larger computational cost ($O(N^6)$ vs $O(N^4)$). MP3:$\omega$B97M-V performs even more poorly, having an RMSE larger than that of the local B97M-V functional. This appears to indicate that there is little practical sense in applying (s)MP3 to problems that are considered to be ``easy" for DFT.  The $\omega$B97M(2) functional\cite{mardirossian2018survival} performs the best with an RMSE of 1.8 kcal/mol, representing a very reliable approach for thermochemistry calculations with only $O(N^5)$ scaling cost. Nonetheless, (s)MP3:$\omega$B97M-V represents a significant increase in accuracy over (s)MP3:HF. 

Presence of delocalization error makes barrier heights considerably more challenging for DFT. Most functionals consequently show significant systematic underestimation of DBH24 barrier heights, bar the range separated (double) hybrid functionals. sMP3:$\omega$B97M-V has significantly lower RMSE than the reference $\omega$B97M-V functional, although MP3:$\omega$B97M-V fares slightly worse. It is also evident that MP:DFT leads to a considerable reduction in RMSE vs MP:HF. However, $\omega$B97M(2) performs equally well as sMP3:$\omega$B97M-V and is the more practical route for accurate barrier height computations for large systems. 

The TA13 dataset of radical-closed shell non-covalent interaction energies offers a clear scenario where MP:DFT is superior to existing hybrid DFT approaches, with sMP2:$\omega$B97M-V improving upon the RMSE of $\omega$B97M-V by nearly a factor of 3. Interestingly, MP3:$\omega$B97M-V is the best performer for this dataset, with sMP3 faring slightly worse (similar to behavior for aug-cc-pVTZ, as shown by Table \ref{tab:ta13_data}). Nonetheless, all MP:DFT methods perform very well, with even unscaled MP2:$\omega$B97M-V being marginally better than the reference functional. Of the DFT methods, only the $\omega$B97M(2) functional is competitive, having essentially the same RMSE as sMP3:$\omega$B97M-V. This is perhaps unsurprising on account of the good performance of sMP2:$\omega$B97M-V, as $\omega$B97M(2) is a more general xDH functional employing the same reference orbitals. 
The challenges faced by other DFT methods for this dataset are delocalization driven (as multireference character appears small, based on the small magnitude of post CCSD(T) corrections for this dataset\cite{Tentscher2013}), permitting MP (and double hybrid) approaches to be more effective. Indeed, routine MP:HF performs reasonably well for this dataset, having RMSEs that are considerably smaller than those predicted by hybrid functionals (though larger than sMP2/(s)MP3:$\omega$B97M-V). It thus appears that (s)MP3:DFT methods are mostly likely to be useful for problems with significant delocalization error, where even modern hybrid density functionals are significantly challenged.  However, the great accuracy of MP2 based modern double hybrid functionals likely makes them the more computationally efficient route for investigating such problems than (s)MP3:DFT. On the other hand, these results do seem to suggest that MP3 based double hybrid xDH functionals could potentially be even more accurate. 

\section{Conclusions and future directions}
In this work we have shown that the use of DFT orbitals yields significant improvement to MP2 and MP3 theory, over all functionals and datasets tested. In fact, the choice of the reference functional had surprisingly little overall impact on the error, although hybrid functionals with lower delocalization error appear to have an edge (especially for barrier heights).
The exception to this general rule is MP2:DFT, which overcorrelates more than standard MP2 due to presense of non-Brillouin singles, and thus performs worse in most cases (with improvements mostly arising from cases with significant spin-contamination in HF, such as in RSE43). Scaling of the MP2 correlation energy is however adequate for ameliorating the overcorrelation problem, with sMP2:DFT providing significant improvement relative to MP2:DFT over the studied datasets (often reducing RMSE by a factor of 2-3). Nonetheless, modern double hybrid density functionals\cite{mardirossian2018survival,santra2019minimally} are likely to offer even better accuracy than sMP2:DFT at the same asymptotic cost (as hinted at by Table \ref{tab:cbs}). It is also quite interesting that the $c_2$ scaling parameter obtained from fitting to W4-11 thermochemistry proved quite transferable across all datasets, indicating similar levels of MP2 overcorrelation throughout.

MP3:DFT also proved quite robust, with a typical 2-3 fold reduction in RMSE  over MP3:HF, even without any scaling. The $c_3$ scaling parameter obtained from W4-11 was  not highly transferable, with slight degradation of performance on going from MP3:DFT to sMP3:DFT in several cases (such as HTBH38). However, the degradation is typically small (0.1-0.2 kcal/mol) while the improvements for datasets like W4-11 and BH76RC are larger, making sMP3:DFT preferable to MP3:DFT for general use. Indeed, MP3:DFT and sMP3:DFT were found to reproduce the CCSD(T) benchmark better than CCSD for several datasets, suggesting that they are quite attractive as $O(N^6)$ scaling wave function methods (especially as multiple $O(N^6)$ iterations are not required, unlike CCSD). (s)MP3:DFT also significantly improves dipole moment predictions relative to MP3:HF, showing success at predicting molecular properties as well. The overall behavior remains mostly functional agnostic (on account of cancellation between MP3 doubles and non-Brillouin MP2 singles), with an edge for hybrid functionals with low delocalization error for challenging problems like barrier heights.  

It thus appears that DFT orbitals are better suited than HF orbitals for many practical applications of MP theory. Self-interaction error free $\kappa-$OOMP2 orbitals offer a similar improvement in performance, but the lower iterative cost of DFT ($O(N^4)$ for hybrid functionals) would reduce the overall computation time. Furthermore, the near universal availability of DFT and MP features in quantum chemistry packages (relative to OOMP2) would permit wide applicability of any DFT orbital based MP approach. We therefore recommend use of range-separated hybrid DFT or $\kappa-$OOMP2 orbitals over HF orbitals for practical use of MP theory. Such reference states would have low delocalization error and thus are likely to be widely applicable. 
While there have been many studies showing slow or erratic convergence of the MP series\cite{handy1985convergence,nobes1987slow,gill1988does,olsen1996surprising,leininger2000mo,cremer1996sixth,cremer2011moller}, it is noteworthy that all of them have used HF orbitals. It may well be interesting to revisit such problems using DFT orbitals to explore whether or not our promising results at 2nd and 3rd order are sustained to higher orders as well.

A purely wave function based MP approach however is unlikely to be competitive with density functional approaches for ground state computations, even with improved reference orbitals. Indeed, Table \ref{tab:cbs} indicates that MP:DFT is likely to only significantly improve upon hybrid DFT for problems with significant delocalization error, making DFT preferable for most problems. Furthermore, modern double hybrid functionals like $\omega$B97M(2) are quite competitive with (s)MP3:DFT for even the challenging cases, making them a computationally more efficient route. However, the good performance of sMP3 over sMP2 for many datasets seem to suggest that xDH double hybrid functionals with MP3 correlation could potentially significantly improve upon the best performing modern double hybrids.Alternatively, MP:DFT could also be employed in composite methods as a more accurate alternative to standard MP2/MP3\cite{semidalas2020canonical}.
MP2 geometries are also considered to be quite accurate\cite{curtiss1998gaussian}, indicating that MP:DFT could be a promising route for even more accurate geometries and frequencies at similar cost. This would entail development of analytical nuclear gradients that account for non-Hellman-Feynman terms stemming from lack of self-consistency, which we are currently exploring.

State-specific excited state computations offer another potential application for MP:DFT. Density functional theory based state-specific excited state approaches\cite{gilbert2008self,kowalczyk2011assessment,kowalczyk2013excitation,barca2018simple,hait2020excited} are challenged by the single-determinant nature of Kohn-Sham theory\cite{kohn1965self}. While reasonable recoupling protocols can be devised in certain cases\cite{ziegler1977calculation,frank1998molecular,hait2020accurate}, many states with significantly multiconfigurational nature remain inaccessible with DFT alone. However, excited state-specific MP2 approaches (based on HF orbitals) have also been fairly successful for problems with single configuration state functions\cite{carter2020state,ye2020self}.  
MP:DFT thus appears to be a promising route that could be employed for multiconfigurational problems, via employing non-orthogonal configuration interaction\cite{thom2009hartree} based recoupling between single reference states generated via MP\cite{yost2016size,yost2018efficient} from DFT optimized orbitals. Work along these directions is presently in progress.  

\section*{Acknowledgment} 
	This research was supported by the Director, Office of Science, Office of Basic Energy Sciences, of the U.S. Department of Energy under Contract No. DE-AC02-05CH11231.
	
\section*{Supporting Information}
Supporting information containing the data used above, as well as further studies not included in the main manuscript is provided:
\begin{enumerate}
    \item \textbf{mp3dft\_supporting\_information.pdf} - A document containing results for a fifth scaling model, results for sMP3 without non-Brilloin singles, and correlation coefficients between errors of sMP3:DFT and wave function theories.
    \item \textbf{mp3dft\_data.xslx} - A spreadsheet containing individual reaction energies for the each dataset and method considered.
\end{enumerate}

\bibliography{mp3}
\end{document}


\newpage

\section{Results without non-Brillouin MP2 singles}

\begin{table}[hbt!]
\vspace{-10pt}
    \centering

    \caption{Scaling coefficients for MP2:DFT and MP3:DFT (with no singles) fit to the non MR subset of W4-11 dataset. }
    \label{tab:nbscoeffs_data}
    \vspace{-25pt}
\end{table}

\begin{table}[hbt!]
    \centering
    \begin{tabular}{| c | S S S | S S S | S S S | S S S |}
        \hline
        \textbf{Name} & \textbf{RMSE} & \textbf{MSE} & \textbf{MAX} & \textbf{RMSE} & \textbf{MSE} & \textbf{MAX} & \textbf{RMSE} & \textbf{MSE} & \textbf{MAX} & \textbf{RMSE} & \textbf{MSE} & \textbf{MAX}
        \csvreader[before line=\ifthenelse{\equal{\Name}{CCSD}}{
                                    \\\hline
                                } { \ifthenelse{\equal{\Name}{HF}}{
                                            \\\hline
                                            & \multicolumn{3}{c|}{\textbf{MP2 (NS)}} & \multicolumn{3}{c|}{\textbf{sMP2 (NS)}} & \multicolumn{3}{c|}{\textbf{MP3 (NS)}} & \multicolumn{3}{c|}{\textbf{sMP3 (NS)}}\\\hline
                                        }{
                                            \\
                                        }
                                },
                    filter not strcmp={\Name}{$\kappa$-OOMP2},
                   head to column names]{newTables/W4-nbs.csv}{}{\ifthenelse{\equal{\Name}{CCSD}}{\Name &  \nbsmptwormse & \nbsmptwomse & \nbsmptwomaxae & & & & & & & & & }{\Name & \nbsmptwormse & \nbsmptwomse & \nbsmptwomaxae &  \fitnbsmptwormse & \fitnbsmptwomse & \fitnbsmptwomaxae & \nbsmpthreermse & \nbsmpthreemse & \nbsmpthreemaxae & \fitnbsmpthreermse & \fitnbsmpthreemse & \fitnbsmpthreemaxae}}\\
        \hline
    \end{tabular}
    \caption{Root mean square errors (RMSE), mean signed errors (MSE), and maximum absolute errors (MAX) in kcal/mol for MP:DFT (with no singles) on the non MR subset of W4-11 dataset (training set). }
    \label{tab:w4nbs_data}
\end{table}

\begin{table}[hbt!]
    \centering
    \begin{tabular}{| c | S S S | S S S | S S S | S S S |}
        \hline
        \textbf{Name} & \textbf{RMSE} & \textbf{MSE} & \textbf{MAX} & \textbf{RMSE} & \textbf{MSE} & \textbf{MAX} & \textbf{RMSE} & \textbf{MSE} & \textbf{MAX} & \textbf{RMSE} & \textbf{MSE} & \textbf{MAX}
        \csvreader[before line=\ifthenelse{\equal{\Name}{CCSD}}{
                                    \\\hline
                                } { \ifthenelse{\equal{\Name}{HF}}{
                                            \\\hline
                                            & \multicolumn{3}{c|}{\textbf{MP2 (NS)}} & \multicolumn{3}{c|}{\textbf{sMP2 (NS)}} & \multicolumn{3}{c|}{\textbf{MP3 (NS)}} & \multicolumn{3}{c|}{\textbf{sMP3 (NS)}}\\\hline
                                        }{
                                            \\
                                        }
                                },
                    filter not strcmp={\Name}{$\kappa$-OOMP2},
                   head to column names]{newTables/BH76RC.csv}{}{\ifthenelse{\equal{\Name}{CCSD}}{\Name &  \nbsmptwormse & \nbsmptwomse & \nbsmptwomaxae & & & & & & & & & }{\Name & \nbsmptwormse & \nbsmptwomse & \nbsmptwomaxae &  \fitnbsmptwormse & \fitnbsmptwomse & \fitnbsmptwomaxae & \nbsmpthreermse & \nbsmpthreemse & \nbsmpthreemaxae & \fitnbsmpthreermse & \fitnbsmpthreemse & \fitnbsmpthreemaxae}}\\
        \hline
    \end{tabular}
    \caption{Root mean square errors (RMSE), mean signed errors (MSE), and maximum absolute errors (MAX) in kcal/mol for MP:DFT (with no singles) on the BH76RC dataset. }
    \label{tab:bh76rc_data}
\end{table}

\begin{table}[hbt!]
    \centering
    \begin{tabular}{| c | S S S | S S S | S S S | S S S |}
        \hline
        \textbf{Name} & \textbf{RMSE} & \textbf{MSE} & \textbf{MAX} & \textbf{RMSE} & \textbf{MSE} & \textbf{MAX} & \textbf{RMSE} & \textbf{MSE} & \textbf{MAX} & \textbf{RMSE} & \textbf{MSE} & \textbf{MAX}
        \csvreader[before line=\ifthenelse{\equal{\Name}{CCSD}}{
                                    \\\hline
                                } { \ifthenelse{\equal{\Name}{HF}}{
                                            \\\hline
                                            & \multicolumn{3}{c|}{\textbf{MP2 (NS)}} & \multicolumn{3}{c|}{\textbf{sMP2 (NS)}} & \multicolumn{3}{c|}{\textbf{MP3 (NS)}} & \multicolumn{3}{c|}{\textbf{sMP3 (NS)}}\\\hline
                                        }{
                                            \\
                                        }
                                },
                    filter not strcmp={\Name}{$\kappa$-OOMP2},
                   head to column names]{newTables/RSE43.csv}{}{\ifthenelse{\equal{\Name}{CCSD}}{\Name &  \nbsmptwormse & \nbsmptwomse & \nbsmptwomaxae & & & & & & & & & }{\Name & \nbsmptwormse & \nbsmptwomse & \nbsmptwomaxae &  \fitnbsmptwormse & \fitnbsmptwomse & \fitnbsmptwomaxae & \nbsmpthreermse & \nbsmpthreemse & \nbsmpthreemaxae & \fitnbsmpthreermse & \fitnbsmpthreemse & \fitnbsmpthreemaxae}}\\
        \hline
    \end{tabular}
    \caption{Root mean square errors (RMSE), mean signed errors (MSE), and maximum absolute errors (MAX) in kcal/mol for MP:DFT (with no singles) on the RSE43 dataset. }
    \label{tab:rse43_data}
\end{table}

\begin{table}[hbt!]
    \centering
    \begin{tabular}{| c | S S S | S S S | S S S | S S S |}
        \hline
        \textbf{Name} & \textbf{RMSE} & \textbf{MSE} & \textbf{MAX} & \textbf{RMSE} & \textbf{MSE} & \textbf{MAX} & \textbf{RMSE} & \textbf{MSE} & \textbf{MAX} & \textbf{RMSE} & \textbf{MSE} & \textbf{MAX}
        \csvreader[before line=\ifthenelse{\equal{\Name}{CCSD}}{
                                    \\\hline
                                } { \ifthenelse{\equal{\Name}{HF}}{
                                            \\\hline
                                            & \multicolumn{3}{c|}{\textbf{MP2 (NS)}} & \multicolumn{3}{c|}{\textbf{sMP2 (NS)}} & \multicolumn{3}{c|}{\textbf{MP3 (NS)}} & \multicolumn{3}{c|}{\textbf{sMP3 (NS)}}\\\hline
                                        }{
                                            \\
                                        }
                                },
                    filter not strcmp={\Name}{$\kappa$-OOMP2},
                   head to column names]{newTables/HTBH38.csv}{}{\ifthenelse{\equal{\Name}{CCSD}}{\Name &  \nbsmptwormse & \nbsmptwomse & \nbsmptwomaxae & & & & & & & & & }{\Name & \nbsmptwormse & \nbsmptwomse & \nbsmptwomaxae &  \fitnbsmptwormse & \fitnbsmptwomse & \fitnbsmptwomaxae & \nbsmpthreermse & \nbsmpthreemse & \nbsmpthreemaxae & \fitnbsmpthreermse & \fitnbsmpthreemse & \fitnbsmpthreemaxae}}\\
        \hline
    \end{tabular}
    \caption{Root mean square errors (RMSE), mean signed errors (MSE), and maximum absolute errors (MAX) in kcal/mol for MP:DFT (with no singles) on the HTBH38 dataset. }
    \label{tab:htbh38_data}
\end{table}

\begin{table}[hbt!]
    \centering
    \begin{tabular}{| c | S S S | S S S | S S S | S S S |}
        \hline
        \textbf{Name} & \textbf{RMSE} & \textbf{MSE} & \textbf{MAX} & \textbf{RMSE} & \textbf{MSE} & \textbf{MAX} & \textbf{RMSE} & \textbf{MSE} & \textbf{MAX} & \textbf{RMSE} & \textbf{MSE} & \textbf{MAX}
        \csvreader[before line=\ifthenelse{\equal{\Name}{CCSD}}{
                                    \\\hline
                                } { \ifthenelse{\equal{\Name}{HF}}{
                                            \\\hline
                                            & \multicolumn{3}{c|}{\textbf{MP2 (NS)}} & \multicolumn{3}{c|}{\textbf{sMP2 (NS)}} & \multicolumn{3}{c|}{\textbf{MP3 (NS)}} & \multicolumn{3}{c|}{\textbf{sMP3 (NS)}}\\\hline
                                        }{
                                            \\
                                        }
                                },
                    filter not strcmp={\Name}{$\kappa$-OOMP2},
                   head to column names]{newTables/NHTBH38.csv}{}{\ifthenelse{\equal{\Name}{CCSD}}{\Name &  \nbsmptwormse & \nbsmptwomse & \nbsmptwomaxae & & & & & & & & & }{\Name & \nbsmptwormse & \nbsmptwomse & \nbsmptwomaxae &  \fitnbsmptwormse & \fitnbsmptwomse & \fitnbsmptwomaxae & \nbsmpthreermse & \nbsmpthreemse & \nbsmpthreemaxae & \fitnbsmpthreermse & \fitnbsmpthreemse & \fitnbsmpthreemaxae}}\\
        \hline
    \end{tabular}
    \caption{Root mean square errors (RMSE), mean signed errors (MSE), and maximum absolute errors (MAX) in kcal/mol for MP:DFT (with no singles) on the NHTBH38 dataset. }
    \label{tab:nhtbh38_data}
\end{table}

\begin{table}[hbt!]
    \centering
    \begin{tabular}{| c | S[round-precision=2] S[round-precision=2] S[round-precision=2] | S[round-precision=2] S[round-precision=2] S[round-precision=2] | S[round-precision=2] S[round-precision=2] S[round-precision=2] | S[round-precision=2] S[round-precision=2] S[round-precision=2] |}
        \hline
        \textbf{Name} & \textbf{RMSE} & \textbf{MSE} & \textbf{MAX} & \textbf{RMSE} & \textbf{MSE} & \textbf{MAX} & \textbf{RMSE} & \textbf{MSE} & \textbf{MAX} & \textbf{RMSE} & \textbf{MSE} & \textbf{MAX}
        \csvreader[before line=\ifthenelse{\equal{\Name}{CCSD}}{
                                    \\\hline
                                } { \ifthenelse{\equal{\Name}{HF}}{
                                            \\\hline
                                            & \multicolumn{3}{c|}{\textbf{MP2 (NS)}} & \multicolumn{3}{c|}{\textbf{sMP2 (NS)}} & \multicolumn{3}{c|}{\textbf{MP3 (NS)}} & \multicolumn{3}{c|}{\textbf{sMP3 (NS)}}\\\hline
                                        }{
                                            \\
                                        }
                                },
                    filter not strcmp={\Name}{$\kappa$-OOMP2},
                   head to column names]{newTables/TA13.csv}{}{\ifthenelse{\equal{\Name}{CCSD}}{\Name &  \nbsmptwormse & \nbsmptwomse & \nbsmptwomaxae & & & & & & & & & }{\Name & \nbsmptwormse & \nbsmptwomse & \nbsmptwomaxae &  \fitnbsmptwormse & \fitnbsmptwomse & \fitnbsmptwomaxae & \nbsmpthreermse & \nbsmpthreemse & \nbsmpthreemaxae & \fitnbsmpthreermse & \fitnbsmpthreemse & \fitnbsmpthreemaxae}}\\
        \hline
    \end{tabular}
    \caption{Root mean square errors (RMSE), mean signed errors (MSE), and maximum absolute errors (MAX) in kcal/mol for MP:DFT (with no singles) on the TA13 dataset. }
    \label{tab:ta13_data}
\end{table}

\begin{table}[hbt!]
    \centering
    \begin{tabular}{| c | S[round-precision=2] S[round-precision=2] S[round-precision=2] | S[round-precision=2] S[round-precision=2] S[round-precision=2] | S[round-precision=2] S[round-precision=2] S[round-precision=2] | S[round-precision=2] S[round-precision=2] S[round-precision=2] |}
        \hline
        \textbf{Name} & \textbf{RMSE} & \textbf{MSE} & \textbf{MAX} & \textbf{RMSE} & \textbf{MSE} & \textbf{MAX} & \textbf{RMSE} & \textbf{MSE} & \textbf{MAX} & \textbf{RMSE} & \textbf{MSE} & \textbf{MAX}
        \csvreader[before line=\ifthenelse{\equal{\Name}{CCSD}}{
                                    \\\hline
                                } { \ifthenelse{\equal{\Name}{HF}}{
                                            \\\hline
                                            & \multicolumn{3}{c|}{\textbf{MP2 (NS)}} & \multicolumn{3}{c|}{\textbf{sMP2 (NS)}} & \multicolumn{3}{c|}{\textbf{MP3 (NS)}} & \multicolumn{3}{c|}{\textbf{sMP3 (NS)}}\\\hline
                                        }{
                                            \\
                                        }
                                },
                    filter not strcmp={\Name}{$\kappa$-OOMP2},
                   head to column names]{newTables/A24.csv}{}{\ifthenelse{\equal{\Name}{CCSD}}{\Name &  \nbsmptwormse & \nbsmptwomse & \nbsmptwomaxae & & & & & & & & & }{\Name & \nbsmptwormse & \nbsmptwomse & \nbsmptwomaxae &  \fitnbsmptwormse & \fitnbsmptwomse & \fitnbsmptwomaxae & \nbsmpthreermse & \nbsmpthreemse & \nbsmpthreemaxae & \fitnbsmpthreermse & \fitnbsmpthreemse & \fitnbsmpthreemaxae}}\\
        \hline
    \end{tabular}
    \caption{Root mean square errors (RMSE), mean signed errors (MSE), and maximum absolute errors (MAX) in kcal/mol for MP:DFT (with no singles) on the A24 dataset. }
    \label{tab:a24_data}
\end{table}

\newpage 
\ 
\newpage
\ 
\newpage
\ 
\newpage

\section{Behavior of Model 5 $(E=E_\textrm{HF}+c_2E_\textrm{MP2}+c_3E_\textrm{MP3})$}
\vspace{-30pts}
\begin{table}[hbt!]
    \centering
    \begin{tabular}{| c | S[round-precision=4] S[round-precision=4] |}
        \hline
        \textbf{Name} & \textbf{c$_2$} & \textbf{c$_3$}
        \csvreader[before line=\ifthenelse{\equal{\Name}{$\kappa$-OOMP2}}{\\\hline}{\\}, filter test= \ifcsvnotstrcmp{\Name}{CCSD}, head to column names]{newTables/W4.csv}{}{\Name &  \scamptwothreectwo & \scamptwothreecthree}\\
        \hline
    \end{tabular}
    \caption{Scaling coefficients for model 5 fit from the non MR subset of W4-11 dataset. }
    \label{tab:nbscoeffs_data}
\end{table}

\DTLloaddb{w4}{newTables/W4.csv}
\DTLloaddb{bh76rc}{newTables/BH76RC.csv}
\DTLloaddb{rse43}{newTables/RSE43.csv}
\dtlexpandnewvalue

\begin{table}[hbt!]
    \centering
    \begin{tabular}{| c | SSS | SSS | SSS |@{}c@{}}
        \cline{1-10}
         & \multicolumn{3}{c}{\textbf{W4-11}} & \multicolumn{3}{|c}{\textbf{BH76RC}} & \multicolumn{3}{|c|}{\textbf{RSE43}}& \\\cline{1-10}
        \textbf{Name} & \textbf{RMSE} & \textbf{MSE} & \textbf{MAX} & \textbf{RMSE} & \textbf{MSE} & \textbf{MAX} &  \textbf{RMSE} & \textbf{MSE} & \textbf{MAX} &
        \DTLforeach*{bh76rc}{\name=Name, \rmsetwo=scamptwothreermse, \msetwo=scamptwothreemse,  \maxtwo=scamptwothreemaxae}{
            \DTLgetvalueforkey{\rmseone}{scamptwothreermse}{w4}{Name}{\name}
            \DTLgetvalueforkey{\mseone}{scamptwothreemse}{w4}{Name}{\name}
            \DTLgetvalueforkey{\maxone}{scamptwothreemaxae}{w4}{Name}{\name}            \DTLgetvalueforkey{\rmsethree}{scamptwothreermse}{rse43}{Name}{\name}
            \DTLgetvalueforkey{\msethree}{scamptwothreemse}{rse43}{Name}{\name}
            \DTLgetvalueforkey{\maxthree}{scamptwothreemaxae}{rse43}{Name}{\name}
            \\\name & \rmseone & \mseone & \maxone & \rmsetwo & \msetwo & \maxtwo & 
            \rmsethree & \msethree & \maxthree &
        }
        \\\cline{1-10}
    \end{tabular}
    \caption{Root mean square errors (RMSE), mean signed errors (MSE), and maximum absolute errors (MAX) in kcal/mol for MP:DFT Model 5 and CCSD on the thermochemistry datasets. }
    \label{tab:a24_data}
\end{table}

\DTLloaddb{htbh38}{newTables/HTBH38.csv}
\DTLloaddb{nhtbh38}{newTables/NHTBH38.csv}
\dtlexpandnewvalue

\begin{table}[hbt!]
    \centering
    \begin{tabular}{|c|SSS|SSS|@{}c@{}}
        \cline{1-7}
         & \multicolumn{3}{c}{\textbf{HTBH38}} & \multicolumn{3}{|c|}{\textbf{NHTBH38}}& \\\cline{1-7}
        \textbf{Name} & \textbf{RMSE} & \textbf{MSE} & \textbf{MAX} & \textbf{RMSE} & \textbf{MSE} & \textbf{MAX} &
        \DTLforeach*{htbh38}{\name=Name, \rmseone=scamptwothreermse, \mseone=scamptwothreemse,  \maxone=scamptwothreemaxae}{
            \DTLgetvalueforkey{\rmsetwo}{scamptwothreermse}{nhtbh38}{Name}{\name}
            \DTLgetvalueforkey{\msetwo}{scamptwothreemse}{nhtbh38}{Name}{\name}
            \DTLgetvalueforkey{\maxtwo}{scamptwothreemaxae}{nhtbh38}{Name}{\name}
            \\\name & \rmseone & \mseone & \maxone & \rmsetwo & \msetwo & \maxtwo &
        }
        \\\cline{1-7}
    \end{tabular}
    \caption{Root mean square errors (RMSE), mean signed errors (MSE), and maximum absolute errors (MAX) in kcal/mol for MP:DFT Model 5 and CCSD  on the kinetics datasets. }
    \label{tab:a24_data}
\end{table}

\DTLloaddb{a24}{newTables/A24.csv}
\DTLloaddb{ta13}{newTables/TA13.csv}
\dtlexpandnewvalue

\begin{table}[hbt!]
    \centering
    \begin{tabular}{| c | S[round-precision=2] S[round-precision=2] S[round-precision=2] | S[round-precision=2] S[round-precision=2] S[round-precision=2]|@{}c@{}}
        \cline{1-7}
         & \multicolumn{3}{c}{\textbf{TA13}} & \multicolumn{3}{|c|}{\textbf{A24}}& \\\cline{1-7}
        \textbf{Name} & \textbf{RMSE} & \textbf{MSE} & \textbf{MAX} & \textbf{RMSE} & \textbf{MSE} & \textbf{MAX} &
        \DTLforeach*{ta13}{\name=Name, \rmseone=scamptwothreermse, \mseone=scamptwothreemse,  \maxone=scamptwothreemaxae}{
            \DTLgetvalueforkey{\rmsetwo}{scamptwothreermse}{a24}{Name}{\name}
            \DTLgetvalueforkey{\msetwo}{scamptwothreemse}{a24}{Name}{\name}
            \DTLgetvalueforkey{\maxtwo}{scamptwothreemaxae}{a24}{Name}{\name}
            \\\name & \rmseone & \mseone & \maxone & \rmsetwo & \msetwo & \maxtwo &
        }
        \\\cline{1-7}
    \end{tabular}
    \caption{Root mean square errors (RMSE), mean signed errors (MSE), and maximum absolute errors (MAX) in kcal/mol for MP:DFT Model 5 and CCSD on the non-covalent interaction datasets. }
    \label{tab:a24_data}
\end{table}

\DTLloaddb{dipolesnsp}{newTables/DIPoles_nsp.csv}
\DTLloaddb{dipolessp}{newTables/DIPoles_sp.csv}
\dtlexpandnewvalue

\begin{table}[hbt!]
    \centering
    \begin{tabular}{| c | S S | S S |@{}c@{}}
        \cline{1-5}
         & \multicolumn{2}{c}{\textbf{NSP}} & \multicolumn{2}{|c|}{\textbf{SP}}& \\\cline{1-5}
        \textbf{Name} & \textbf{RMSRE} & \textbf{MSRE} & \textbf{RMSRE} & \textbf{MSRE} &
        \DTLforeach*{dipolesnsp}{\name=Name, \rmseone=scamptwothreermse, \mseone=scamptwothreemse}{
            \DTLgetvalueforkey{\rmsetwo}{scamptwothreermse}{dipolessp}{Name}{\name}
            \DTLgetvalueforkey{\msetwo}{scamptwothreemse}{dipolessp}{Name}{\name}
            \\\name & \rmseone & \mseone &  \rmsetwo & \msetwo & 
        }
        \\\cline{1-5}
    \end{tabular}
    \caption{Root mean square regularized errors (RMSRE) and mean signed regularized errors (MSRE) in $\%$ for MP:DFT Model 5 and CCSD on the dipoles datasets. }
    \label{tab:a24_data}
\end{table}

\newpage  

\section{sMP3 Correlation Coefficients}

\begin{table}[htb!]
    \centering
    \begin{tabular}{|c|SSS|}
        \hline
        \textbf{Name} & \textbf{CCSD} & \textbf{HF} & \textbf{HFLYP} \\
\end{tabular}
    \caption{Correlation coefficients ($r$) computed between sMP3:DFT and CCSD, sMP3:HF, sMP3:HFLYP errors for the nonMR subset of the W4-11 dataset.}
    \label{tab:w4_corr}
\end{table}


\DTLloaddb[noheader, omitlines=2]{rsecorr}{fullCorrelationPlots/RSE43_corr.csv}
\dtlexpandnewvalue

\begin{table}[hbt!]
    \centering
    \begin{tabular}{| c | SS | SS |@{}c@{}}
        \cline{1-5}
         & \multicolumn{2}{c}{\textbf{BH76RC}} & \multicolumn{2}{|c|}{\textbf{RSE43}}& \\\cline{1-5}
        \textbf{Name} & \textbf{CCSD} & \textbf{HF} & \textbf{CCSD} & \textbf{HF} &
        \DTLforeach*{bh76corr}{\name=Column1,\ccsdbh=Column2, \hfbh=Column4}{
            \DTLgetvalueforkey{\ccsdrse}{Column2}{rsecorr}{Column1}{\name}
            \DTLgetvalueforkey{\hfrse}{Column4}{rsecorr}{Column1}{\name}
            \\\name & \ccsdbh & \hfbh & \ccsdrse & \hfrse & 
        }
        \\\cline{1-5}
    \end{tabular}
    \caption{Correlation coefficients ($r$) computed between sMP3:DFT and CCSD,sMP3:HF errors for the BH76RC and RSE43 datasets. }
    \label{tab:thermo_corr}
\end{table}


\DTLloaddb[noheader, omitlines=2]{nhtbhcorr}{fullCorrelationPlots/NHTBH38_corr.csv}
\dtlexpandnewvalue

\begin{table}[hbt!]
    \centering
    \begin{tabular}{| c | SS | SS |@{}c@{}}
        \cline{1-5}
         & \multicolumn{2}{c}{\textbf{HTBH38}} & \multicolumn{2}{|c|}{\textbf{NHTBH38}}& \\\cline{1-5}
        \textbf{Name} & \textbf{CCSD} & \textbf{HF} & \textbf{CCSD} & \textbf{HF} &
        \DTLforeach*{htbhcorr}{\name=Column1,\ccsdhtbh=Column2, \hfhtbh=Column4}{
            \DTLgetvalueforkey{\ccsdnhtbh}{Column2}{nhtbhcorr}{Column1}{\name}
            \DTLgetvalueforkey{\hfnhtbh}{Column4}{nhtbhcorr}{Column1}{\name}
            \\\name & \ccsdhtbh & \hfhtbh & \ccsdnhtbh & \hfnhtbh & 
        }
        \\\cline{1-5}
    \end{tabular}
    \caption{Correlation coefficients ($r$) computed between sMP3:DFT and CCSD,sMP3:HF errors for the HTBH38 and NHTBH38 datasets. }
    \label{tab:kinetics_corr}
\end{table}


\DTLloaddb[noheader, omitlines=2]{a24corr}{fullCorrelationPlots/A24_corr.csv}
\dtlexpandnewvalue

\begin{table}[hbt!]
    \centering
    \begin{tabular}{| c | SS | SS |@{}c@{}}
        \cline{1-5}
         & \multicolumn{2}{c}{\textbf{TA13}} & \multicolumn{2}{|c|}{\textbf{A24}}& \\\cline{1-5}
        \textbf{Name} & \textbf{CCSD} & \textbf{HF} & \textbf{CCSD} & \textbf{HF} &
        \DTLforeach*{ta13corr}{\name=Column1,\ccsdbh=Column2, \hfbh=Column4}{
            \DTLgetvalueforkey{\ccsdrse}{Column2}{a24corr}{Column1}{\name}
            \DTLgetvalueforkey{\hfrse}{Column4}{a24corr}{Column1}{\name}
            \\\name & \ccsdbh & \hfbh & \ccsdrse & \hfrse & 
        }
        \\\cline{1-5}
    \end{tabular}
    \caption{Correlation coefficients ($r$) computed between sMP3:DFT and CCSD,sMP3:HF errors for the TA13 and A24 datasets. }
    \label{tab:nci_corr}
\end{table}


\DTLloaddb[noheader, omitlines=2]{dipspcorr}{fullCorrelationPlots/DIPoles_sp.csv}
\dtlexpandnewvalue

\begin{table}[hbt!]
    \centering
    \begin{tabular}{| c | SS | SS |@{}c@{}}
        \cline{1-5}
         & \multicolumn{2}{c}{\textbf{NSP}} & \multicolumn{2}{|c|}{\textbf{SP}}& \\\cline{1-5}
        \textbf{Name} & \textbf{CCSD} & \textbf{HF} & \textbf{CCSD} & \textbf{HF} &
        \DTLforeach*{dipnspcorr}{\name=Column1,\ccsdbh=Column2, \hfbh=Column4}{
            \DTLgetvalueforkey{\ccsdrse}{Column2}{dipspcorr}{Column1}{\name}
            \DTLgetvalueforkey{\hfrse}{Column4}{dipspcorr}{Column1}{\name}
            \\\name & \ccsdbh & \hfbh & \ccsdrse & \hfrse & 
        }
        \\\cline{1-5}
    \end{tabular}
    \caption{Correlation coefficients ($r$) computed between sMP3:DFT and CCSD,sMP3:HF errors for the not spin-polarized (NSP) and spin-polarized (SP) subsets of the dipoles datasets. }
    \label{tab:dips_corr}
\end{table}

\newpage
\
\newpage
\
\newpage

\section{RSE43 Spin Contamination}

\begin{table}[htb!]
    \centering
    
    %
    \caption{Root mean square errors (RMSE), mean signed errors (MSE), and maximum absolute errors (MAX) in kcal/mol for MP:DFT on the RSE43 dataset with spin-contaminated reactions removed. }
    \label{tab:rse43_data}
\end{table}

\subsection{Spin-contaminated cases in RSE43}

\begin{enumerate}
    \item Singlet species with UHF $\langle S^2 \rangle >0$: toluene, acrylonitrile and nitromethane. 
    \item Doublet radicals with UHF $\langle S^2 \rangle > 0.9$: allyl, acetaldehyde radical (vinoxy), acetonitrile radical, nitromethyl radical, benzyl, acrylonitrile radical, aminoacetonitrile radical, propargyl and 1-fluorovinyl. 
\end{enumerate}